\documentclass[fleqn,usenatbib]{mnras}

\usepackage[T1]{fontenc}
\usepackage{ae,aecompl}

\DeclareRobustCommand{\VAN}[3]{#2}
\let\VANthebibliography\thebibliography
\def\thebibliography{\DeclareRobustCommand{\VAN}[3]{##3}\VANthebibliography}

\usepackage{graphicx}	
\usepackage{amsmath}	
\usepackage{epstopdf}
\usepackage{adjustbox}
\usepackage{textcomp, gensymb}
\usepackage{subcaption} 
\usepackage{tabu}
\usepackage{float}
\usepackage{rotfloat}
\usepackage{rotating}
\usepackage{multirow}
\usepackage{multicol}
\usepackage{pdflscape} 
\usepackage{breqn} 
\usepackage[flushleft]{threeparttable}
\usepackage{orcidlink}
\usepackage{array}
\usepackage{amsmath}
\newcommand{\plus}{\scalebox{1.1}{+}}
\newcommand{\ie}{\it i.e.,}
\newcommand{\eg}{\it e.g.,}

\title[Short title, max. 45 characters]{X-ray spectral properties of dust$-$obscured galaxies}

\title{X$-$ray spectral properties of dust$-$obscured galaxies in the XMM$-$SERVS coverage of the XMM-LSS field}

\author[Kayal et al.]{
Abhijit Kayal $^{\orcidlink{0000-0001-9851-8243}}$,$^{1,2}$\thanks{E-mail: abhijitk@prl.res.in}
Veeresh Singh $^{\orcidlink{0000-0002-6040-4993}}$,$^{1}$
\\ \\
$^{1}$Physical Research Laboratory, Navrangpura, Ahmedabad, Gujarat-380 009, India\\
$^{2}$Indian Institute of Technology Gandhinagar, Palaj, Gandhinagar, Gujarat-382 355, India \\
}

\date{Accepted 2024 May 1. Received 2024 April 19; in original form 2023 September 1}

\pubyear{2024}

\begin{document}
\label{firstpage}
\pagerange{\pageref{firstpage}--\pageref{lastpage}}
\maketitle

\begin{abstract}
With an aim to unveil the population of obscured AGN hosted in high$-z$ dust$-$obscured galaxies (DOGs),
we performed X-ray spectral study of 34 DOGs (0.59 $\leq$ $z$ $\leq$ 4.65)
lying within 5.3~deg$^{2}$ of the XMM$-$SERVS coverage in the XMM-LSS field.
To improve the spectral quality of individual sources,
we combined all the existing {\em XMM$-$Newton} data and also included {\em Chandra}/ACIS data, whenever available.
We find that the X-ray spectra of our DOGs can be fitted with a simple absorbed power law or with a physically-motivated {\scshape borus02} model. The line-of-sight column densities ($N_{\rm H}$) in our sources span across
a wide range (1.02 $\times$ 10$^{22}$~cm$^{-2}$ $\leq$ $N_{\rm H}$ $\leq$ 1.21 $\times$ 10$^{24}$~cm$^{-2}$), with a substantial
fraction ($\sim$ 17.6 per cent) of them being heavily obscured ($N_{\rm H}$ $\geq$ 10$^{23}$~cm$^{-2}$).
We also identified one new CT-AGN candidate, yielding the CT-AGN fraction in our sample to be only 3 per cent.
The absorption-corrected 2.0$-$10 keV X-ray luminosities of our sources
($2.00~\times~10^{43}$ erg~s$^{-1}$ $\leq$ $L_{\rm 2-10~keV}^{\rm int}$ $\leq$ $6.17~\times~10^{45}$ erg~s$^{-1}$) suggest them to be luminous quasars.
The $N_{\rm H}$ versus Eddington ratio diagnostic plot infers that our sample consists of a heterogeneous population
that includes a small fraction ($\sim$ 12 per cent) of DOGs belonging to an early phase (Hot DOGs) during which accretion and obscuration peaks, while the remaining DOGs belong to an intermediate or late phase during which radiative feedback from the dominant
AGN blows away surrounding obscuring material.
\end{abstract}
\begin{keywords}
galaxies: active --- quasars: general --- X-rays: galaxies --- galaxies: evolution --- methods: observational
\end{keywords}



\section{Introduction}
\label{sec:intro}
Active Galactic Nuclei (AGN) in the local universe are known to be obscured by circumnuclear material distributed in
the form of a torus \citep{Bianchi12,Ricci17,Zhao21} as envisaged by the AGN unification model \citep{Antonucci85,Urry95}.
Depending on the orientation of the obscuring torus, AGN can be broadly classified into type~1 (pole-on view) and type~2
(edge-on view). As expected, type~2 AGN exhibit higher line-of-sight column densities ($N_{\rm H,~LOS}$ $\sim$ 10$^{22}$ $-$ 10$^{24}$ cm$^{-2}$ or even higher) at X-ray wavelengths \citep{Singh11,Ricci15} and they constitute most of
the obscured population of AGN in the local universe ($z$~$\leq$~0.05) \citep{Matt2000,Bianchi12}.
Hard X-ray ($>$ 10~keV) spectral studies have also revealed the presence of Compton-thick (CT) AGN with $N_{\rm H,~LOS}$ $\geq$ $1.5~\times~10^{24}$ cm$^{-2}$ but
their fraction is likely to be small ($\sim$ 5$-$10 per cent) among the local type~2 AGN \citep{Comastri04,Burlon11,Torres21}.
Although, a much higher fraction of CT-AGN is inferred to be present at higher redshifts ($z$ $\sim$ 0.5$-$1.5) \citep{Akylas12,Ananna19}.
The modelling of X-ray background (XRB) spectrum peaking at 20$-$30~keV requires 10 to 40 per cent of CT-AGN \citep{Gilli07,Treister09}.
The local super-massive black hole (SMBH) mass function derived from the AGN luminosity function can also
be reconciled with a significantly large population of CT-AGN at $z$ $\sim$ 1$-$2,
the epoch during which AGN activity peaked \citep{Marconi04}.
Moreover, the population of obscured AGN at high$-z$ is poorly explored and the exact fraction of CT-AGN at higher redshifts
is still a subject of debate.
Therefore, it is important to detect and constrain the population of obscured AGN, especially at higher redshifts.
\par
Dust-obscured galaxies (DOGs) containing large reservoirs of gas and dust, are arguably thought to be potential hosts of
obscured AGN \citep{Narayanan10,Suleiman22}.
DOGs are defined to be bright in mid-IR but faint in optical with the flux ratio of 24~$\mu$m mid-IR band to $R-$band optical
($\frac{f_{24~\mu m}}{f_{R}}$) $\geq$ 1000 \citep{Dey08}.
In general, DOGs represent the population of optically-faint high-redshift ($z$ $\sim$ 1.5$-$2.5) galaxies with their
total 8$-$1000 $\mu$m IR luminosities (10$^{11}$~$-$~10$^{14}$~$L{\odot}$) similar to local luminous IR galaxies
(LIRGs) and ultra-luminous IR galaxies (ULIRGs) \citep{Sanders96,Melbourne12}.
The high IR luminosity of DOGs is explained by invoking gas-rich major merger
leading to an intensely star-forming dusty merged system \citep{Hopkins08,Yutani22}.
The large gas reservoir is also likely to fuel as well as obscure the accreting SMBH.
In fact, it is believed that the star-forming (SF) DOGs can evolve into AGN-dominated DOGs, which can eventually turn into
quasars \citep{Granato04,Hopkins06,Alexander12}.
The obscured AGN are commonly detected in the extreme population of DOGs
that are known as the `Hot DOGs’ characterized by hotter (up to hundreds of K) dust emission and extremely high
IR luminosity \citep[$L_{\rm IR}$~$\geq$~10$^{13}$~$L_{\odot}$;][]{Tsai15,Farrah17}.
However, the presence of obscured AGN in less extreme AGN-SF composite DOGs cannot be ruled out as their spectral energy distributions (SEDs)
can be well-fitted with or without AGN component \citep{Bussmann09,Lanzuisi09,Teng10}.
\par
In the literature, there have been several attempts to exploit the X-ray observations of DOGs.
Although, most of the previous studies have been limited either to highly luminous DOGs, {\ie} Hot DOGs
\citep{Stern14,Ricci17} or hampered with poor quality X-ray spectra \citep{Martinez07,Vito18,Zou20}.
The good quality X-ray observations of DOGs are limited only to a few relatively nearby sources
\citep[see][]{Zappacosta18,Assef20,Toba20}.
Considering the non-detection of X-ray emission in a large fraction of DOGs, as they fall below the detection limit of
X-ray surveys, there have been attempts to examine the X-ray spectrum of the stacked emission of undetected DOGs \citep[e.g.,][]{Fiore08,Treister09a}.
The stacked emission exhibiting flat X-ray spectrum is considered as an indication of a large fraction of CT-AGN
in DOGs \citep{Fiore09}.
However, the flat X-ray spectrum can also be produced from a low-luminosity AGN having a
moderate obscuration \citep[see][]{Georgantopoulos08} or due to a significant population of non-AGN DOGs among
the X-ray undetected DOGs \citep{Pope08}.
Therefore, deep and large-area X-ray surveys are required to examine the prevalence of obscured AGN in DOGs.
\par
We point out that the X-ray observations from {\em XMM-Newton}
and {\em Chandra} limited up to 10 keV pose a challenge to detect nearby CT-AGN
as it becomes difficult to accurately determine {the} photoelectric absorption cut-off seen at higher energies ($\geq$ 4.0 keV).
However, for high$-z$ AGN, X-ray spectra are redshifted to lower energies, which in turn makes {\em XMM-Newton} and
{\em Chandra} observations useful to determine the absorption cut-off and constrain the absorbing column density in CT-AGN \citep{Lanzuisi15,Koss16}.
In this paper, we unveil and study the nature of AGN hosted in DOGs using deep {\em XMM-Newton} and {\em Chandra}
observations available in the XMM Large-Scale Structure (XMM-LSS) field. Our study is limited within
the coverage of XMM-{\em Spitzer} Extragalactic Representative Volume Survey (XMM-SERVS) performed in the XMM-LSS field.
The choice of XMM-LSS field is based on the fact that X-ray studies of DOGs utilising deep {\em XMM-Newton} and {\em Chandra} observations
have already been performed in some of the deep fields, {\eg} Chandra Deep Field-South (CDF-S), Chandra Deep Field-North (CDF-N),
Extended Chandra Deep Field-South (ECDF-S) and Cosmic Evolution Survey (COSMOS) \citep[see][]{Fiore09,Treister09a,Georgantopoulos11,Corral16}.
\par
This paper is structured as follows. In Section~\ref{sec:sample}, we describe our DOGs sample, selection criteria,
redshifts and multi-wavelength data. In Section~\ref{sec:obs}, we provide the details of X-ray observations and data reduction.
In Section~\ref{sec:modelling}, we discuss X-ray spectral modelling.
Section~\ref{sec:Discussion} is devoted to the discussion on the nature and plausible evolutionary scenario of our DOGs.
In Section~\ref{sec:Conclusions}, we present the conclusions of our study. \\
In this paper, we adopt $\Lambda$CDM cosmology with $H_{0}$ = 70~km~s$^{-1}$~Mpc$^{-1}$, ${\Omega}_{\rm M}$ = 0.27, ${\Omega}_{\Lambda}$ = 0.73.
Errors quoted on parameters are of 90 per cent confidence level unless stated otherwise.
\section{The sample, selection criteria and redshifts}
\label{sec:sample}
We identified 34 DOGs with sufficiently good X-ray spectral quality in the 5.3~deg$^{2}$ area of the XMM-SERVS coverage
(34$^{\circ}$.2 $\leq$ RA (J2000) $\leq$ 37$^{\circ}$.125; and $-5^{\circ}$.72 $\leq$ DEC (J2000) $\leq$ $-3^{\circ}$.87) in the XMM-LSS field.
In Table~\ref{tab:DOGsSample}, we list our sample sources and other parameters,
{\ie} source name, including RA and DEC information, 24~$\mu$m flux, $r-$band magnitude, flux ratio of 24~$\mu$m to $r-$band,
redshift, and the X-ray source ID from the XMM-SERVS catalogue reported by \cite{Chen18}.
In the following sub-section, we describe our method to identify DOGs and provide the details of the multi-wavelength data available
in the XMM-SERVS coverage.
\begin{table*}
\centering
\caption{The sample of X-ray detected DOGs in the XMM-SERVS coverage of the XMM-LSS field}
\begin{threeparttable}
\begin{tabular}{ccccccc}
\hline
Source & $S_{24~{\mu}m}$ &  $m_{r}$  & $S_{24~{\mu}m}/S_{r}$ & Redshift    &  Ref. & XID   \\
Name   & (mJy)           &   (mag)   &                       &             &       &        \\
(1)   & (2)           &   (3)   &      (4)                 &     (5)        &   (6)    &   (7)     \\ \hline
J02:16:57-04:02:02 & 0.806$\pm$0.026 & 24.96$\pm$0.10 & 2141.7 & 1.155$_{-0.122}^{+0.146}$ & D23 & XMM00059  \\
J02:17:05-04:56:54 & 0.601$\pm$0.023 & 25.64$\pm$0.03 & 3005.0 & 1.732$_{-0.098}^{+0.284}$ & N23 & XMM00131  \\
J02:17:06-05:25:47 & 0.648$\pm$0.022 & 24.53$\pm$0.02 & 1159.3 & 1.843$_{-1.143}^{+0.121}$ & N23 & XMM00136  \\
J02:17:15-04:01:17 & 0.708$\pm$0.028 & 25.12$\pm$0.09 & 2187.2 & 2.188$_{-1.822}^{+0.401}$ & D23 & XMM00191  \\
J02:17:16-04:30:09 & 0.923$\pm$0.027 & 24.06$\pm$0.01 & 1074.4 & 0.5862$\pm$0.0079 (s)     & PR & XMM00205   \\
J02:17:22-04:36:55 & 1.282$\pm$0.024 & 25.71$\pm$0.06 & 6788.8 & 4.648$_{-0.251}^{+0.091}$ & N23 & XMM00250  \\
J02:17:24-04:18:44 & 0.412$\pm$0.027 & 25.94$\pm$0.05 & 2706.8 & 2.948$_{-1.000}^{+0.080}$ & N23 & XMM00267   \\
J02:17:33-04:06:13 & 0.436$\pm$0.026 & 24.85$\pm$0.03 & 1045.5 &  1.287$_{-0.033}^{+0.033}$ & D23 & XMM00359 \\
J02:17:36-04:59:11 & 0.391$\pm$0.026 & 25.80$\pm$0.04 & 2255.6 & 3.067$_{-0.547}^{+0.165}$  & N23 & XMM00393 \\
J02:17:36-05:01:07 & 0.544$\pm$0.027 & 24.71$\pm$0.02 & 1144.4 & 1.4210$\pm$0.0182 (s)      & UDSz & XMM00395 \\
J02:17:40-04:32:43 & 0.391$\pm$0.024 & 24.99$\pm$0.02 & 1071.9 & 1.732$_{-0.108}^{+0.338}$  & N23 & XMM00421  \\
J02:17:49-05:23:07 & 8.104$\pm$0.024 & 22.63$\pm$0.01 & 2534.3 & 0.8420$\pm$0.0092 (s)      & PR & XMM00497 \\
J02:18:37-04:29:50 & 0.579$\pm$0.027 & 24.68$\pm$0.01 & 1193.1 & 1.473$_{-0.345}^{+0.115}$ & N23 & XMM00860  \\
J02:19:01-04:24:41 & 0.438$\pm$0.023 & 25.40$\pm$0.03 & 1756.8 & 1.759$_{-0.091}^{+0.241}$ & N23 & XMM01034  \\
J02:19:31-04:49:41 & 1.040$\pm$0.027 & 24.29$\pm$0.01 & 1501.2 & 0.7953$\pm$0.0089 (s) & PR  & XMM01279  \\
J02:19:56-05:05:01 & 1.217$\pm$0.024 & 25.06$\pm$0.02 & 3548.3 & 1.652$_{-0.0269}^{+3.768}$ & N23 & XMM01464 \\
J02:20:32-04:50:02 & 4.436$\pm$0.040 & 22.68$\pm$0.01 & 1454.4 & 1.084$\pm$0.0011 (s) & VI & XMM01723   \\
J02:20:33-05:08:55 & 1.346$\pm$0.028 & 24.20$\pm$0.01 & 1775.1 & 2.048$_{-0.076}^{+0.383}$ & N23 & XMM01731  \\
J02:20:34-05:06:58 & 0.833$\pm$0.022 & 24.74$\pm$0.01 & 1808.1 & 2.870$_{-0.258}^{+0.136}$ & N23 & XMM01740 \\
J02:21:30-04:02:03 & 1.013$\pm$0.022 & 25.84$\pm$0.10 & 6070.7 & 1.096$_{-0.335}^{+0.387}$ & D23 & XMM02186 \\
J02:21:50-05:23:58 & 0.647$\pm$0.024 & 24.45$\pm$0.02 & 1077.9 & 1.900$_{-0.166}^{+0.153}$ & N23 & XMM02347   \\
J02:22:32-04:49:09 & 0.682$\pm$0.021 & 24.39$\pm$0.02 & 1080.9 & 1.815$_{-0.099}^{+1.106}$ & N23 & XMM02660  \\
J02:23:30-04:34:42 & 0.668$\pm$0.022 & 24.47$\pm$0.02 & 1132.8 & 2.300$_{-0.289}^{+0.319}$ & N23 & XMM03098  \\
J02:23:38-04:05:13 & 1.684$\pm$0.027 & 24.39$\pm$0.04 & 2660.3 & 3.275$_{-0.162}^{+0.292}$ & N23 & XMM03153 \\
J02:24:01-04:05:28 & 0.919$\pm$0.029 & 25.41$\pm$0.08 & 3687.1 & 1.678$_{-0.053}^{+0.148}$ & N23 & XMM03342  \\
J02:24:59-04:14:14 & 3.239$\pm$0.029 & 22.71$\pm$0.01 & 1092.0 & 1.900$_{-0.118}^{+1.025}$ & N23 & XMM03798 \\
J02:25:12-04:19:11 & 0.805$\pm$0.025 & 24.16$\pm$0.01 & 1031.1 & 1.759$_{-0.141}^{-0.065}$ & N23 & XMM03900  \\
J02:25:14-04:34:21 & 2.376$\pm$0.040 & 23.09$\pm$0.01 & 1131.5 & 3.538$_{-0.212}^{+0.069}$ & N23 & XMM03916  \\
J02:26:06-04:44:19 & 1.259$\pm$0.028 & 24.62$\pm$0.03 & 2460.1 & 1.732$_{-0.082}^{+0.295}$ & N23 & XMM04259 \\
J02:26:24-04:13:43 & 0.647$\pm$0.027 & 25.10$\pm$0.04 & 1957.8 & 1.652$_{-0.074}^{+0.075}$ & N23 & XMM04404  \\
J02:26:33-04:43:07 & 0.756$\pm$0.024 & 24.42$\pm$0.03 & 1228.2 & 2.333$_{-0.557}^{+0.748}$ & N23 & XMM04475  \\
J02:26:47-05:31:13 & 1.112$\pm$0.025 & 24.67$\pm$0.15 & 2256.0 & 1.070$_{-0.185}^{+0.231}$ & D23  & XMM04583  \\
J02:27:16-04:32:42 & 0.887$\pm$0.028 & 24.59$\pm$0.01 & 1683.1 & 1.759$_{-0.055}^{+0.325}$ & N23 & XMM04804  \\
J02:27:29-04:48:58 & 0.620$\pm$0.027 & 25.08$\pm$0.04 & 1854.3 & 1.652$_{-0.078}^{+0.123}$ & N23 & XMM04899 \\
\hline
\end{tabular}
\label{tab:Sample}
\begin{tablenotes}
\normalsize
\item
Notes - Column (1): Source name based on its RA (J2000) and DEC (J2000), Column (2): 24~$\mu$m flux from the SWIRE survey,
Column (3): $r-$band magnitude from the HSC-SSP catalogue, Column (4): flux ratio of 24~$\mu$m band to $r-$band, Column (5):
Redshift, all are photo$-z$ except those marked with `(s)' for spectroscopic redshift, Column (6): Reference for the redshift
(N23 - \cite{Nyland23}; D23 - \cite{Desprez23}, PR - PRIsm MUlti-Object Survey (PRIMUS), VI - VIMOS Public Extragalactic
Redshift Survey (VIPERS) and UDSz - UKIDSS Ultra-Deep Survey, Column (7): X-ray source ID from the XMM-SERVS
catalogue reported by \cite{Chen18}.
\end{tablenotes}
\end{threeparttable}  
\label{tab:DOGsSample}
\end{table*}
\subsection{Identification of DOGs and multi-wavelength data in the XMM-SERVS coverage}
The SERVS, a deep near-IR photometric survey, was
performed with the post-cryogenic Spitzer in 3.6~$\mu$m and 4.5~$\mu$m Infrared Array Camera (IRAC) bands
in the XMM-LSS region \citep{Mauduit12}.
The XMM-SERVS provides the {\em XMM-Newton} survey of the SERVS coverage of 5.3 deg$^{2}$ \citep{Chen18}.
The XMM-SERVS region is also covered with other multi-wavelength surveys that include UV observations from
the GALEX Deep Imaging Survey\footnote{\url{http://www.galex.caltech.edu/researcher/techdoc-ch2.html}},
optical photometric observations in  $g$, $r$, $i$, $z$, $y$, and four narrow-band filters
from the Hyper Suprime-Cam Subaru Strategic Program survey \citep[HSC-SSP;][]{Aihara18},
$Y$, $J$, $H$ and $K_{\rm s}$ bands photometric observations from the VISTA Deep Extragalactic Observations Survey
\citep[VIDEO;][]{Jarvis13},
mid-IR (3.6~$\mu$m, 4.5~$\mu$m, 5.8~$\mu$m and 8.0~$\mu$m from IRAC
and 24~$\mu$m, 70~$\mu$m and 160~$\mu$m from the Multi-band Imaging Photometer for {\em Spitzer} (MIPS)) observations
from the {\em Spitzer} Wide-area IR Extragalactic Survey \citep[SWIRE;][]{Lonsdale03}; far-IR (100~$\mu$m, 160~$\mu$m
from PACS and 250~$\mu$m, 350~$\mu$m and 500~$\mu$m from SPIRE) observations from the
{\em Herschel} Multi-tiered Extragalactic Survey \citep[HerMES;][]{Oliver12},
and radio surveys at various frequencies \citep[see][]{Singh14,Heywood20}.
\par
To identify DOGs in the XMM-SERVS region, we utilised 24~$\mu$m SWIRE and wide HSC-SSP optical data and followed the method described in \cite{Kayal22}.
We began with the 24~$\mu$m source catalogue (signal-to-noise ratio (SNR) $\geq$ 5) and
identified their optical counterparts.
Since the positional uncertainty of 24~$\mu$m sources (2$^{\prime\prime}$.0) is relatively larger
than that of the optical sources (0$^{\prime\prime}$.2), we used positions of 3.6~$\mu$m IRAC counterparts
of 24~$\mu$m sources from the SWIRE band-merged catalogue\footnote{\url{https://irsa.ipac.caltech.edu/data/SPITZER/SWIRE/overview.html}} to increase the reliability of the positional cross-matching.
The optical and band-merged 3.6~$\mu$m $-$ 24~$\mu$m source catalogues
were cross-matched using a radius of 1$^{\prime\prime}$.0
and the nearest match was considered as a true counterpart.
We found optical as well as 3.6~$\mu$m counterparts for all our 24~$\mu$m sources, owing to
a much higher sensitivity in the optical (5$\sigma$ limiting magnitude in $i-$band ($m_{ i}$) 26.2)
and 3.6~$\mu$m band (7.3 $\mu$Jy beam$^{-1}$ at 5$\sigma$)
than that of the 24~$\mu$m band (0.45 mJy beam$^{-1}$ at 5$\sigma$).
\par
To select DOGs, we used the conventional criterion of flux ratio of 24~$\mu$m to $r-$band optical
($\frac{f_{24~{\mu}m}}{f_{r}}$) $\geq$ 1000, which corresponds to the colour cut of $r~ - ~[24]$ $\geq$ 7.5,
where magnitudes are in the AB system \citep[see][]{Dey08,Fiore08,Toba15}.
Using the aforementioned criterion, we found a total of 1239 DOGs within 5.3 deg$^{2}$ of the XMM-SERVS region.
To study the X-ray properties of DOGs, we searched for their X-ray counterparts using the XMM-SERVS X-ray point source
catalogue. The availability of the optical$-$NIR$-$X-ray band-merged catalogue from \cite{Chen18}, which contains
optical (from CFHTLS and HSC-SSP) and near-IR counterparts (from VIDEO and SERVS) of X-ray point
sources, allows us to identify the X-ray counterparts of the DOGs by simply cross-matching
the optical positions of it with our optical$-$3.6~${\mu}$m$-$24~${\mu}$m band-merged catalogue using a tolerance radius
of 1$^{\prime\prime}$.0.
This exercise gives us only 89/1239 (7.2 per cent) DOGs with X-ray counterparts.
The fraction of X-ray detected DOGs is similar to that reported in our previous study \citep{Kayal22} that used data of similar
sensitivities.
\par
We note that, for 89 DOGs, the total X-ray counts in 0.5$-$10~keV full band ranges from 44 to 1038 with a median value of 124.
The total count value listed in the XMM-SERVS catalogue represents the sum of the counts from all the
three imaging detectors (PN, MOS1 and MOS2) in the {\em XMM-Newton} observations.
The total X-ray counts of a source determines its X-ray spectral quality.
Therefore, to obtain reasonably good-quality X-ray spectra, we restricted our sample to 34/89 sources
with total counts $\geq$ 150 in 0.5$-$10 keV energy band or those having additional {\em Chandra} observations.
We note that, the cut-off limit on the total counts is based on our qualitative assessment of spectral quality.
It ensures that our X-ray spectra are better than those used in several previous studies
\citep[e.g.,][]{Martinez07,Vito18,Zou20,Glikman24} which reported X-ray spectra of DOGs with ${\leq}$ 50 total counts in
0.5$-$10 keV energy band.
The better spectral quality allows us to obtain reliable constraints on the continuum spectral
shape and absorption column density. We would like to point out that the total X-ray counts found in our
analysis may differ from the value listed in the XMM-SERVS catalogue, as we have added more recent data, whenever available,
and excluded observations in which source$-$of$-$interest falls within CCD gaps or in the peripheral region
(see Section~\ref{sec:obs}).
\par
Further, our final sample of 34 sources represents a faint population of DOGs
having 24~$\mu$m flux (S$_{24~{\mu}m}$) in the range of 0.39 mJy to 8.11 mJy
with a median value of 0.81 mJy.
As expected, our DOGs are also faint in the optical wavelengths with $m_{r}$ distributed across 22.64 to 25.94, with
a median value of 24.68. The faintness of DOGs can be attributed to their high redshifts
(0.59 $\leq$ $z$ $\leq$ 4.65, $z_{\rm median}$ = 1.75).
Therefore, unlike some of the previous studies limited to relatively bright DOGs \citep[e.g.,][]{Lanzuisi09,Zou20}, our sample allows us to probe
a fainter population of DOGs residing at higher redshifts.

\subsection{Spectroscopic and photometric redshifts of DOGs}
Redshift estimates of DOGs are essential for their X-ray spectral fittings and luminosity measurements.
To obtain the redshifts of our DOGs, we utilised various redshift measurement campaigns
performed in the XMM-SERVS region. For instance, we used the spectroscopic redshift catalogue from the HSC-SSP PDR-3
that includes spectroscopic redshifts from different campaigns, {\ie} PRIsm MUlti-Object Survey \citep[PRIMUS;][]{Coil11},
VIMOS Public Extragalactic Redshift Survey \citep[VIPERS;][]{Garilli14},
VIMOS VLT Deep Survey \citep[VVDS;][]{LeFevre13}, SDSS Baryon Oscillation Spectroscopic Survey (SDSS-BOSS)
programme \citep{Dawson13,Menzel16}. We have also used \cite{Chen18} catalogue that lists
spectroscopic redshifts for X-ray sources, whenever available. They used
UKIDSS Ultra-Deep Survey \citep[UDSz;][]{Bradshaw13,McLure13},
3D-HST Survey \citep{Skelton14,Momcheva16} and other publicly available spectroscopic redshift
catalogues{\footnote{\url{https://www.nottingham.ac.uk/astronomy/UDS/data/data.html}}}.
From these spec$-z$ catalogues, we find spectroscopic redshifts for only 05 of our 34 X-ray detected DOGs (see Table~\ref{tab:DOGsSample}).
\par
For the remaining 29 sources, we obtained photometric redshifts using  multi-band photometry-based photo$-z$ catalogues reported by
\cite{Nyland23} and \cite{Desprez23}. We note that several photo$-z$ campaigns are available in the XMM-SERVS region,
{\eg} CFHTLS-based photo$-z$ estimates \citep{Ilbert06,Coupon09},
HSC-SSP-based photo$-z$ estimates \citep{Tanaka18,Schuldt21}.
We prefer to use more accurate photo$-z$ estimates derived from the multi-band photometry spanning across optical to near-IR.
\cite{Nyland23} estimated photometric redshifts using the Tractor image-modelling software-based
de-blended multi-band forced photometry across 13
optical to near-IR bands (${u}^{{\prime}}$ band from the CFHTLS, $g$, $r$, $i$, $z$ and $y$ bands from the HSC-SSP,
$Z$, $Y$, $J$, $H$, and $Ks$ bands from the VIDEO and 3.6~$\mu$m and 4.5~${\mu}$m bands from the {\em Spitzer}/DeepDrill).
The photo$-z$ estimates based on the forced photometry are more accurate and supersede previous estimates based on
the traditional position-matched multi-band photometry \citep[e.g.,][]{Chen18,Ni21,Zou21,Zou22}.
Previous test$-$bed study on the Tractor-based photo$-z$ estimate limited to 1.0 deg$^{2}$ in the XMM-SERVS,
showed accurate photometric redshifts with normalized median absolute deviation (NMAD), ${\sigma}_{\rm NMAD}$ $\leq$ 0.08 and
an outlier fraction of only $\leq$ 1.5 per cent \citep[see][]{Nyland17}.
\par
We note that the photo$-z$ estimates reported in \cite{Nyland23} are limited only to the VIDEO/{\em Spitzer} DeepDrill region of
4.5~deg$^{2}$, which does not fully cover the XMM-SERVS field of 5.3 deg$^{2}$.
We find that only 24/29 of our DOGs falling within the VIDEO region have photo$-z$ estimates
from \cite{Nyland23}. For the remaining 05/29 DOGs, we considered photo$-z$ estimates from \cite{Desprez23}, who
used multi$-$band photometric data ($u$, $u^{\star}$ from the CFHT-MegaCam, $g$, $r$, $i$, $z$, and $y$ from the HSC-SSP,
and $Y$, $J$, $H$, and $K_s$ data from the VIDEO, whenever available). The photo$-z$ for sources falling outside the VIDEO
region are based on only seven-band optical data. The photo$-z$ estimates are precise with ${\sigma}_{\rm NMAD}$ $\leq$ 0.04 down to $m_{i}$ = 25, and an outlier fraction of $\leq$ 6 per cent.
Further, we point out that \cite{Desprez23} reported two different versions of the photo$-z$ catalogues, {\ie} one version
used the HSC pipeline for the photometric extraction and Phosphoros code for the redshift estimation, while
the second version used Source Extractor for the photometry and Le PHARE code \citep{Arnouts99, Ilbert06} for the redshift measurements.
We used the second version of the photo$-$z catalogue due to its slightly better performance.
We found photo$-z$ for all remaining 05 sources from \cite{Desprez23} photo$-z$ catalogue.
Thus, in total, we found redshift estimates for all 34 DOGs, {\ie} spec$-z$ for 05 sources and photo$-z$ for 29 sources (see Table~\ref{tab:DOGsSample}).
\section{X-ray observations and data reduction}
\label{sec:obs}
To study the X-ray spectral properties of our DOGs, we used archival {\em XMM-Newton} observations, which were
performed mainly under the XMM-SERVS project.
Using a total of 155 pointings, XMM-SERVS observations provide
a nearly uniform coverage of 5.3~deg$^{2}$ sky-area in the XMM-LSS field \citep{Chen18}.
The XMM-SERVS project used a total of 2.7 Ms of flare-filtered exposure time, which
includes 1.3 Ms observations carried out in AO-15 and all the available archival {\em XMM-Newton}
data from other surveys, {\eg} XMM-LSS survey \citep{Pacaud06,Pierre16},
{\em XMM-Newton} Medium Deep Survey \citep[XMDS,][]{Chiappetti05}, Subaru {\em XMM-Newton} Deep Survey \citep[SXDS,][]{Ueda08}
and XMM-XXL-North field \citep{Pierre16}.
The combination of different epochs of archival data makes XMM-SERVS survey the deepest X-ray survey
in the XMM-LSS field. The depth of the XMM-SERVS survey is comparable to some of the deep small$-$area
surveys, such as SXDS covering only 1.14 deg$^{2}$ \citep{Ueda08} and XMM-COSMOS covering only 2.0 deg$^{2}$ \citep{Cappelluti09}.
\par
For each of our sample sources, we downloaded all the available X-ray data from the {\em XMM-Newton} Science
Archive\footnote{\url{https://nxsa.esac.esa.int/nxsa-web/search}}.
In order to improve the X-ray spectral quality, we also checked the availability of {\em Chandra} observations of
our sample sources using {\em Chandra} Source Catalog Release 2.0 (CSA 2.0\footnote{\url{https://cxc.cfa.harvard.edu/csc/}}).
Three of our sample sources have {\em Chandra} archival data with total counts $>$ 10.
In Table~\ref{tab:Sample}, we provide the details of X-ray observations, {\ie} observation ID, observation date,
observation time and X-ray source ID from the XMM-SERVS catalogue.
To gain further improvement in the X-ray spectral quality of our faint sources, we added all
the available {\em XMM-Newton} data, for a given source, taken at different epochs.
We caution that the addition of multi-epoch data renders average spectral
properties, if a source is variable across different epochs.
\subsection{{\em XMM-Newton} data reduction}
We reduced the {\em XMM-Newton} pn and MOS data using Science Analysis System ({\scshape sas}) v21.0.0.
The Observation Data Files (ODFs) were processed using {\scshape epicproc} tasks ({\scshape epproc} and {\scshape emproc} tasks for pn and
MOS, respectively) to create pn, MOS1, and MOS2, event files for each observation ID.
From each event file, we created good time interval (GTI) event file using the {\scshape evselect} task.
The flaring background time intervals were identified from the single-event light curves of high
(10$-$12 keV) and low (0.3$-$10 keV) energies and the time intervals exceeding count rates 3$\sigma$
above the mean value were removed.
The GTI files were calibrated using the most recent calibration files.
Further, we filtered the event files at the energy ranges that overlap with the instrumental background lines,
{\ie} Cu lines at 7.2$-$7.6 and 7.8$-$8.2 keV for pn.
\par
For each observation ID, we extracted the spectrum from each detector using a circular region centred on the source position. The radius of the source extraction region is in the range of 15$^{\prime\prime}$ to 25$^{\prime\prime}$ depending on the offset of the target source from the on-axis.
The background spectrum was extracted from a neighbouring source-free region using a circular aperture of 40$^{\prime\prime}$ radius
(see Figure~\ref{fig:SourceBkgExtImgXMM00267}).
We note that despite moderately deep XMM-SERVS observations, our DOGs tend to suffer from poor photon statistics.
Therefore, for a source with multi-epoch observations, we added the source spectra from
individual observation IDs. The background spectra from individual observation IDs were also added.
To increase the spectral quality further, we combined the spectra from pn, MOS1 and MOS2 using
{\scshape epicspeccombine} task (see Figure~\ref{fig:BestFitsXMM00267} and Figure~\ref{fig:CombinedSpecXMM00267}).
To combine the spectra from all three detectors, we ensured a common energy range of 0.5$-$10~keV.
The response and auxiliary files were computed by averaging the individual files of each detector.
\subsection{{\em Chandra} data reduction}

We reduced {\em Chandra}/ACIS data using Chandra Interactive Analysis of Observations ({\scshape ciao}) software version 4.15
(CALDB version 4.10.4).
For data reduction, we followed the standard procedure, which includes the removal of hot pixels, cosmic afterglows and background
flaring time intervals. The cleaned event files were calibrated using the most recent calibration files.
From each individual observation, we have extracted the source spectrum using the {\scshape specextract} task and
considered a circular extraction region centred on the X-ray source. The radius of extraction region includes 90 per cent Encircled Energy Fraction (EEF) and is around 1$^{\prime\prime}$.0 $-$  3$^{\prime\prime}$.0 depending on the off-axis angle.
The background spectrum was extracted
from a source-free circular region of 12$^{\prime\prime}$.0 $-$ 15$^{\prime\prime}$.0 radius selected in the same CCD chip.
The {\scshape specextract} task also generates auxiliary and response matrix files.
In case of sources having multiple {\em Chandra}/ACIS observations,
we added all individual spectra of a source.
We created combined source spectrum, background spectrum, response, and auxiliary matrices using the {\scshape combine\_spectra} task
in the {\scshape ciao}.
\section{X-ray spectral analysis}
\label{sec:modelling}
We performed X-ray spectral fittings of our sample sources using {\scshape xspec} v12.13.0c \citep{Arnaud96}.
In order to get reliable goodness-of-fit statistics with moderate to low-count spectra, we preferred to use Cash statistics \citep{Cash79}
instead of ${\chi}^{2}$ statistics.
We attempted to fit the X-ray spectra of our sample sources using a simple absorbed power law model as well as with a physically motivated model.
\subsection{X-ray spectral fitting with a simple absorbed power law}
\label{sec:PL}
To fit the X-ray spectra, we began with a baseline model characterised by a simple absorbed power law, which can be expressed
as {\scshape tbabs $\times$ (ztbabs $\times$ powerlaw)}. The first absorption component ({\scshape tbabs}) accounts for
the Galactic absorption column density ($N_{\rm H,~Gal}$), which is fixed to the value
obtained from the NASA's HEASARC $N_{\rm H}$
calculator\footnote{\url{https://heasarc.gsfc.nasa.gov/cgi-bin/Tools/w3nh/w3nh.pl}}. The galactic absorption in the direction
of the XMM-LSS field is found to be in the range of 1.86 $\times$~10$^{20}$~cm$^{-2}$ $-$ 2.45 $\times$~10$^{20}$~cm$^{-2}$.
The second absorption component accounts for the photoelectric absorption at the source redshift.
To constrain the absorbing column density accurately, we kept the power law photon index ($\Gamma$) as a free parameter.
However, we caution that 0.5$-$10~keV X-ray spectra of heavily obscured AGN can also be mimicked with
a flat power law absorbed with a low column density \citep{George91,Georgantopoulos11}.
Therefore, to avoid the degeneracy between photon index ($\Gamma$) and {$N_{\rm H}$}, and to place reliable constraints on $N_{\rm H}$, we
fixed $\Gamma$ to 2.0, whenever spectral fitting rendered a flat ($\Gamma$ $<$ 1.7) photon index.
Considering the fact that AGN generally exhibit a steep photon index in the range of 1.8
to 2.1, it is a common practice to fix the photon index to a typical value (1.8$-$2.1) when dealing with low-count
X-ray spectra \citep[see][]{Tozzi06,Corral16,Zou20}.
We note that the fit statistics do not show any appreciable change if the photon index is altered from 2.0 to 1.9 or 1.8.
The best-fitted spectral parameters obtained from the absorbed power law model are listed in Table~\ref{tab:XrayPOW}.
To our baseline model (simple absorbed power law), we attempted to add a reflection component
using {\scshape pexrav}, a phenomenological model,
which considers reflection from a semi-infinite slab of neutral medium \citep{Magdziarz95}. However, we found either no or
insignificant improvement in the fit statistics. Thus, a simple absorbed power law seems adequate for fitting the X-ray spectra of
the most of our sample sources.
\par
In one of our sample sources (XMM01723), residuals seen at the soft energies ($<$ 2.0~keV) can be
accounted for with an additional power law which is interpreted as the scattered AGN emission reaching directly to the observer without piercing through
the obscuring material. The photon index of the scattered power law is tied with that of the transmitted power law component.
The addition of a scattered power law improves the fit statistics from Cstat/d.o.f. = 206.2/172 (1.20)
to 200.5/171 (1.17).
In  XMM01723, the ratio of normalizations of the scattered power law component to the direct power law component is found to be only 0.07,
which shows that the leaked scattered emission is much weaker than the directly transmitted component, as expected in case of heavily obscured
sources.
We note that the presence of the scattered power law component in XMM01723 is consistent with the previous X-ray studies of DOGs
and dust-reddened quasar samples \citep[e.g.,][]{Corral16,LaMassa16,Glikman17}.
\par
The Fe K$\alpha$ emission line, a ubiquitous feature in nearby AGN spectra, is detected in five of our sample sources (see Table~\ref{tab:XrayPOW}).
The presence of the Fe K${\alpha}$ emission line is apparent from the residuals. The addition of a line component
yields an improvement in the fit statistics.
For instance, in the case of XMM03916, the addition of a line component representing Fe K$\alpha$ line emission
improves the fit statistics from Cstat/d.o.f. = 176.75/167 (1.06) to Cstat/d.o.f. = 171.49/165 (1.04).
The Fe K$\alpha$ line is fitted with a narrow unresolved ($<$100 eV) Gaussian by fixing the line-width to 1 eV and keeping the line energy
as a free parameter. In all five sources, the energy of the emission line in the rest-frame corresponds to 6.4 keV, {\ie} the energy
of the neutral Fe K$\alpha$ line.
The detection of the Fe K$\alpha$ line also provides confirmation of redshift accuracy.
We note that the detection of the Fe K$\alpha$ emission line in the remaining sample sources might be hindered due to
low counts in their spectra.
\par
The equivalent widths (EWs) of Fe K$\alpha$ lines in our sample sources are found to be in the range of 0.06 keV to 0.26 keV.
We point out that, in general, heavily obscured AGN show high EWs ($\geq$ 0.5 keV) of Fe K$\alpha$ emission line
\citep{Maiolino98,Matt2000,Singh11}.
However, in our sample, heavily obscured sources (XMM00497, XMM01723 and XMM03916) show low EWs,
which can be explained from the well-known Baldwin effect, also known as Iwasawa–Taniguchi
effect \citep{Baldwin77,Iwasawa93}, an anti-correlation between EW of Fe K$\alpha$ line and X-ray luminosity \citep[see][]{Boorman18,Matt19}.
The heavily obscured sources in our sample have high X-ray luminosities
($L_{\rm 2.0-10~keV}^{\rm int}$ $\sim$ 10$^{44}$ $-$ 10$^{45}$ erg s$^{-1}$) which can give rise to systematically lower EWs of Fe K$\alpha$ lines.
In fact, several CT-AGN of high X-ray luminosities ($\geq$ 10$^{44}$ erg s$^{-1}$) are known to exhibit moderate EWs of a few hundred eVs or lower  \citep[see][]{Fukazawa11,Iwasawa12,Boorman18}.
Although, we note that the Fe K$\alpha$ line EW estimates of our sample DOGs ought to be treated with caution due to their low-count spectra.
\begin{table*}
\centering
\caption{The best-fitted spectral parameters using absorbed power law model}
\renewcommand{\arraystretch}{1.2}
\begin{adjustbox}{width=\textwidth}
\begin{threeparttable}
\begin{tabular}{cccccccccc}
\hline
XID &  Model  & $N_{\rm H}$ & $\Gamma$ & ${\Gamma}_{\rm norm}$ & pl$^{\rm sct}_{\rm norm}$ & $E_{\rm Fe}$ &  $EW_{\rm Fe}$ & Cstat (dof) & $P_{\rm CT}$ \\
&  & ($10^{22}$ cm$^{-2}$)  &  & ($10^{-5}$) & ($10^{-5}$) & (keV) & (keV) & & \\
(1) & (2) & (3) & (4) & (5) & (6) & (7) & (8) & (9) & (10)  \\
 \hline
XMM00059 & {abs*pl+L} & ${1.77^{+0.50}_{-0.33}}$ & $2.0^f$ & ${1.74^{+0.23}_{-0.16}}$ &  & $2.94^{{+0.05}}_{{-0.60}}$ & $0.18^{{+0.14}}_{{-0.16}}$ & ${596.8~(656)}$ & ${0.0}$ \\
XMM00131 & {abs*pl} & ${1.94^{+0.85}_{-0.60}}$ & $2.0^f$ & ${0.91^{+0.14}_{-0.11}}$ &  &  &  & ${531.9~(530)}$ & ${0.0}$ \\
XMM00136 & {abs*pl} & ${1.83^{+1.36}_{-0.85}}$ & $2.0^f$ & ${0.38^{+0.09}_{-0.07}}$ &  &  &  & ${174.9~(209)}$ & ${0.0}$ \\
XMM00191 & {abs*pl} & ${44.22^{+12.16}_{-8.86}}$ & ${2.0^f}$ & ${3.19^{+0.74}_{-0.55}}$ &  &  &  & ${490.0~(490)}$ & ${0.0}$ \\
XMM00205 & abs*pl & $2.94^{{+1.56}}_{{-0.82}}$ & $2.0^f$ & $0.61^{{+0.20}}_{{-0.12}}$ &  &  &  & $185.2~(189)$ & ${0.0}$ \\
XMM00250 & abs*pl & $31.66^{{+56.51}}_{{-11.18}}$ & $2.15^{{+1.25}}_{{-0.32}}$ & $0.53^{{+1.70}}_{{-0.15}}$ &  &  &  & $218.7~(216)$ & ${0.0}$ \\
XMM00267 & {abs*pl+L} & ${3.29^{+1.97}_{-1.33}}$ & ${1.72^{+0.23}_{-0.17}}$ & $1.15^{{+0.27}}_{{-0.17}}$ & & $1.49^{{+0.32}}_{{-0.43}}$ & $0.06^{{+0.04}}_{{-0.06}}$ & ${655.3~(696)}$ & ${0.0}$ \\
XMM00359 & abs*pl & $1.43^{+0.41}_{-0.41}$ & $2.52^{+0.06}_{-0.44}$ & $0.36^{+0.11}_{-0.05}$ &  &  &  & $252.4~(292)$ & $0.0$ \\
XMM00393 & abs*pl & ${9.72^{+11.73}_{-3.91}}$ & ${2.14^{+0.43}_{-0.36}}$ & $0.40^{{+0.32}}_{{-0.12}}$ &  &  &  & $234.4~(228)$ & ${0.0}$ \\
XMM00395 & {abs*pl} & ${1.10^{+0.66}_{-0.46}}$ & $2.0^f$ & ${0.35^{+0.07}_{-0.05}}$ &  &  &  & ${333.5~(369)}$ & ${0.0}$ \\
XMM00421 & {abs*pl} & ${0.49^{+0.74}_{-0.38}}$ & ${2.0^f}$ & ${0.40^{+0.09}_{-0.06}}$ &  &  &  & ${281.2~(312)}$ & ${0.0}$ \\
XMM00497 & {abs*pl+L} & ${13.12^{+2.29}_{-1.53}}$ & ${2.0^f}$ & ${5.58^{+0.82}_{-0.58}}$ &  & ${3.19^{+0.05}_{-0.05}}$ & ${0.17^{+0.09}_{-0.10}}$ & ${637.6~(656)}$ & ${0.0}$ \\
XMM00860 & {abs*pl} & ${3.37^{+2.74}_{-1.42}}$ & ${2.04^{+0.51}_{-0.43}}$ & ${0.36^{+0.26}_{-0.12}}$ &  &  &  & ${226.8~(233)}$ & ${0.0}$ \\
XMM01034 & {abs*pl} & ${1.40^{+1.42}_{-0.83}}$ & $2.0^f$ & ${0.25^{+0.07}_{-0.05}}$ &  &  &  & ${225.0~(232)}$ & ${0.0}$ \\
XMM01279 & {abs*pl} & ${1.17^{+0.36}_{-0.25}}$ & $2.0^f$ & ${1.29^{+0.19}_{-0.15}}$ &  &  &  & ${500.9~(616)}$ & ${0.0}$ \\
XMM01464 & abs*pl & ${0.86^{+0.87}_{-0.56}}$ & $1.91^{{+0.28}}_{{-0.20}}$ & ${0.45^{+0.12}_{-0.07}}$ &  &  &  & ${412.3~(479)}$ & ${0.0}$ \\
XMM01723 & {abs*pl+pl+L} & ${18.62^{+88.91}_{-5.26}}$ & $2.0^f$ & ${0.83^{+2.33}_{-0.22}}$ & ${0.06^{+0.06}_{-0.02}}$ & ${3.09^{+0.40}_{-0.05}}$ & ${0.14^{+0.63}_{-0.12}}$ & ${200.5~(171)}$ & ${0.01}$ \\
XMM01731 & {abs*pl} & ${7.03^{+8.10}_{-2.89}}$ & ${2.42^{+1.75}_{-0.63}}$ & ${0.75^{+1.61}_{-0.30}}$ &  &  &  & ${239.3~(238)}$ & ${0.0}$ \\
XMM01740 & {abs*pl} & ${6.17^{+18.97}_{-3.30}}$ & $2.0^f$ & ${0.19^{+0.19}_{-0.06}}$ &  &  &  & ${110.0~(127)}$ & ${0.0}$ \\
XMM02186 & {abs*pl} & ${0.89^{+0.85}_{-0.44}}$ & ${2.0^f}$ & ${0.54^{+0.18}_{-0.12}}$ &  &  &  & ${115.1~(155)}$ & ${0.0}$ \\
XMM02347 & {abs*pl} & ${3.97^{+2.53}_{-1.31}}$ & ${2.0^f}$ & ${0.37^{+0.11}_{-0.07}}$ &  &  &  & ${265.2~(264)}$ & ${0.0}$ \\
XMM02660 & {abs*pl} & ${5.62^{+4.42}_{-1.76}}$ & ${1.89^{+0.70}_{-0.31}}$ & ${0.79^{+0.75}_{-0.21}}$ &  &  &  & ${245.3~(287)}$ & ${0.0}$ \\
XMM03098 & abs*pl & ${0.92^{+2.04}_{-0.71}}$ & $2.0^f$ & $0.28^{{+0.09}}_{{-0.04}}$ &  &  &  & ${210.0}~(238)$ & ${0.0}$ \\
XMM03153 & {abs*pl} & ${2.31^{+3.60}_{-1.70}}$ & ${1.76^{+0.36}_{-0.20}}$ & ${0.37^{+0.13}_{-0.06}}$ &  &  &  & ${356.8~(376)}$ & ${0.0}$ \\
XMM03342 & abs*pl & ${1.18^{+4.13}_{-0.77}}$ & $2.22^{{+0.74}}_{{-0.43}}$ & $0.21^{{+0.28}}_{{-0.06}}$ &  &  &  & $141.1~(130)$ & ${0.0}$ \\
XMM03798 & abs*pl & ${4.86^{+3.34}_{-1.66}}$ & $2.32^{{+0.60}}_{{-0.39}}$ & $0.67^{{+0.46}}_{{-0.19}}$ &  &  &  & $221.7~(248)$ & ${0.0}$ \\
XMM03900 & {abs*pl} & ${5.49^{+1.52}_{-1.14}}$ & $2.0^f$ & ${1.64^{+0.24}_{-0.19}}$ &  &  &  & ${572.0~(637)}$ & ${0.0}$ \\
XMM03916 & {abs*pl+L} & ${114.84^{+49.89}_{-28.94}}$ & $2.0^f$ & ${0.97^{+0.42}_{-0.24}}$ &  & ${1.49^{+0.22}_{-0.17}}$ & ${0.26^{+0.60}_{-0.22}}$ & ${171.5~(166)}$ & ${0.12}$ \\
XMM04259 & {abs*pl} & ${13.79^{+16.32}_{-5.31}}$ & ${1.79^{+1.04}_{-0.47}}$ & ${0.49^{+1.15}_{-0.23}}$ &  &  &  & ${192.0~(212)}$ & ${0.0}$ \\
XMM04404 & abs*pl & $1.55^{+0.53}_{-0.53}$ & $1.87^{+0.68}_{-0.29}$ & $0.54^{+0.17}_{-0.07}$ &  &  &  & $327.4~(387)$ & $0.0$ \\
XMM04475 & {abs*pl} & ${0.79^{+8.99}_{-0.48}}$ & $2.0^f$ & ${0.12^{+0.10}_{-0.02}}$ &  &  &  & ${308.0~(305)}$ & ${0.0}$ \\
XMM04583 & {abs*pl} & ${3.49^{+1.26}_{-0.82}}$ & ${2.0^f}$ & ${1.08^{+0.25}_{-0.19}}$ &  &  &  & ${412.3~(462)}$ & ${0.0}$ \\
XMM04804 & {abs*pl} & ${3.86^{+3.17}_{-1.64}}$ & ${2.07^{+0.56}_{-0.39}}$ & ${0.50^{+0.35}_{-0.15}}$ &  &  &  & ${224.8~(234)}$ & ${0.0}$ \\
XMM04899 & {abs*pl} & ${9.00^{+3.71}_{-2.24}}$ & $2.0^f$ & ${0.75^{+0.18}_{-0.12}}$ &  &  &  & ${295.3~(328)}$ & ${0.0}$ \\
\hline
\end{tabular}
\begin{tablenotes}
\item Notes:
{Column (1): Source XIDs, Column (2): the best-fit model where `abs' denotes absorption component, `pl' denotes power law, `L' denotes emission
line and an additional `pl' denotes the scattered power law, Column (3): line-of-sight column density, Column (4): photon index, Column(5): power law normalisation, Column (6): normalisation of the scattered power law component, Column (7): Fe K$\alpha$ line energy in the observed$-$frame,
Column (8): equivalent width of Fe K$\alpha$ line, Column (9): Cash statistics and degrees of freedom of the spectral fit, Column (10): MCMC-based probability of source being a CT-AGN.
}
\end{tablenotes}
\end{threeparttable}
  \end{adjustbox}
\label{tab:XrayPOW}
\end{table*}
\begin{table*}
 \centering
 \caption{The best-fitted spectral parameters using {\scshape borus02} model}
 \renewcommand{\arraystretch}{1.2}
 \begin{adjustbox}{width=0.95\textwidth}
 \begin{threeparttable}
\begin{tabular}{cccccccccccc}
 \hline
 XID &  Model & $N_{\rm H}$ & $\Gamma$ & ${\Gamma}_{\rm norm}$ & pl$^{\rm sct}_{\rm norm}$ &  $E_{\rm Fe}$ &  $EW_{\rm Fe}$ & Cstat (dof) & $P_{\rm CT}$  \\
 &    & ($10^{22}$ cm$^{-2}$) &  & ($10^{-5}$)  & ($10^{-5}$)  & (keV) & (keV) & &  \\
 (1) & (2) & (3) & (4) & (5) & (6) & (7) & (8) & (9) & (10)  \\
 \hline
XMM00059 & B02+L &  $1.81^{{+0.45}}_{{-0.37}}$ & $2.0^f$ & $1.81^{{+0.23}}_{{-0.19}}$ &  & $2.94^{{+0.05}}_{{-0.49}}$ & $0.18^{{+0.15}}_{{-0.16}}$ & $597.9~(656)$ & ${0.0}$ \\
XMM00131 & B02  & $1.99^{{+0.73}}_{{-0.70}}$ & $2.0^f$ & $0.95^{{+0.13}}_{{-0.13}}$ &  &  &  & $532.6~(530)$ & ${0.0}$ \\
XMM00136 & B02 & $1.89^{{+1.14}}_{{-0.77}}$ & $2.0^f$ & $0.39^{{+0.09}}_{{-0.07}}$ & &  &  & $175.5~(209)$ & ${0.0}$  \\
XMM00191 & {B02}  & ${45.00^{+13.97}_{-8.50}}$ & ${2.0^f}$ & ${4.63^{+1.75}_{-0.96}}$ &  &  &  & ${490.3~(490)}$ & ${0.0}$ \\
XMM00205 & {B02} & ${2.87^{+1.54}_{-0.85}}$ & ${2.0^f}$ & ${0.62^{+0.22}_{-0.14}}$ &  &  &  & ${185.4~(189)}$ & ${0.0}$  \\
XMM00250 & B02  & ${31.00^{+23.32}_{-18.79}}$ & $2.12^{{+0.43}}_{{-0.49}}$ & $0.66^{{+0.73}}_{{-0.34}}$ &  &  &  & $218.7~(216)$ & ${0.0}$ \\
XMM00267 & B02+L & $3.21^{{+1.73}}_{{-1.53}}$ & $1.69^{{+0.22}}_{{-0.18}}$ & $1.18^{{+0.26}}_{{-0.19}}$ &  & $1.49^{{+0.38}}_{{-0.06}}$ & $0.06^{{+0.04}}_{{-0.06}}$ & $655.6~(696)$ & ${0.0}$  \\
XMM00359 & B02 & $1.24^{+0.29}_{-0.24}$ & $2.51^{+0.07}_{-0.40}$ & $0.37^{+0.09}_{-0.05}$ &  &  &  & $252.5~(291)$ & $0.0$  \\
XMM00393 & B02 & $9.57^{{+9.94}}_{{-5.60}}$ & $2.10^{{+0.44}}_{{-0.47}}$ & $0.43^{{+0.35}}_{{-0.17}}$ &  &  &  & $234.3~(228)$ & ${0.0}$  \\
XMM00395 & B02 & $1.63^{{+0.15}}_{{-0.60}}$ & $2.0^f$ & ${0.37^{+0.07}_{-0.04}}$ &  &  &  & ${337.6}~(369)$ & ${0.0}$  \\
XMM00421 & B02 & $1.63^{{+0.25}}_{{-0.61}}$ & ${{2.03}}^{{+0.51}}_{{-0.09}}$ & ${0.43^{+0.19}_{-0.01}}$ &  &  &  & ${{291.7}}~(311)$ & ${0.0}$  \\
XMM00497 & B02+L  & ${12.78^{+3.11}_{-0.87}}$ & ${2.0^f}$ & ${6.12^{+1.32}_{-0.44}}$ & & ${3.19^{+0.04}_{-0.05}}$ & ${0.17^{+0.10}_{-0.10}}$ & ${638.0~(656)}$ & ${0.0}$ \\
XMM00860 & B02  & $3.37^{{+2.39}}_{{-1.70}}$ & $2.02^{{+0.51}}_{{-0.48}}$ & $0.37^{{+0.25}}_{{-0.14}}$ &  &  &  & $226.8~(233)$ & ${0.0}$ \\
XMM01034 & B02 & $1.44^{{+1.25}}_{{-0.38}}$ & $2.0^f$ & $0.26^{{+0.07}}_{{-0.04}}$ &  &  &  & $225.2~(232)$ & ${0.0}$  \\
XMM01279 & {B02}  & ${1.63^{+0.09}_{-0.60}}$ & ${2.0^f}$ & ${1.51^{+0.02}_{-0.31}}$ &  &  &  & ${506.4~(616)}$ & ${0.0}$ \\
XMM01464 & {B02} & ${1.02^{+0.78}_{-0.02}}$ & ${1.93^{+0.28}_{-0.09}}$ & ${0.47^{+0.12}_{-0.02}}$ &  &  &  & ${412.3~(479)}$ & ${0.0}$  \\
XMM01723 & B02+cpl+L  & $17.91^{{+39.73}}_{{-9.71}}$ & $2.0^f$ & $0.95^{{+1.64}}_{{-0.46}}$ & $0.06^{{+0.08}}_{{-0.04}}$ & $3.08^{{+0.01}}_{{-0.07}}$ & $0.13^{{+0.45}}_{{-0.11}}$ & $200.6~(171)$ & ${0.0}$ \\
XMM01731 & B02  & ${{7.02}}^{{+2.48}}_{{-4.25}}$ & ${{2.40}}^{{+0.16}}_{{-0.83}}$ & ${{0.80}}^{{+0.25}}_{{-0.42}}$ &  &  &  & $239.3~(238)$ & ${0.0}$ \\
XMM01740 & B02  & $6.23^{{+12.03}}_{{-4.60}}$ & $2.0^f$ & $0.20^{{+0.17}}_{{-0.08}}$ &  &  &  & $110.0~(127)$ & ${0.0}$ \\
XMM02186 & {B02}  & $1.63^{{+0.46}}_{{-0.60}}$ & ${2.06^{+0.43}_{-0.37}}$ & ${0.66^{+0.25}_{-0.17}}$ &  &  &  & ${118.0~(154)}$ & ${0.0}$ \\
XMM02347 & B02 & ${4.07^{+2.12}_{-1.61}}$ & ${2.0^f}$ & ${0.39^{+0.10}_{-0.09}}$ &  &  &  & ${265.6~(264)}$ & ${0.0}$  \\
XMM02660 & {B02}  & ${5.64^{+3.51}_{-2.09}}$ & ${1.87^{+0.55}_{-0.33}}$ & ${0.83^{+0.62}_{-0.26}}$ &  &  &  & ${245.4~(287)}$ & ${0.0}$ \\
XMM03098 & B02  & $1.63^{{+1.20}}_{{-0.59}}$ & $2.0^f$ & ${0.29^{+0.08}_{-0.04}}$ &  &  &  & ${211.1}~(238)$ & ${0.0}$ \\
XMM03153 & B02  & $2.23^{{+2.90}}_{{-1.12}}$ & $1.73^{{+0.31}}_{{-0.18}}$ & $0.37^{{+0.12}}_{{-0.05}}$ &  &  &  & $356.8~(376)$ & ${0.0}$ \\
XMM03342 & B02  & ${1.63^{+2.28}_{-0.57}}$ & ${2.27^{+0.30}_{-0.50}}$ & ${0.23^{+0.16}_{-0.06}}$ &  &  &  & ${141.3}~(130)$ & ${0.0}$ \\
XMM03798 & B02  & ${4.83^{+2.05}_{-2.28}}$ & ${2.29^{+0.27}_{-0.50}}$ & ${0.69^{+0.25}_{-0.26}}$ &  &  &  & $221.7~(248)$ & ${0.0}$ \\
XMM03900 & {B02}  & ${5.45^{+1.53}_{-1.11}}$ & $2.0^f$ & ${1.73^{+0.28}_{-0.21}}$ &  &  &  & ${573.3~(637)}$ & ${0.0}$ \\
XMM03916 & B02+L & ${120.86^{+110.09}_{-30.26}}$ & $2.0^f$ & ${2.56^{+5.29}_{-1.00}}$ &  & $1.49^{{+0.20}}_{{-0.13}}$ & $0.26^{{+0.76}}_{{-0.22}}$ & ${171.5}~(165)$ & ${0.30}$  \\
XMM04259 & B02  & ${13.56^{+16.78}_{-4.24}}$ & $1.76^{{+0.78}}_{{-0.25}}$ & ${0.54^{+1.36}_{-0.17}}$ &  &  &  & $192.0~(212)$ & ${0.0}$ \\
XMM04404 & B02  & $1.03^{+0.03}_{-0.03}$ & $1.89^{+0.05}_{-0.01}$ & $0.55^{+0.01}_{-0.01}$ &  &  &  & $327.7~(386)$ & $0.0$ \\
XMM04475 & {B02}  & ${1.63^{+5.84}_{-0.54}}$ & ${2.0^f}$ & ${0.13^{+0.09}_{-0.02}}$ &  &  &  & ${308.4~(305)}$ & ${0.0}$ \\
XMM04583 & B02  & ${3.55^{+1.14}_{-0.93}}$ & ${2.0^f}$ & ${1.14^{+0.26}_{-0.22}}$ &  &  &  & ${412.6~(462)}$ & ${0.0}$ \\
XMM04804 & {B02}  & ${3.73^{+2.67}_{-2.00}}$ & ${2.02^{+0.50}_{-0.44}}$ & ${0.50^{+0.31}_{-0.18}}$ &  &  &  & ${224.9~(234)}$ & ${0.0}$ \\
XMM04899 & {B02}  & ${8.81^{+3.83}_{-2.15}}$ & ${2.0^f}$ & ${0.81^{+0.23}_{-0.14}}$ &  &  &  & ${295.4~(328)}$ & ${0.0}$ \\
\hline
\end{tabular}
\begin{tablenotes}
\item Notes -
{Column (1): Source XIDs, Column (2): the best-fit model where B02 represents {\scshape tbabs$\times$(borus02 + ztbabs$\times$cabs$\times$cutoffpl)}, and cpl represents an additional {\scshape cutoffpl} accounting for the scattered X-ray emission, Column (3): line-of-sight column
density, Column (4): photon index, Column(5): normalisation of the transmitted cutoff power law component, Column(6): normalisation of the scattered cutoff power law component, Column (7) Fe K$\alpha$ line energy at the observed$-$frame, Column (8): equivalent width of Fe K$\alpha$ line,
Column (9): the best-fit Cash statistics and degrees of freedom, Column (10): MCMC-based probability of source being a CT-AGN.
}
\end{tablenotes}
\end{threeparttable}
\end{adjustbox}
\label{tab:B02}
\end{table*}
\subsection{X-ray spectral fitting with {\scshape BORUS02} model}
In addition to modelling the X-ray spectra of our sample sources with a simple absorbed power law, we also attempted to use a
physically motivated model which considers reprocessing of X-ray emission from the circumnuclear material in AGN.
It is widely accepted that the circumnuclear material in AGN is distributed in the form of a torus \citep{Elitzur06,Netzer15,Zhao21}.
The torus-based models have commonly been used to model the X-ray spectra of DOGs, even those with low counts in their
spectra \citep[see][]{Vito18,Laloux23,Yan23}.
Therefore, we used {\scshape borus02} \citep{Balokovic18}, a torus-based model, which assumes a smooth distribution of matter
in a spherical geometry with two polar conical cuts. This model also accounts for Compton scattering of X-ray photons
and Fe K emission lines in a self-consistent manner.
\par
To fit the X-ray spectra of our DOGs, we used a model that can be expressed as
$c_1 ~\times$ {\scshape tbabs} $\times$ ( {\scshape borus02} + {\scshape ztbabs} $\times$ {\scshape cabs} $\times$ {\scshape cutoffpl}),
where $c_1$ represents cross-calibration factor in case of fitting multi-instrument spectra.
The transmitted component is represented by
a cut-off power law ({\scshape cutoffpl}) which includes the effects of absorption ({\scshape ztbabs}) and Compton-scattering
losses along the line-of-sight ({\scshape cabs}). The {\scshape borus02} model accounts for the reprocessed AGN emission component.
We used the {\scshape borus02} model in its standard form by keeping all the parameters free.
Although, we tied line-of-sight column density to the average torus column density considering low-count spectra.
Often, we were unable to constrain the covering factor ($f_c$) and we fixed it to 0.5, whenever it remained
unconstrained. In case of unusually flat ($<$ 1.7) photon index, we fixed it to the value obtained from the absorbed power law model, if constrained,
otherwise, we fixed it to 2.0.
\par
We note that the {\scshape borus02} model inherently includes reflection component, and, for the most of our sample sources,
the reflection component is insignificant in comparison to the transmitted component (see Figure~\ref{fig:BestFits}).
Due to a very weak or nearly absent reflection component, the {\scshape borus02} model is unable to reproduce the Fe K$\alpha$ line self-consistently.
Therefore, we added the Fe K$\alpha$ emission line in five of our sample sources in which the presence of the emission line was evident
from the absorbed power law model fittings. We kept the line width fixed to 1 eV and the line energy as a free parameter.
The Fe K$\alpha$ line parameters (line energy and EW) are similar to those found in the absorbed power law model fittings.
In XMM01723, an additional cutoff power law representing a leaked or scattered component of intrinsic emission marginally improves
the fit-statistics, which is consistent with the spectral fitting performed with the absorbed power law model (see Section~\ref{sec:PL}).
In Table~\ref{tab:B02}, we list the best-fitted parameters provided by the {\scshape borus02}.
In Figure~\ref{fig:BestFits}, we show the best-fitted spectra of our sample sources modelled with the {\scshape borus02} model.
Based on the comparison of fit statistics listed in Table~\ref{tab:XrayPOW} and Table~\ref{tab:B02}, we find that
both simple absorbed power law as well as {\scshape borus02} model provide reasonably good fits to the X-ray spectra of our sample sources.
\section {Results and Discussion}
\label{sec:Discussion}
\begin{table*}
\centering
\caption{Comparison of various parameters}
\renewcommand{\arraystretch}{1.2}
\begin{adjustbox}{width=\textwidth}
\begin{threeparttable}
\begin{tabular}{ccccccc}
\hline
Reference & No. of & Redshift & $S_{24~{\mu}m}$ &  $m_{\rm r}$ & $N_{\rm H}$          & log$L^{\rm int}_{\rm 2-10~keV}$ \\
          & DOGs   &  ( $z$)   &       (mJy)         &     (mag)    & (10$^{22}$~cm${^{-2}}$) & (erg~s$^{-1}$)            \\
(1) & (2) & (3) & (4) & (5) & (6) & (7) \\
\hline
1    & 34 & 0.586$-$4.65 (1.75)   & 0.39$-$8.11 (0.81)  & 22.64$-$25.94 (24.68) & $1.02-120.86 ~(3.29)$ & ${43.30-45.79 (44.45})$  \\
2    & 14 & 1.22$-$5.22~(2.29)    & 0.08$-$1.08~(0.24) & 24.51$-$27.01~(26.34) & 0.80$-$900.0~(17.0)  & 41.50$-$44.82~(43.57)  \\
3    & 10 & 2.085$-$2.658~(2.503) & 1.92$-$19.16 (7.50)& 20.76$-$22.37 (22.07) & 1.00$-$8.0~(1.95)   & 44.10$-$45.60~(45.10)  \\
4    & 6  & 0.282$-$1.023~(0.775) &9.02$-$16.19 (16.07)& 18.22$-$21.56 (21.29) & 4.10$-$67.20~(18.10) & 43.30$-$45.30~(44.20)  \\
\hline
\end{tabular}
\begin{tablenotes}
\item Note: Reference - 1 - This work; 2 - \citet{Corral16}; 3 - \citet{Lansbury20}; 4 - \citet{Zou20}. The IR fluxes given in column (4)
for references 3 and 4 are from WISE $22~\mu$m band.
\end{tablenotes}
\end{threeparttable}
\end{adjustbox}
\label{tab:Results}
\end{table*}
\subsection{Absorbing column densities}
\label{sec:NH}
\begin{figure}
\centering
\includegraphics[angle=0, width=\columnwidth, trim={0.0cm 0.0cm 0.0cm 0.0cm}, clip]{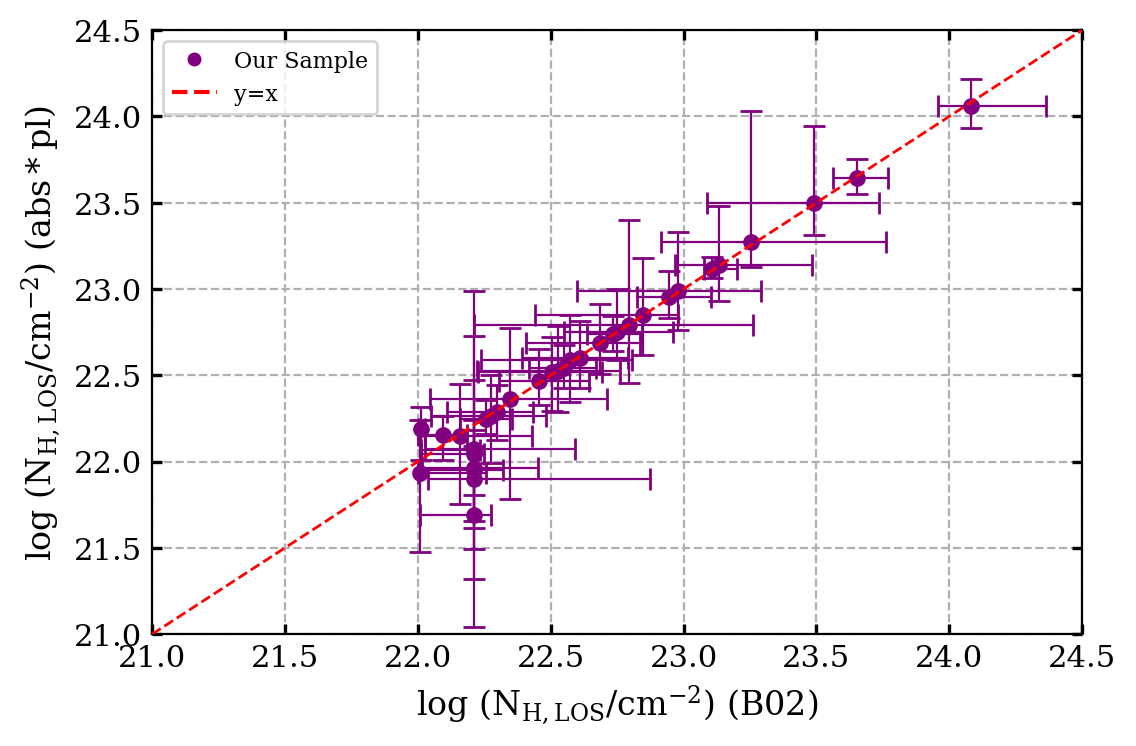}
\caption{The comparison of $N_{\rm H}$ obtained from two different models, {\ie} absorbed power law
and {\scshape borus02} models.}
\label{fig:NHLXcor}
\end{figure}

One of the main objectives of our study is to constrain the line-of-sight absorbing column densities ($N_{\rm H}$) in DOGs and estimate the fraction
of CT-AGN in them. For our DOGs, we obtained $N_{\rm H}$ by modelling their 0.5$-$10 keV X-ray spectra using two different models,
{\ie} absorbed power law and {\scshape borus02}.
We find that both models provide nearly similar  $N_{\rm H}$ estimates (see Figure~\ref{fig:NHLXcor}).
We prefer to use $N_{\rm H}$ estimates derived from the {\scshape borus02} model, considering the fact that it accounts for Compton scattering, as
well as reprocessing of X-ray emission.
We find that our sample DOGs have $N_{\rm H}$ in the range
of $1.02$ $\times$ 10$^{22}$~cm$^{-2}$ to 1.21 $\times$ 10$^{24}$~cm$^{-2}$ with a median value of
$3.29$ $\times$ 10$^{22}$~cm$^{-2}$ (see Table~\ref{tab:Results}).
There are only 06/34 (17.6 per cent) sources which can be classified as heavily obscured AGN with $N_{\rm H}$ $>$ 10$^{23}$~cm$^{-2}$,
and the remaining DOGs show only moderate level obscuration (10$^{22}$~cm$^{-2}$ $\leq$ $N_{\rm H}$ $\leq$ 10$^{23}$~cm$^{-2}$).
We note that the $N_{\rm H}$ estimates in our sample DOGs are broadly consistent with other samples of DOGs reported in the literature
\citep[e.g.,][]{Corral16,Lansbury20,Zou20}.
\par
In Table~\ref{tab:Results}, we present a comparison of $N_{\rm H}$ and other parameters,
{\it i.e.,} redshift, 24~$\mu$m flux, optical $r-$band magnitude, and absorption corrected 2.0$-$10~keV X-ray luminosity
of our DOGs with those reported in the literature.
We caution that despite apparently similar $N_{\rm H}$ estimates found in different samples of DOGs, we need to account for
the inherent biases in the samples and differences in the spectral quality.
For instance, \cite{Corral16} derived $N_{\rm H}$ for a sample of 14 DOGs in the CDFS using 6 Ms deep {\em Chandra} observations
and an additional 3 Ms {\em XMM-Newton} data for 07/14 sources.
They demonstrated that, for the same set of sources, better quality spectra often render  significantly
higher $N_{\rm H}$ than that derived from their low counts spectra presented in \cite{Georgantopoulos11}.
We find that $N_{\rm H}$ estimates for relatively bright DOGs reported in \cite{Zou20} and \cite{Lansbury20} are somewhat lower than that found
in our sample. We point out that \cite{Zou20} performed X-ray spectral study of six relatively bright nearby (0.3 $<$ $z$ $<$ 1.0) DOGs
using {\em Chandra} snapshot observations with only 3 ks exposure time for each source, and they found only moderately obscured AGN
(4.1 $\times$ 10$^{22}$~cm$^{-2}$ $\leq$ $N_{\rm H}$ $\leq$ 6.7 $\times$ 10$^{23}$~cm$^{-2}$; see Table~\ref{tab:Results}).
\cite{Lansbury20} also used relatively low exposure (nearly 10$-$ 30 ks exposure time) {\em XMM-Newton} observations for a sample of ten
heavily dust-reddened quasars and found $N_{\rm H}$ in the range of 1 $\times$ 10$^{22}$ cm$^{-2}$ $-$ 8 $\times$ 10$^{22}$ cm$^{-2}$.
We caution that a robust comparison of $N_{\rm H}$ across different samples of DOGs warrants sufficiently good-quality spectra.
Although, lower obscuration can also be attributed to sample biases such as sources belonging to late evolutionary stages (see Section~\ref{sec:Edd}).
\subsection{Compton-thick AGN in DOGs}
In our sample, we found only one source (XMM03916) with $N_{\rm H}$ $=$ 1.21 $\times$ 10$^{24}$~cm$^{-2}$, which
can be classified as CT-AGN candidate.
To confirm the CT nature, we applied the Markov Chain Monte Carlo (MCMC) parameter estimation techniques on the best fits of all our sample sources and obtained the probability distribution function (PDF) of $N_{\rm H}$. To perform the MCMC, we used the Goodman–Weare algorithm
\citep{Goodman10} within the {\scshape xspec}. For robust estimation of the parameter space, we used 20 walkers and 10$^{4}$ steps in the MCMC and derived $N_{\rm H}$ PDFs for each sample source.
Using the MCMC on both the absorbed power law and the {\scshape borus02} model best fits, we derived the probability of being
CT-AGN ($P_{\rm CT}$) for each source (see Table~\ref{tab:XrayPOW} and Table~\ref{tab:B02}).
We find that, as expected, only XMM03916 shows non-zero probability of being CT-AGN with
$P_{\rm CT}$ values of 0.12 and 0.30 obtained from the power law and {\scshape borus02} best fits, respectively.
The $N_{\rm H}$ PDF of XMM03916, a CT-AGN candidate, is shown in Figure~\ref{fig:PDFs}.
The fraction of CT-AGN candidates in our sample is only (01/34) 3.0 per cent.
We note that the comparison of CT-AGN fraction among the different samples of DOGs may not be straightforward due to the small
number statistics and inherent sample biases.
We point out that using low-count X-ray spectra \cite{Zou20} and \cite{Lansbury20} found no CT-AGN in their small samples
of relatively bright DOGs. However, using deep {\em Chandra} and {\em XMM-Newton} observations, \cite{Corral16} found (3/14) 21 per cent of
their sample sources as CT-AGN.
The fraction of CT-AGN in deep X-ray surveys has been limited to nearly 10 per cent \citep[see][]{Lanzuisi15,Lanzuisi18,Marchesi16,Masini18,Li19},
which is also higher than that found in our study.
\par
Further, we compare our heavily obscured sources with those reported by \cite{Yan23} who
used Bayesian spectral analysis and identified 12 CT-AGN and 58 heavily obscured AGN (23.70 $\leq$ log$N_{\rm H}$ $<$ 24.17) among
all the X-ray detected sources in the XMM-SERVS coverage of the XMM-LSS.
Since our DOGs are gleaned from the X-ray detected sources in the same field, one can expect a significant overlap between the two samples.
Surprisingly, none except one (XMM00497) of our sample sources is found among the CT-AGN and heavily obscured AGN reported in \cite{Yan23}.
Our CT-AGN candidate XMM03916 is not identified as either CT-AGN or heavily obscured AGN in their study.
This discrepancy can be attributed to several factors that include differences in spectral model, redshift, spectral quality
and posterior $N_{\rm H}$ probability cutoff limit.
We point out that, to all their sample sources, \cite{Yan23} uniformly used a model
which can be expressed as {\scshape a$\star$phabs$\star$(borus02 + zphabs$\star$cabs$\star$cutoffpl + b$\star$cutoffpl)} in the {\scshape xspec}.
This model is fairly similar to our {\scshape borus02} model except for the additional scattered component modelled with cutoff power law considered for
all the sources. In our study, we find that the addition of the scattered power law component provides marginal improvement in the fit statistics of
only one of our sample sources (see Table~\ref{tab:B02}).
Further, unlike \cite{Yan23}, who used photo$-z$ based on the forced photometry \citep{Zou22}, we used photo$-z$ derived from the
Tractor image-modelling software-based de-blended multi-band forced photometry across 13 optical to near-IR bands \citep{Nyland23},
which are considered to be more accurate than previous estimates \citep[e.g.,][]{Ni21,Zou22}.
For some of our sources, photo$-z$ estimates from \cite{Nyland23} and \cite{Zou22} are substantially different.
Also, to get the better spectral quality, we have added {\em Chandra} data, whenever available.
It is worth pointing out that \cite{Yan23} identified a source as CT-AGN or heavily obscured AGN only if its posterior $N_{\rm H}$ probability is
$>$ 50 per cent for the threshold $N_{\rm H}$ values set to 1.5 $\times$ 10$^{24}$ cm$^{-2}$
and 5 $\times$ 10$^{23}$~cm$^{-2}$ for CT-AGN and heavily obscured AGN, respectively.
We note that the cutoff limit placed on the posterior $N_{\rm H}$ probability in \cite{Yan23} is somewhat arbitrary and strict.
Sources with posterior $N_{\rm H}$ probability $<$ 50 per cent have only lesser chances of being CT-AGN/ heavily obscured AGN but
cannot be completely ruled out.
\cite{Lanzuisi18} demonstrated the existence of a large number of CT-AGN with only 5 per cent of their posterior $N_{\rm H}$ probability
above 10$^{24}$ cm$^{-2}$. {\cite{Akylas16} identified 53/604 CT-AGN considering posterior $N_{\rm H}$ probability only $\geq$ 3.0 per cent.
Therefore, it is fairly possible that our CT-AGN candidate and heavily obscured AGN are missed by \cite{Yan23} due
to a much higher cutoff limit (50 per cent) placed on the posterior $N_{\rm H}$ probability.}
\subsection{The 2.0$-$10 X-ray luminosity versus 6~${\mu}$m MIR luminosity diagnostic plot}
\begin{figure*}
\centering
\includegraphics[angle=0, width=\columnwidth, trim={0.0cm 0.0cm 0.0cm 0.0cm}, clip]{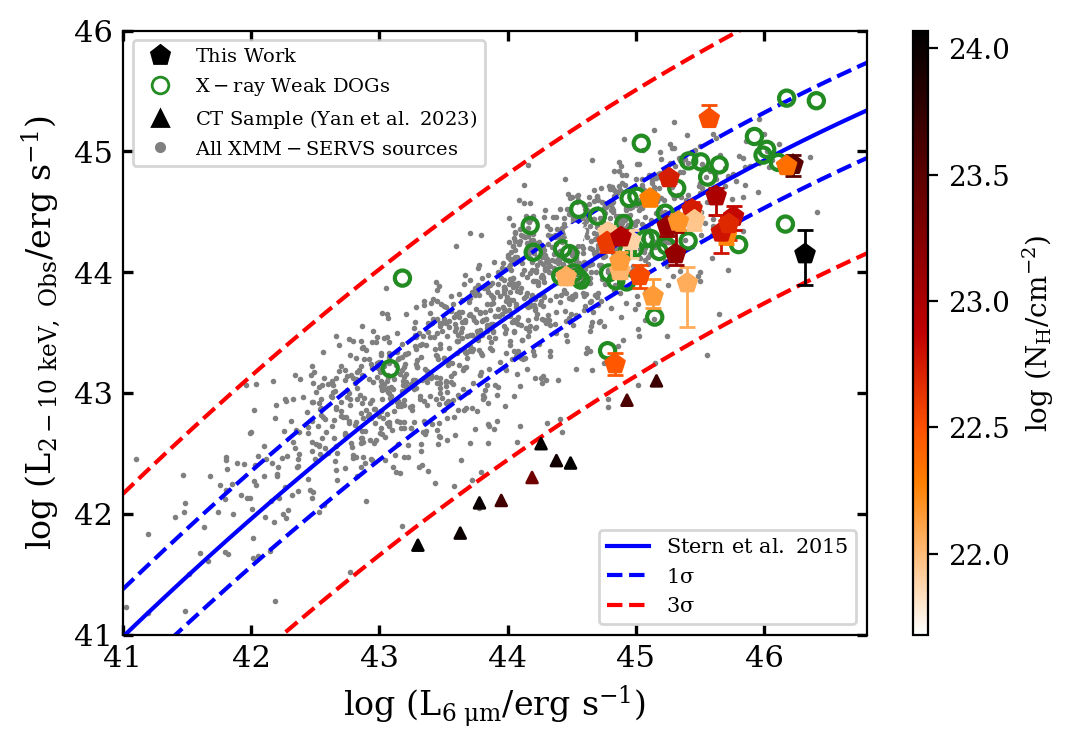}
\includegraphics[angle=0, width=\columnwidth, trim={0.0cm 0.0cm 0.0cm 0.0cm}, clip]{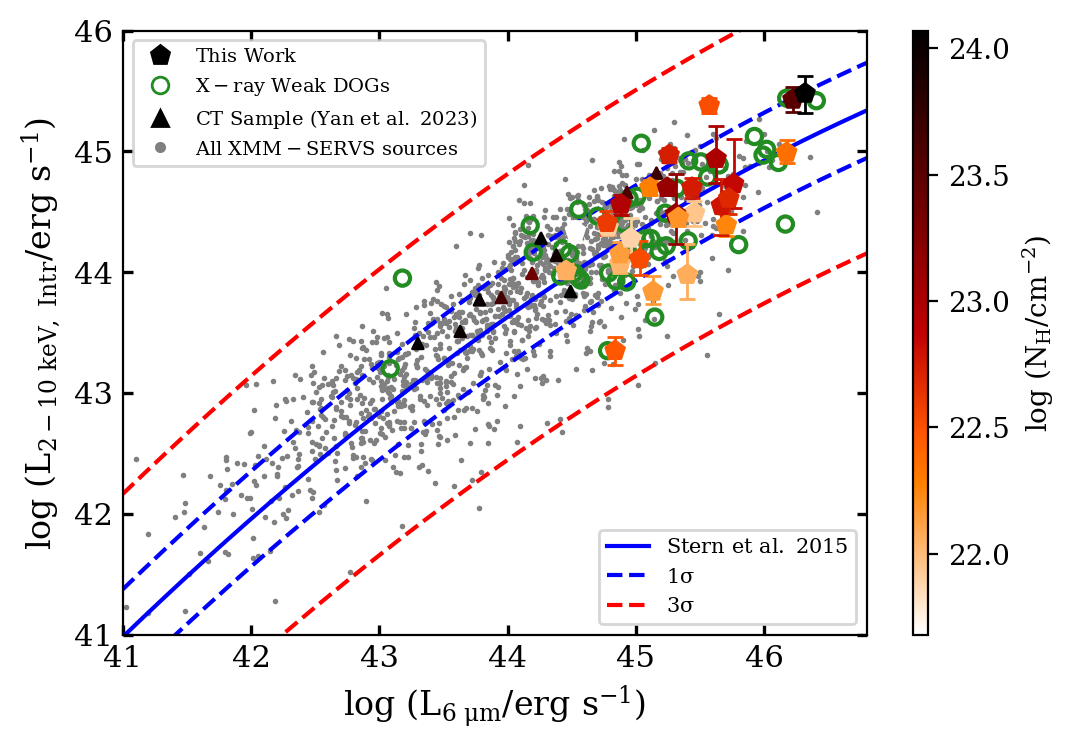}
\caption{{\it Left panel} : The plot of observed 2.0$-$10~keV X-ray luminosity ($L_{\rm 2.0-10~keV}$) versus mid-IR 6~${\mu}m$ luminosity ($L_{6.0~{\mu}m}$) for various types of X-ray detected AGN.
{\it Right panel} : Same as the left panel plot except for absorption-corrected 2.0$-$10~keV X-ray luminosities ($L_{\rm 2.0-10~keV}$) are plotted for
our sample sources and CT-AGN from \citet{Yan23}. Our sample DOGs are shown with filled pentagon symbols. The X-ray faint DOGs with no X-ray spectral
analysis are shown with green open circles. The X-ray sources from \citet{Chen18} are shown with grey dots.
The CT-AGN reported in \citet{Yan23} are shown with triangles. The line-of-sight column densities, whenever available, are indicated with a colour bar.
The blue solid curve represents the $L_{\rm 2.0-10~keV}$ $-$ $L_{6.0~{\mu}m}$ correlation reported by Stern et al. 2015. The blue dashed and red dashed curves on either side represent 1$\sigma$ and 3$\sigma$ dispersion, respectively, for the X-ray source population reported by \citet{Chen18}.}
\label{fig:LxVsL6}
\end{figure*}

To further investigate the nature of the obscured AGN hosted in our DOGs, we exploit the
correlation between 2.0$-$10~keV X-ray luminosity ($L_{\rm 2.0-10~keV}$) and 6.0~${\mu}m$ mid-IR luminosity ($L_{ 6~{\mu}m}$).
Due to the heavy absorption, the observed X-ray luminosities of CT-AGN and obscured AGN are expected to get suppressed while their
mid-IR luminosities remain nearly unaffected by the obscuring medium. Thus, $L_{\rm 2.0-10~keV}$ $-$ $L_{6~{\mu}m}$ correlation is
commonly used as a diagnostic to identify CT-AGN, which shows a large deviation with respect to the unobscured or less obscured AGN \citep{Lanzuisi18,Guo21}.
In Figure~\ref{fig:LxVsL6} ({\it left panel}), we plot observed $L_{\rm 2.0-10~keV}$ versus $L_{6~{\mu}m}$ for our sample sources, as well as
other different kinds of X-ray sources, {\eg} X-ray faint DOGs with no X-ray spectral analysis, CT-AGN from \cite{Yan23}, and
X-ray detected sources from \cite{Chen18} in the XMM-LSS field.
To maintain uniformity with other X-ray sources, we obtained 2.0$-$10~keV luminosities from the absorbed power law model for our
sample sources. For X-ray faint DOGs and X-ray sources, we obtained 2.0$-$10~keV flux from
the XMM-SERVS catalogue \citep{Chen18}.
The 6.0~${\mu}m$ luminosities are taken from \cite{Zou22} and these are available only for
30/34 of our sample DOGs, 42/55 X-ray faint DOGs, ${1401}$ X-ray sources with spec$-z$, and 10 CT-AGN from \cite{Yan23}.
\par
We compared our sources with the $L_{\rm 2.0-10~keV}$ $-$ $L_{6~{\mu}m}$ correlation reported by \cite{Stern15}, who probed it for a sample of
radio-quiet AGN distributed across a wide range of X-ray luminosities (10$^{42}$ erg~s$^{-1}$ $-$ 10$^{46}$ erg~s$^{-1}$).
The correlation can be expressed as $L_{\rm 2.0-10~keV}$ = $40.981 + 1.024x - 0.047x^{2}$, where $x$ = log$({\nu}L_{\nu}$(6~${\mu}$m)/10$^{41}$~erg~s$^{-1}$), and $L_{\rm 2.0-10~keV}$ is in the units of erg~s$^{-1}$.
We find that our CT-AGN candidate XMM03916 as well as CT-AGN reported in \cite{Yan23} show
3$\sigma$ or larger deviation from the $L_{\rm 2.0-10~keV}$ $-$ $L_{6~{\mu}m}$ correlation curve.
Our DOGs with moderate absorption seem to be consistent with the $L_{\rm 2.0-10~keV}$ $-$ $L_{6~{\mu}m}$ correlation within 1$\sigma$ deviation.
Thus, the CT-AGN candidate identified in our sample is consistent with the $L_{\rm 2.0-10~keV}$ $-$ $L_{6~{\mu}m}$ correlation diagnostic.
We note that, in comparison to the CT-AGN reported in \cite{Yan23}, our DOGs are systematically more luminous both in X-ray as well as at 6.0
$\mu$m (see Figure~\ref{fig:LxVsL6}).
The systematically high luminosities of our sample DOGs can be attributed to their much higher redshifts
(0.586 $\leq$ $z$ $\leq$ 4.65 with a median value of 1.75) than that for CT-AGN (0.058 $\leq$ $z$ $\leq$ 1.98 with a median value of 1.03).
\par
We also overplotted the absorption-corrected X-ray luminosities of our sample sources and CT-AGN on
the $L_{\rm 2.0-10~keV}$ $-$ $L_{6~{\mu}m}$ correlation (see Figure~\ref{fig:LxVsL6}, right panel).
As expected, all of our sample DOGs, as well as CT-AGN lie on the $L_{\rm 2.0-10~keV}$ $-$ $L_{6~{\mu}m}$ correlation curve mostly
within 1$\sigma$ deviation once their X-ray luminosities are corrected for absorption.
This further confirms the veracity of $N_{\rm H}$ estimates in our sample sources.
Based on the absorption-corrected X-ray luminosities, which span across 2.00 $\times$ 10$^{43}$ erg~s$^{-1}$ $\leq$ $L_{\rm 2.0-10~keV}$
$\leq$ 6.17 $\times$ 10$^{45}$ erg~s$^{-1}$, our sample DOGs can be classified as highly luminous quasars with varying levels of obscuration
around them.
\subsection{$N_{\rm H}$ versus Eddington ratios: Evolutionary scenario of AGN in DOGs }
\label{sec:Edd}
\begin{figure}
\centering
\includegraphics[angle=0, width=\columnwidth, trim={0.0cm 0.0cm 0.0cm 0.0cm}, clip]{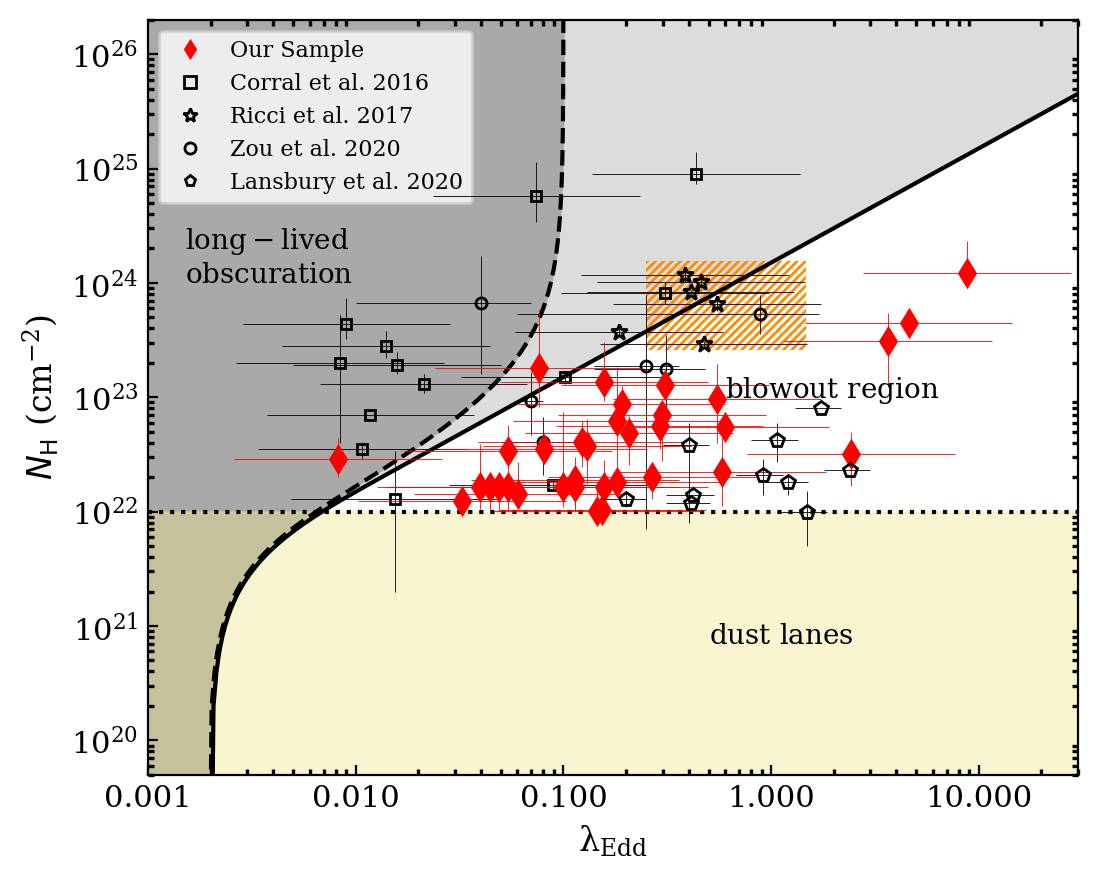}
\caption{The $N_{\rm H}$ versus Eddington ratio plot for our sample DOGs. The DOGs and reddened quasars reported in the literature
\citep{Corral16,Ricci17,Zou20,Lansbury20} are also plotted. The grey-shaded region represents region for long-lived obscuration.
The solid curve and dashed curve segregating long-lived obscuration and short-lived obscuration
are based on single-scattering limit and radiation-trapping limit, respectively \citep{Ishibashi18}.
The yellow-shaded region with low absorption represents obscuration caused by host galaxy dust lanes.
The orange-shaded region belongs to Hot DOGs reported in \citet{Vito18,Wu18}.}
\label{fig:NHVsEdd}
\end{figure}

Since our DOGs host highly luminous AGN, the surrounding obscuring material is likely to be affected by the radiative feedback.
To gain insights into the evolutionary stage of AGN hosted in our DOGs, we exploited
the $N_{\rm H}$ versus Eddington ratio (${\lambda}_{\rm Edd}$)
diagnostic plot. In Figure~\ref{fig:NHVsEdd}, we plot the $N_{\rm H}$ versus ${\lambda}_{\rm Edd}$ for our sample DOGs, as well as
DOGs reported in the literature \citep{Corral16,Ricci17,Zou20,Lansbury20}.
For our sample sources, we estimated ${\lambda}_{\rm Edd}$ = $\frac{L_{\rm bol}}{1.26 {\times} 10^{38}~M_{\rm BH}}$;
where bolometric luminosity ($L_{\rm bol}$) is measured in the units of erg~s$^{-1}$ and black hole mass ($M_{\rm BH}$) in the units of $M_{\odot}$.
We obtained $L_{\rm bol}$ estimates from $L_{\rm 2-10~keV}$ using $L_{\rm bol} = K_xL_{\rm 2 - 10~keV}$ correlation \citep{Duras20}, where the bolometric correction factor $K_x$ is given by
\begin{equation*}
K_x = a \left [ 1 + \left (\frac{{\rm log} (L_{\rm 2 - 10~keV}/L_{\odot})}{b} \right)^c \right ]
\end{equation*}
where, $a$ = 15.33, $b$ = 11.48 and $c$ = 16.20.
Since black hole masses of our DOGs are not available, we assume $M_{\rm BH}$ = 10$^{8.5}$ M$_\odot$, which is an average value for
DOGs reported by \cite{Zou20}. To account for the deviation of the actual value of $M_{\rm BH}$ from the average value,
we consider 10$^{8}$ $M_{\odot}$ $-$ 10$^{9}$ $M_{\odot}$ range for $M_{\rm BH}$.
The large error bars introduced in ${\lambda}_{\rm Edd}$ correspond to the range of $M_{\rm BH}$, and these are much larger than
that contributed by the errors in X-ray luminosities.
\par
In the $N_{\rm H}$ versus ${\lambda}_{\rm Edd}$ plot, we show different regions, {\ie} blowout region having short-lived obscuration region, long-lived obscuration region, and host-galaxy dust lanes caused obscuration region.
The tracks segregating short-lived and long-lived obscuring regions represent the effective column density ($N_{\rm H}$)
around AGN characterised with ${\lambda}_{\rm Edd}$. The solid and dotted tracks are obtained by assuming
single scattering \citep{Fabian09} and radiation trapping \citep{Ishibashi18} of photons emitted from the AGN, respectively.
The circumnuclear material is expected to undergo a fast blowout phase if column densities are lower,
and hence resulting in AGN to lie in the blowout region.
The obscuration is likely to sustain against outflows, which in turn results in long-lived obscuration around AGN, if column densities are much higher.
\par
From Figure~\ref{fig:NHVsEdd}, it is evident that the most of our DOGs, irrespective of their obscuration level,
lie in the blowout region.
Thus, obscuration in our DOGs is short-lived and they are likely to evolve into unobscured AGN.
The 04/34 ($\sim$ 11.7 per cent) of our DOGs showing super Eddington accretion rates and high obscuration ($N_{\rm H}$ $>$ 10$^{23}$~cm$^{-2}$)
can be Hot DOGs, which are known to exhibit higher obscuration than reddened quasars \citep{Vito18}.
Hot DOGs supposedly belong to an early phase of evolution during which accretion peaks, but the feedback is yet to blow out the
surrounding reservoir of gas and dust.
In contrast, reddened quasars belong to a later phase during which feedback is dominantly ongoing.
Further, reddened quasars can show different levels of X-ray obscuration depending upon their evolutionary
stages \citep{Goulding18}. In other words, reddened quasars can represent a heterogeneous population belonging
to an early evolutionary phase just after the Hot DOGs, as well as a late phase during which radiative feedback
has swept away surrounding material.
From the $N_{\rm H}$ versus ${\lambda}_{\rm Edd}$ plot, we conclude that all but four of our DOGs belong to the
intermediate to late evolutionary phase, during which dominant AGN feedback has blown away most of the surrounding obscuring material.
\section{Conclusions}
\label{sec:Conclusions}
In this work, we present the X-ray spectral properties of 34 DOGs using deep {\em XMM-Newton} observations in the XMM-SERVS
coverage of the XMM-LSS field. To achieve better-quality spectra, we combined all the archival {\em XMM-Newton} pn and MOS observations, and utilised
{\em Chandra}/ACIS observations, whenever available. Our conclusions are outlined as below.
\begin{itemize}
\item We find that the 0.5$-$10 keV X-ray spectra of our DOGs can be fitted with a simple absorbed power law as well as with
{\scshape borus02}, a physical model assuming toroidal geometry of the obscuring material around AGN.
Both the models give similar absorbing column densities and photon indices.
In one of our sample sources, an additional scattered component is also required to obtain the best fit.
\item The line-of-sight absorbing column densities in our DOGs span across a wide range from
$1.02~ \times$ 10$^{22}$ cm$^{-2}$ to 1.21 $\times$ 10$^{24}$ cm$^{-2}$ with a median value of $3.29$ $\times$ 10$^{22}$ cm$^{-2}$.
Nearly 06/34 (17.6 per cent) of our sample sources can be classified as heavily obscured AGN with $N_{\rm H}$ $>$ 1.0 $\times$ 10$^{23}$ cm$^{-2}$.
Thus, AGN hosted in DOGs show varying levels of obscuration ranging from moderately obscured AGN to CT-AGN.
\item
In our work, we identified one new CT-AGN candidate XMM03916 which
have $N_{\rm H}$ $>$ 1.1 $\times$ 10$^{24}$ cm$^{-2}$ and posterior probability of being CT-AGN ($P_{\rm CT}$)  0.30.
As expected, our CT-AGN candidate follows $L_{\rm 2.0-10~keV}$ $-$ $L_{6~{\mu}m}$ correlation once the X-ray luminosity
is corrected for absorption. The consistency between our CT-AGN candidate and previously identified CT-AGN provides further confirmation
of its CT nature.

\item
The fraction of CT-AGN candidates in our sample is merely 01/34 = 3.0 per cent, which is lower than that (nearly 10 per cent) found in the deep X-ray surveys \citep{Marchesi16,Lanzuisi18,Li19}. We caution that the moderate to poor spectral quality and inherent sample biases may affect the fraction of CT-AGN.
\item The AGN hosted in our DOGs are found to be highly luminous with their absorption-corrected X-ray luminosities spanning
across 2.00 $\times$ 10$^{43}$ erg~s$^{-1}$ to 6.17 $\times$ 10$^{45}$ erg~s$^{-1}$ with
a median value of 2.82 $\times$ 10$^{44}$ erg~s$^{-1}$, which suggest them to be luminous quasars.
\item In the $N_{\rm H}$ versus ${\lambda}_{\rm Edd}$ diagnostic plot, most of our DOGs belong to the blowout
region, suggesting a short-lived obscuration. We find that all but four of our DOGs show similarity
with reddened quasars. The four DOGs with super Eddington accretion and high obscuration, are likely to be Hot DOGs, which
belong to an early evolutionary phase during which accretion as well as obscuration peaks.
Thus, on the basis of $N_{\rm H}$ versus ${\lambda}_{\rm Edd}$ diagnostic plot, we conclude that
our DOGs are likely to represent a heterogeneous population belonging to the early to late evolutionary phases.
\end{itemize}
\section*{Acknowledgments}
We thank the anonymous reviewers for their useful comments and suggestions which helped us to improve the manuscript.
The research work at the Physical Research Laboratory is funded by the Department of Space, Government of India.
This research used observations obtained with {\em XMM-Newton}, an ESA science mission with instruments
and contributions directly funded by ESA Member States and NASA.
This research has made use of data obtained from the Chandra Data Archive and the Chandra Source Catalog,
and software provided by the Chandra X-ray Center (CXC) in the application packages CIAO.
We acknowledge the use of Spitzer data provided by the Spitzer Science Center.
This research has made use of data obtained through the High Energy Astrophysics Science Archive Research
Center Online Service, provided by the NASA/Goddard Space Flight Center.
%
%
\section*{Facilities}
{\em XMM-Newton}, {\em Chandra}, {\em Subaru} and {\em Spitzer}
%
\section*{Data Availability}
The {\em XMM-Newton} and {\em Chandra} data are publicly available from
the archives of HEASARC maintained by NASA.
%

%
\bibliographystyle{mnras}
\bibliography{RadioDOGPaper2} 


\appendix
\section{The X-ray observation details of our DOGs}
\begin{table*}
\centering
 \caption{The X-ray observations log of our DOGs.}
 \begin{threeparttable}
\begin{tabular}{cccccc}
\hline
 XID  & Obs ID & Obs Date               & $T_{\rm obs}$ & Total $T^{\rm eff}_{\rm exp}$ & ${\rm Total~Counts_{(0.5-10~keV)}}$  \\
      &        & (YYYY-MM-DD:Thh:mm:ss) &  (ks)         &  (ks)                   & (cts)\\
(1) & (2) & (3) & (4) & (5) & (6) \\
\hline
 XMM00059 & 0404967501 & 2007-01-09T18:45:25.0 & $18.9$ & $70.8$ & $695$  \\
          & 0404967901 & 2007-01-10T14:40:38.0 & $14.9$ &        &         \\
          & 0553911601 & 2008-07-03T19:15:54.0 & $13.5$ &        &        \\
          & 0742430301 & 2015-02-06T19:24:58.0 & $100.0$&        &          \\
 XMM00131 & 0112370101 & 2000-07-31T21:49:26.0 & $61.4$ & $86.5$ & $535$ \\
          & 0112371001 & 2000-08-02T20:24:27.0 & $66.0$ &        &         \\
          & 0112370601 & 2002-08-12T05:43:17.0 & $47.9$ &        &         \\
 XMM00136 & 0112370701 & 2002-08-08T15:05:39.0 & $49.6$ & $46.5$ & $176$\\
          & 0404966501 & 2006-08-09T07:21:12.0 & $11.9$ &        &         \\
 XMM00191 & 0404967401 & 2007-01-08T14:06:49.0 & $15.0$ & $70.8$ & $385$ \\
          & 0404967501 & 2007-01-09T18:45:25.0 & $18.9$ &        &         \\
          &  0404967901 & 2007-01-10T14:40:38.0 & $14.9$ &        &         \\
          & 0553911601 & 2008-07-03T19:15:54.0 & $13.5$ &        &         \\
          &  0742430301 & 2015-02-06T19:24:58.0 & $100.0$ &        &        \\
 XMM00205 & 0112371701 & 2000-08-08T04:37:14.0 & $39.6$   & $58.2$ & $125$ \\
          & 0112372001 & 2003-01-07T04:18:37.0 & $28.0$ &        &        \\
          & 0404967001 & 2007-01-08T00:52:16.0 & $14.9$ &        &         \\
XMM00250  & 0112371701 & 2000-08-08T04:37:14.0 & $39.6$ & $53.8$ & $173$\\
          & 0112372001 & 2003-01-07T04:18:37.0 & $28.0$ &        &        \\
          & 0404967001 & 2007-01-08T00:52:16.0 & $14.9$ &        &         \\
 XMM00267 & 0404967001 & 2007-01-08T00:52:16.0 & $14.9$ & $79.7$ & $1037$ \\
          & 0404967401 & 2007-01-08T14:06:49.0 & $15.0$ &        &        \\
          & 0404967501 & 2007-01-09T18:45:25.0 & $18.9$ &        &         \\
          & 0553911601 & 2008-07-03T19:15:54.0 & $13.5$ &        &        \\
          & 0742430301 & 2015-02-06T19:24:58.0 & $100.0$&        &          \\
          & 0785100101 & 2016-07-01T15:53:31.0 & $22.5$ &        &         \\
 XMM00359 & 0404967401 & 2007-01-08T14:06:49.0 & $15.0$ & $46.0$ & $160$ \\
          &  0404967501 & 2007-01-09T18:45:25.0 & $18.9$ &        &        \\
          & 0553911601 & 2008-07-03T19:15:54.0 & $13.5$ &        &         \\
          & 0742430301 & 2015-02-06T19:24:58.0 & $100.0$ &        &        \\
 XMM00393 & 0112370101 & 2000-07-31T21:49:26.0 & $61.4$  & $70.5$ & $130$ \\
          &   0112371001 & 2000-08-02T20:24:27.0 & $66.0$&        &          \\
 XMM00395 & 0112370101 & 2000-07-31T21:49:26.0 & $61.4$ & $80.9$ & $305$ \\
          & 0112371001 & 2000-08-02T20:24:27.0 & $66.0$ &        &        \\
 XMM00421 & 0112371701 & 2000-08-08T04:37:14.0 & $39.6$ & $60.0$ & $242$ \\
          &  0112372001 & 2003-01-07T04:18:37.0 & $28.0$&        &         \\
          & 0404967001 & 2007-01-08T00:52:16.0 & $14.9$ &        &         \\
          &  0785100101 & 2016-07-01T15:53:31.0 & $22.5$&        &         \\
 XMM00497 & 0112370701 & 2002-08-08T15:05:39.0 & $49.6$ & $64.4$ & $742$ \\
          & 0112370801 & 2002-08-09T05:29:19.0 & $50.8$ &        &        \\
          & 0404966401 & 2006-07-31T02:38:25.0 & $11.9$ &        &        \\
          & 0553911301 & 2008-08-10T10:08:41.0 & $13.7$ &        &         \\
 XMM00860 & 0112370401 & 2000-08-06T05:12:57.0 & $46.8$ & $79.5$ & $129$ \\
          & 0112371501 & 2000-08-06T20:08:34.0 & $11.8$ &        &        \\
          & 0404966901 & 2007-01-07T18:38:27.0 & $19.9$ &        &        \\
          & 0553911501 & 2009-01-01T17:29:55.0 & $15.0$ &        &        \\
          & 0785100101 & 2016-07-01T15:53:31.0 & $22.5$ &        &        \\
          & 0785100301 & 2016-07-02T20:09:16.0 & $26.1$ &        &         \\
          & 0793580101 & 2017-01-02T14:22:49.0 & $28.0$ &        &        \\
 XMM01034 & 0112370401 & 2000-08-06T05:12:57.0 & $46.8$ & $62.4$ & $146$  \\
          & 0112371501 & 2000-08-06T20:08:34.0 & $11.8$ &        &        \\
          & 0404966901 & 2007-01-07T18:38:27.0 & $19.9$ &        &        \\
          & 0553911501 & 2009-01-01T17:29:55.0 & $15.0$ &        &        \\
          & 0785100101 & 2016-07-01T15:53:31.0 & $22.5$ &        &        \\
          & 0785100301 & 2016-07-02T20:09:16.0 & $26.1$ &        &        \\
          & 0793580101 & 2017-01-02T14:22:49.0 & $28.0$ &        &        \\
\hline
\end{tabular}
\label{tab:Sample}
\begin{tablenotes}
\normalsize
\item
Notes - Column (1): X-ray source ID from the XMM-SERVS
catalogue reported by \cite{Chen18}, Column (2): observation IDs of {\em XMM-Newton} and {\em Chandra} observations,
`(Ch)' indicates {\em Chandra} observations, Column (3): Observation dates for the corresponding observation IDs,  Column (4): Observation times in ks for each observation IDs; {Column (5): The sum of effective exposure times from each observation IDs; Column (6): The total number of counts in 0.5$-$10 keV band}.
\end{tablenotes}
\end{threeparttable}
\end{table*}
\addtocounter{table}{-1}
\begin{table*}
\centering
 \caption{The X-ray observations log of our DOGs.}
\begin{tabular}{cccccc}
\hline
 XID  & Obs ID & Obs Date               & $T_{\rm obs}$ & Total $T^{\rm eff}_{\rm exp}$ & ${\rm Total~Counts_{(0.5-10~keV)}}$ \\
      &        & (YYYY-MM-DD:Thh:mm:ss) &  (ks)         &  (ks)                   & (cts)\\ 
(1) & (2) & (3) & (4) & (5) & (6) \\
\hline
 XMM01279 & 0112370301 & 2000-08-04T20:16:28.0 & $66.0$ & $76.1$ & $565$  \\
          &  0112370401 & 2000-08-06T05:12:57.0 & $46.8$ &        &         \\
          &  0112371501 & 2000-08-06T20:08:34.0 & $11.8$ &        &         \\
          & 0785100801 & 2016-07-05T01:04:36.0 & $23.0$ &        &         \\
          & 0793580301 & 2017-01-01T10:04:38.0 & $9.0$ &        &        \\
          & 0793580101 & 2017-01-02T14:22:49.0 & $28.0$ &        &         \\
          & 0780452301 & 2017-02-09T11:06:48.0 & $24.1$ &        &         \\
 XMM01464 & 0112370301 & 2000-08-04T20:16:28.0 & $66.0$ & $85.9$ & $531$  \\
          & 0404966601 & 2007-01-06T14:04:02.0 & $13.9$ &        &        \\
          & 0785100801 & 2016-07-05T01:04:36.0 & $23.0$ &        &         \\
          & 0780452401 & 2017-02-09T18:08:28.0 & $23.0$ &        &         \\
          & 14348 (Ch) & 2011-10-03T09:28:05 & $63.7$   &        &         \\
          & 13374 (Ch) & 2011-10-07T05:41:46 & $75.7$   &        &        \\
XMM01723 & 0037982001 & 2002-08-14T13:52:03.0 & $17.8$  & $52.7$ & $101$  \\
          &  0785100501 & 2016-07-03T10:23:26.0 & $22.0$ &        &        \\
          & 0785100801 & 2016-07-05T01:04:36.0 & $23.0$  &        &        \\
          & 0785101201 & 2016-07-06T03:57:56.0 & $25.9$  &        &        \\
          & 0793580301 & 2017-01-01T10:04:38.0 & $9.0$   &        &        \\
          & 0793580801 & 2017-01-16T20:33:01.0 & $18.0$  &        &        \\
 XMM01731 & 0147111301 & 2003-07-24T09:02:34.0 & $12.9$  & $51.9$ & $137$ \\
          & 0404966601 & 2007-01-06T14:04:02.0 & $13.9$  &        &        \\
          & 0785101101 & 2016-07-05T21:22:56.0 & $22.5$  &        &        \\
          & 0793580701 & 2017-01-03T08:02:49.0 & $16.0$  &        &        \\
          & 0780452401 & 2017-02-09T18:08:28.0 & $23.0$  &        &         \\
          & 14348 (ch) & 2011-10-03T09:28:05 & $63.7$    &        &        \\
          & 13374 (ch) & 2011-10-07T05:41:46 & $75.7$    &        &        \\
 XMM01740 & 0147111301 & 2003-07-24T09:02:34.0 & $12.9$  & $60.8$ & $47$ \\
          & 0404966601 & 2007-01-06T14:04:02.0 & $13.9$  &        &        \\
          & 0785100801 & 2016-07-05T01:04:36.0 & $23.0$  &        &        \\
          & 0785101101 & 2016-07-05T21:22:56.0 & $22.5$  &        &        \\
          & 0793580701 & 2017-01-03T08:02:49.0 & $16.0$  &        &        \\
          & 0793580801 & 2017-01-16T20:33:01.0 & $18.0$  &        &        \\
          & 14348 (Ch) & 2011-10-03T09:28:05 & $63.7$    &        &         \\
          & 13374 (Ch) & 2011-10-07T05:41:46 & $75.7$    &        &        \\
 XMM02186 & 0037982501 & 2003-01-25T01:19:09.0 & $14.1$  & $33.7$ & $109$ \\
          & 0037982401 & 2003-01-25T05:51:41.0 & $18.9$  &        &         \\
          & 0037982201 & 2003-01-28T19:15:43.0 & $16.4$  &        &        \\
          & 0404960601 & 2006-07-07T03:36:58.0 & $11.9$  &        &        \\
          & 0785101501 & 2016-07-07T14:16:31.0 & $22.0$  &        &        \\
          & 0793581001 & 2017-01-04T20:54:08.0 & $9.0$   &        &        \\
 XMM02347 & 0147111501 & 2003-07-24T17:12:34.0 & $11.0$  & $65.9$ & $163$  \\
          & 0404960401 & 2006-07-06T19:43:38.0 & $11.9$  &        &         \\
          & 0785101001 & 2016-07-05T14:47:56.0 & $22.5$  &        &        \\
          & 0785101101 & 2016-07-05T21:22:56.0 & $22.5$  &        &        \\
          & 0785101601 & 2016-07-07T20:55:41.0 & $22.0$  &        &        \\
          & 0793580701 & 2017-01-03T08:02:49.0 & $16.0$  &        &        \\
          & 0793581101 & 2017-01-04T23:56:38.0 & $9.0$   &        &        \\
 XMM02660 & 0109520501 & 2001-07-03T22:44:36.0 & $24.8$  & $65.6$ & $226$ \\
          & 0112680801 & 2002-01-31T20:21:47.0 & $15.6$  &        &        \\
          & 0785101701 & 2016-07-08T03:22:21.0 & $27.4$  &        &        \\
          & 0785101801 & 2016-07-29T05:19:31.0 & $22.0$  &        &        \\
          & 0780450201 & 2016-08-13T23:26:35.0 & $17.0$  &        &        \\
          &  0793581201 & 2017-01-01T12:54:38.0 & $35.0$ &        &         \\
 XMM03098 & 0109520601 & 2002-01-31T13:04:39.0 & $23.6$  & $66.0$ & $166$ \\
          & 0112680501 & 2002-07-25T16:24:58.0 & $23.6$  &        &        \\
          & 0780450301 & 2016-08-14T04:29:55.0 & $17.0$  &        &        \\
          & 0780450601 & 2016-08-14T14:36:35.0 & $17.0$  &        &        \\
          &  0780452201 & 2017-01-07T17:07:14.0 & $18.0$ &        &        \\
 XMM03153 & 0109520101 & 2002-01-29T08:46:18.0 & $26.6$  & $104.7$ & $375$  \\
          & 0210490101 & 2005-01-01T19:07:51.0 & $107.5$ &        &        \\
\hline
\end{tabular}
\label{tab:Sample}
\end{table*}
\addtocounter{table}{-1}
\begin{table*}
\centering
 \caption{The X-ray observations log of our DOGs.}
\begin{tabular}{cccccc}
\hline
 XID  & Obs ID & Obs Date               & $T_{\rm obs}$ & Total $T^{\rm eff}_{\rm exp}$ & ${\rm Total~Counts_{(0.5-10~keV)}}$  \\
      &        & (YYYY-MM-DD:Thh:mm:ss) &  (ks)         &  (ks)                   & (cts)\\
(1) & (2) & (3) & (4) & (5) & (6) \\
\hline
 XMM03342 &  0109520101 & 2002-01-29T08:46:18.0 & $26.6$ & $62.8$ & $61$ \\
          & 0112680301 & 2003-01-19T04:19:09.0 & $23.4$  &        &        \\
          & 0210490101 & 2005-01-01T19:07:51.0 & $107.5$ &        &         \\
          & 6390 (Ch) & 2005-09-13T21:44:12 & 11.9       &        &        \\
          & 7182 (Ch) & 2005-10-12T03:55:24 & 22.9       &        &        \\
          & 6394 (Ch) & 2005-10-12T22:57:34 & 17.5       &        &        \\
          & 7184 (Ch) & 2005-10-14T19:16:55 & 22.7       &        &        \\
          & 7183 (Ch) & 2005-10-15T14:18:10 & 19.9       &        &        \\
          & 7185 (Ch) & 2005-11-21T09:59:42 & 32.9       &        &        \\
 XMM03798 & 0112680301 & 2003-01-19T04:19:09.0 & $23.4$  & $55.6$ & $170$\\
          & 0210490101 & 2005-01-01T19:07:51.0 & $107.5$ &        &        \\
          & 0780450701 & 2016-08-14T19:39:55.0 & $17.9$  &        &        \\
          & 0780451101 & 2017-01-07T10:58:53.0 & $20.9$  &        &        \\
 XMM03900 & 0112681001 & 2002-01-30T16:49:27.0 & $41.8$  & $72.0$ & $661$ \\
          & 0112680301 & 2003-01-19T04:19:09.0 & $23.4$  &        &        \\
          & 0780450701 & 2016-08-14T19:39:55.0 & $17.9$  &        &        \\
          & 0780451101 & 2017-01-07T10:58:53.0 & $20.9$  &        &        \\
XMM03916  & 0112681001 & 2002-01-30T16:49:27.0 & $41.8$  & $40.2$ & $96$ \\
          & 0780450601 & 2016-08-14T14:36:35.0 & $17.0$  &        &        \\
          & 0780450701 & 2016-08-14T19:39:55.0 & $17.9$  &        &        \\
          & 0780451001 & 2017-01-07T04:50:33.0 & $20.9$  &        &        \\
          & 6864 (Ch) & 2006-11-12T05:00:26 & $29.7$     &        &        \\
 XMM04259 & 0109520301 & 2002-02-02T11:26:13.0 & $22.6$  & $51.9$ & $103$  \\
          & 0780451001 & 2017-01-07T04:50:33.0 & $20.9$  &        &        \\
          & 0780451301 & 2017-01-08T18:27:03.0 & $20.0$  &        &        \\
          & 0780451401 & 2017-01-09T00:20:23.0 & $20.0$  &        &        \\
          & 0780452601 & 2017-02-10T05:38:28.0 & $16.0$  &        &        \\
          & 9368 (Ch)  & 2007-11-23T16:55:22 & $75.0$    &        &        \\
          & 18264 (Ch) & 2016-09-27T19:47:17 & $22.8$    &        &        \\
 XMM04404 & 0112680101 & 2002-01-28T23:39:09.0 & $30.2$  & $47.0$ & $238$ \\
          & 0112680201 & 2002-07-14T02:10:42.0 & $21.6$  &        &        \\
          & 0780451501 & 2017-01-09T06:13:43.0 & $20.0$  &        &        \\
          & 0780451601 & 2017-01-09T12:07:03.0 & $36.4$  &        &        \\
          & 0780451701 & 2017-01-10T19:11:08.0 & $20.0$  &        &        \\
 XMM04475 & 0109520201 & 2002-01-29T16:53:38.0 & $25.6$  & $101.9$ & $131$  \\
          &  0109520301 & 2002-02-02T11:26:13.0 & $22.6$ &        &        \\
          &   0112681301 & 2002-07-26T08:26:58.0 & $40.4$ &        &         \\
          & 0780451301 & 2017-01-08T18:27:03.0 & $20.0$  &        &        \\
          & 0780451401 & 2017-01-09T00:20:23.0 & $20.0$  &        &        \\
          & 0780452101 & 2017-01-13T09:20:32.0 & $20.0$  &        &        \\
          & 0780452601 & 2017-02-10T05:38:28.0 & $16.0$  &        &        \\
          & 9368 (Ch) & 2007-11-23T16:55:22 & $75.0$     &        &        \\
 XMM04583 & 0404964801 & 2006-07-07T11:42:54.0 & $11.9$  & $67.8$ & $255$  \\
          & 0404964901 & 2006-07-07T15:39:35.0 & $11.7$  &        &        \\
          & 0404969201 & 2006-07-26T20:57:30.0 & $7.9$   &        &        \\
          & 0553910401 & 2008-08-06T22:19:42.0 & $11.9$  &        &        \\
          & 0785102401 & 2016-08-12T12:23:12.0 & $22.0$  &        &        \\
          & 0785102501 & 2016-08-12T18:49:52.0 & $22.0$  &        &        \\
          & 0793581601 & 2017-01-05T08:26:38.0 & $9.0$   &        &        \\
 XMM04804 & 0112680401 & 2002-02-02T18:26:41.0 & $24.9$  & $60.2$ & $151$ \\
          &  0112681301 & 2002-07-26T08:26:58.0 & $40.4$ &        &         \\
          &  0780451801 & 2017-01-11T01:04:28.0 & $20.0$ &        &        \\
          &  0780452101 & 2017-01-13T09:20:32.0 & $20.0$ &        &        \\
 XMM04899 & 0109520201 & 2002-01-29T16:53:38.0 & $25.6$  & $49.7$ & $257$ \\
          &  0780451901 & 2017-01-11T06:57:48.0 & $20.0$ &        &        \\
          &  0780452101 & 2017-01-13T09:20:32.0 & $20.0$ &        &         \\
\hline
\end{tabular}
\label{tab:Sample}
\end{table*}
%
%
\begin{figure*}
\centering
\includegraphics[angle=0, width=2\columnwidth, trim={0.0cm 0.0cm 0.0cm 0.0cm}, clip]{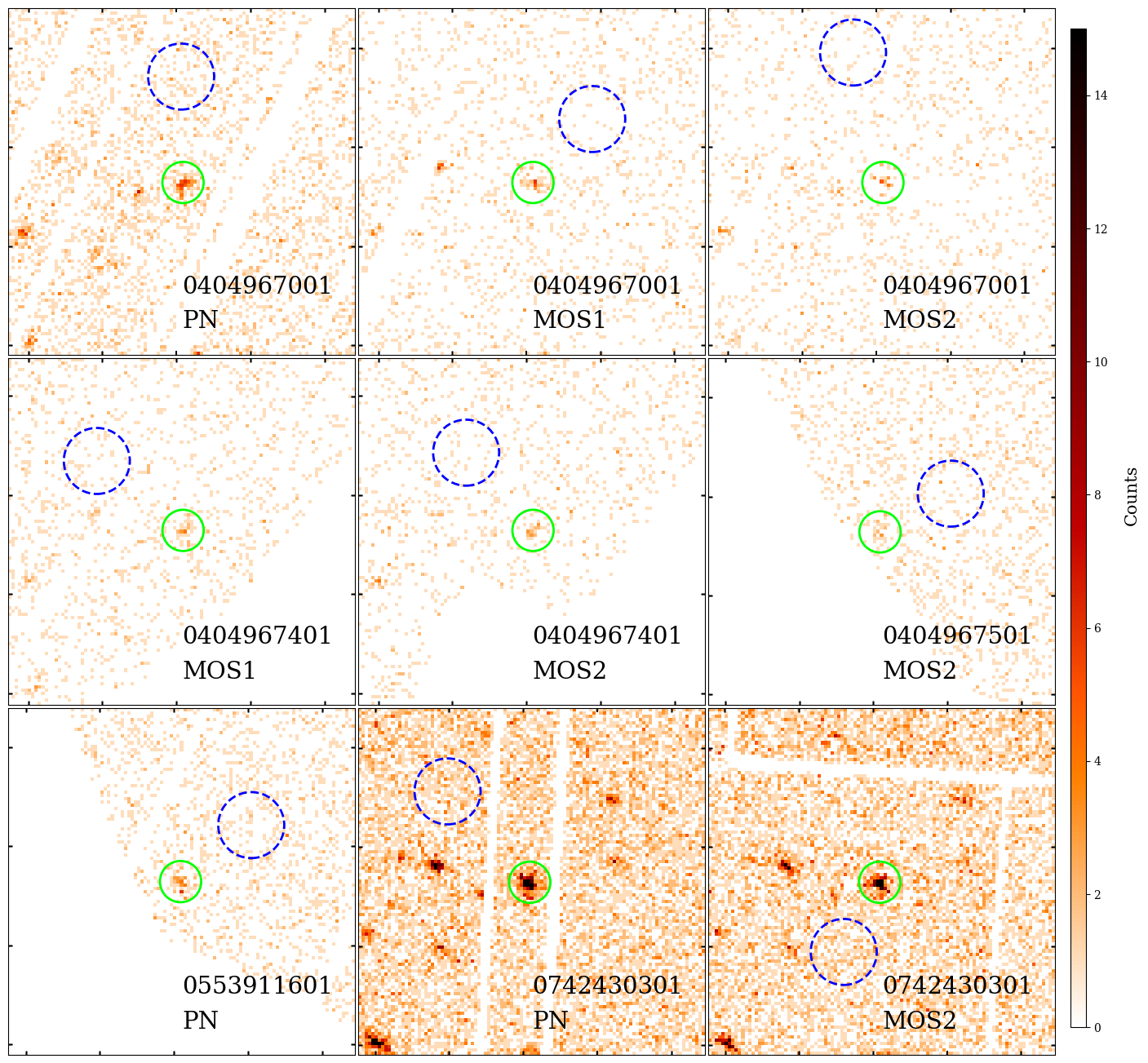}
\caption{The {\em XMM-Newton} pn and MOS images of XMM00267 from different epochs (observation IDs).
The source and background extraction regions are shown with green solid circles and blue dotted circles, respectively.
The epochs (observation IDs) in which the target source fell outside the CCD or in the CCD gaps were discarded. The vertical colour bar indicates
total counts in the 0.5$-$10 keV band. As expected, images from observation ID 0742430301 with a much higher exposure time of 100 ks show a lot more counts.
The extraction regions on XMM00276 images are shown here as an example case, and the same strategy is followed for other sample sources.}
\label{fig:SourceBkgExtImgXMM00267}
\end{figure*}

\begin{figure*}
 \centering
\begin{subfigure}[b]{0.65\columnwidth}
 \centering
 \captionsetup[subfigure]{oneside,margin={1.5cm,0cm}}
 \includegraphics[angle=-90, width=\columnwidth, trim={0 1.3cm 0 0}, clip]{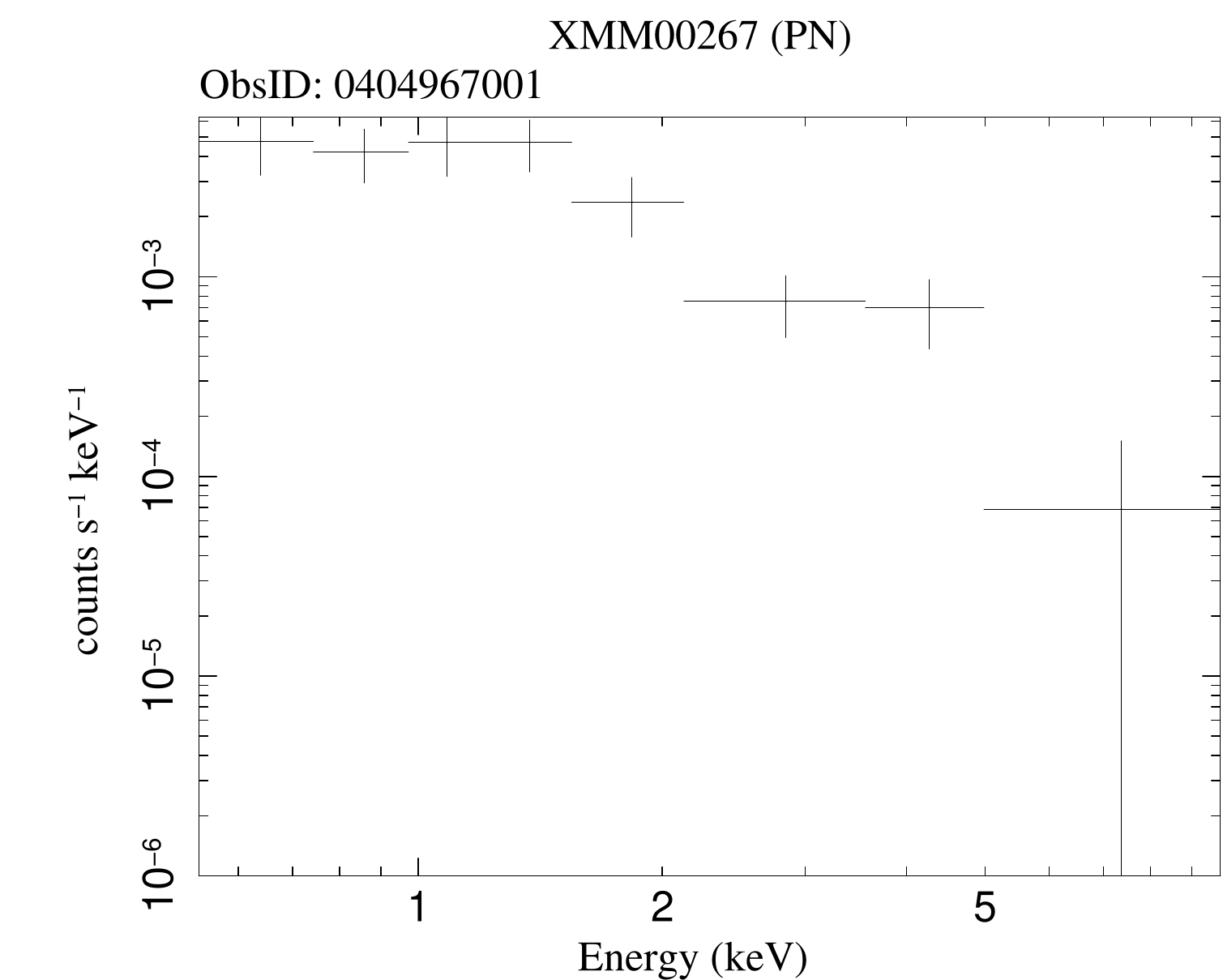}
 \end{subfigure}
 \hfill
 \begin{subfigure}[b]{0.65\columnwidth}
 \centering
 \captionsetup[subfigure]{oneside,margin={1.5cm,0cm}}
 \includegraphics[angle=-90, width=\columnwidth, trim={0 1.3cm 0 0}, clip]{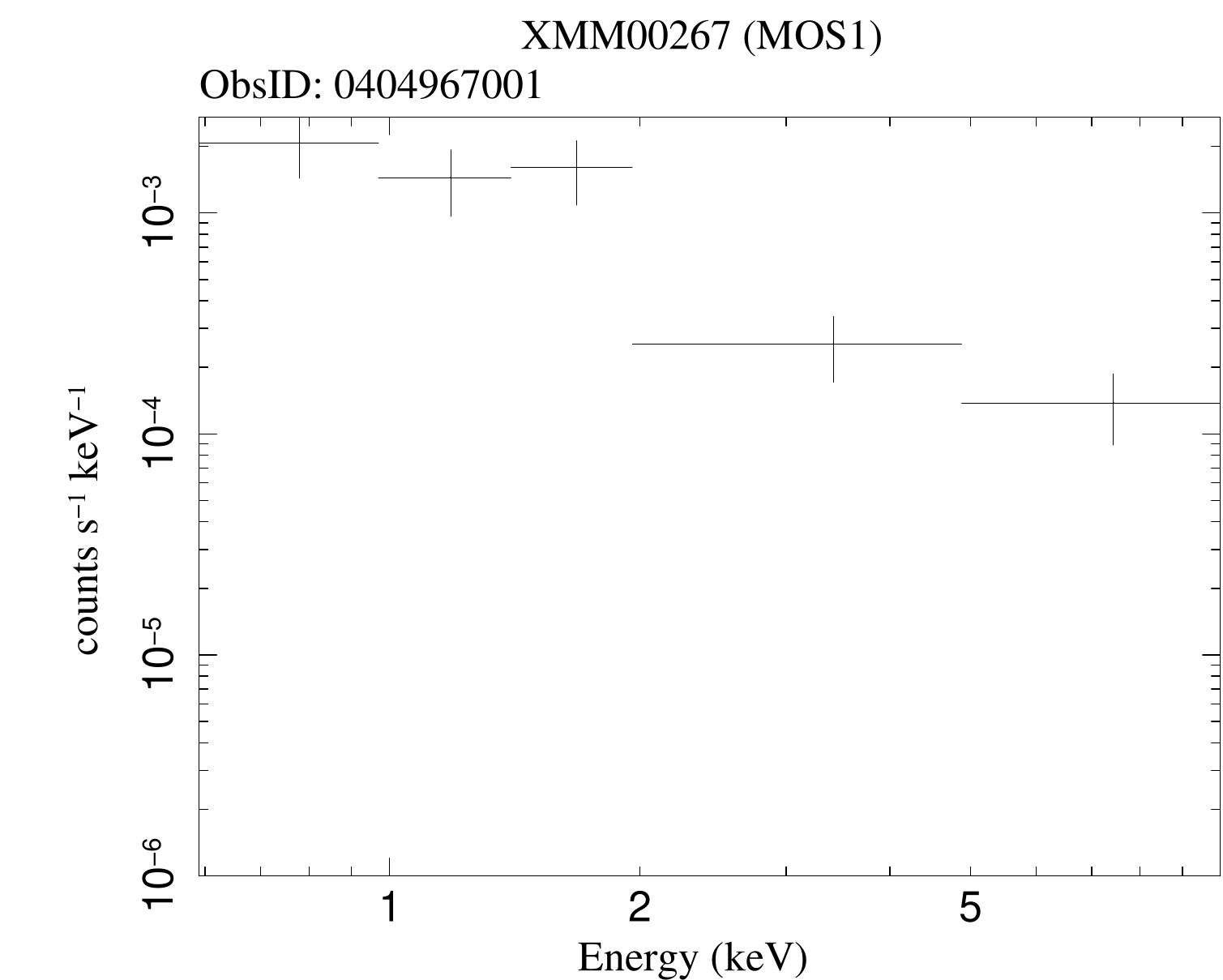}
 \end{subfigure}
 \hfill
  \begin{subfigure}[b]{0.65\columnwidth}
 \centering
 \captionsetup[subfigure]{oneside,margin={1.5cm,0cm}}
 \includegraphics[angle=-90, width=\columnwidth, trim={0 1.3cm 0 0}, clip]{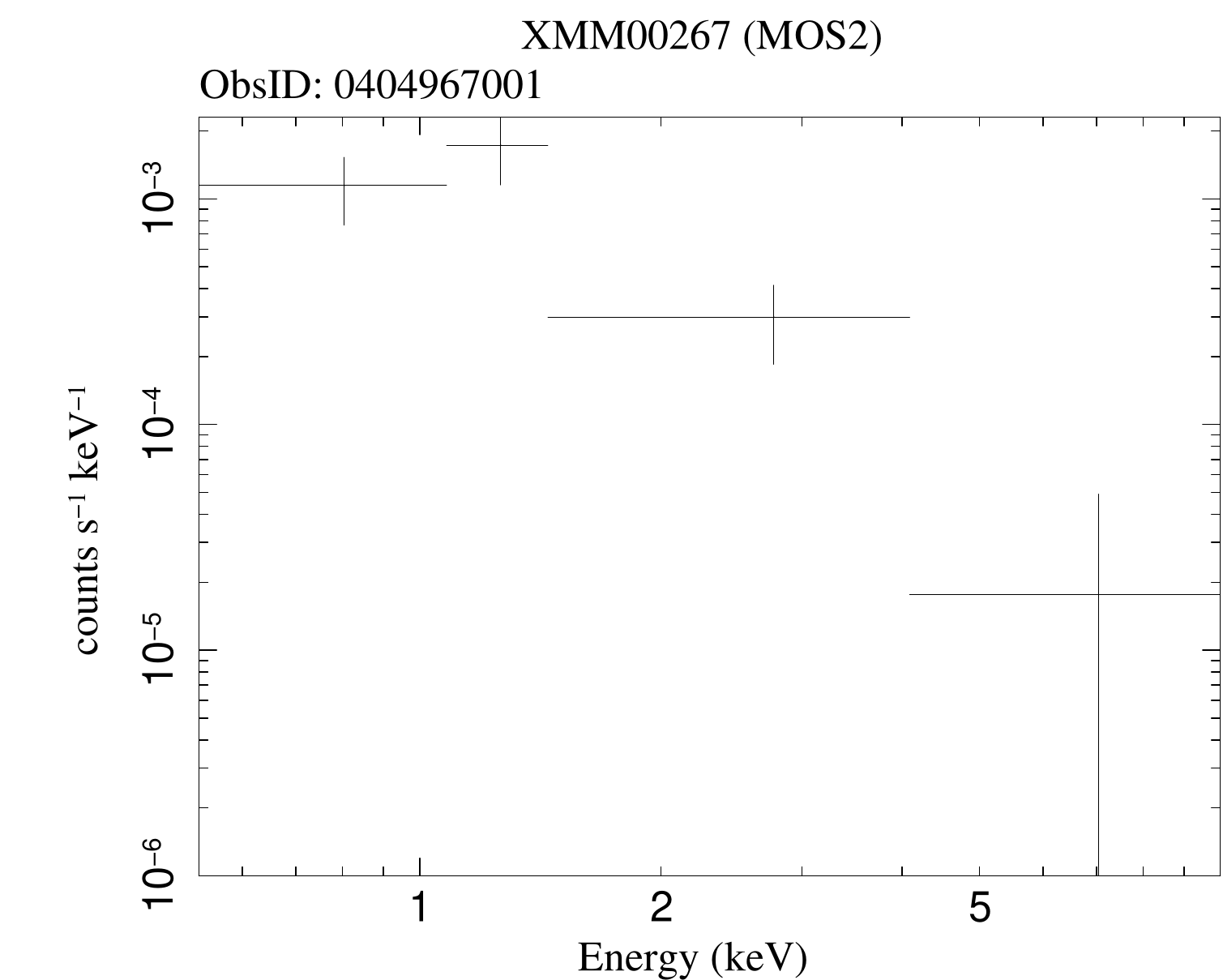}
 \end{subfigure}
 \hfill
  \begin{subfigure}[b]{0.65\columnwidth}
 \centering
 \captionsetup[subfigure]{oneside,margin={1.5cm,0cm}}
 \includegraphics[angle=-90, width=\columnwidth, trim={0 1.3cm 0 0}, clip]{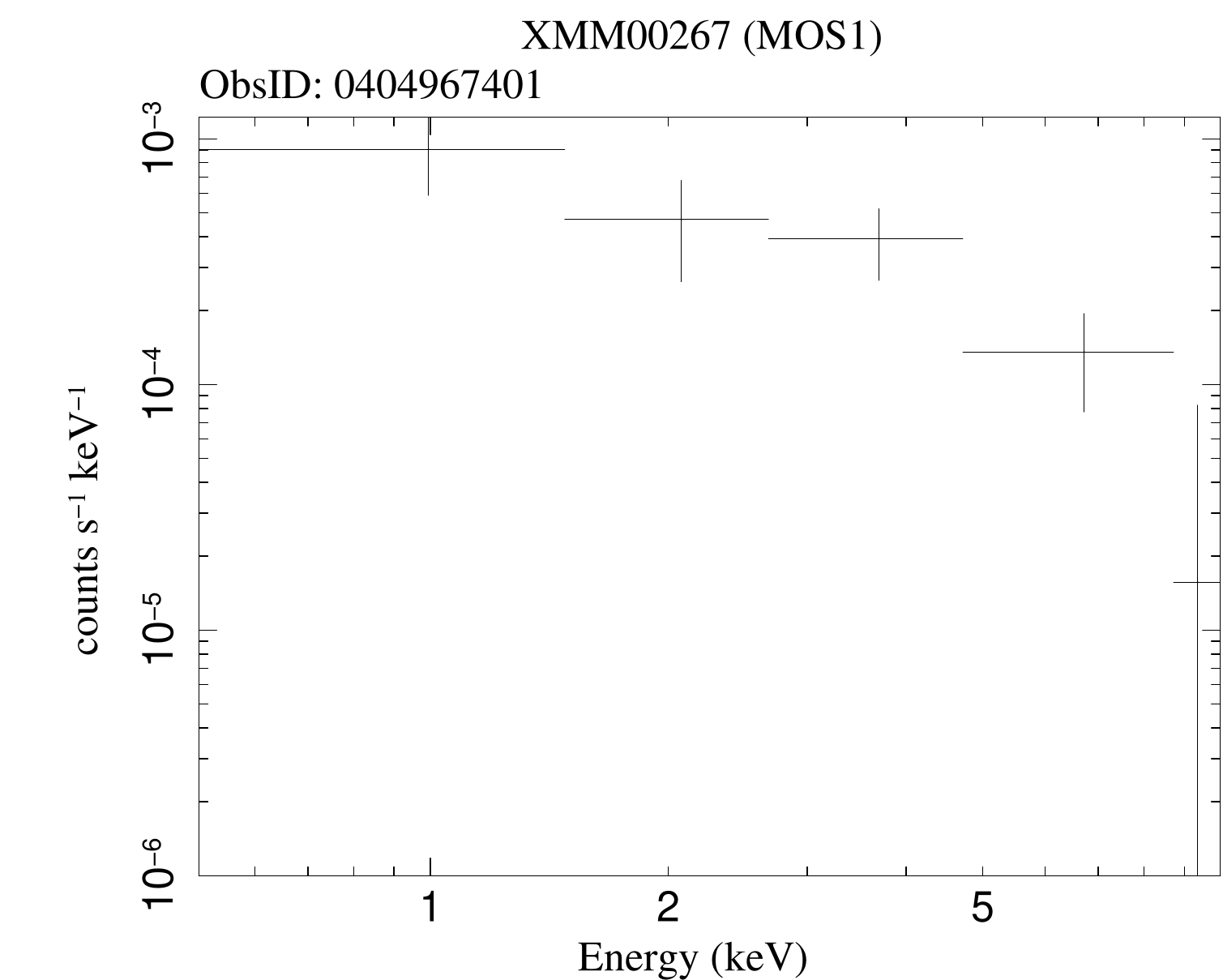}
 \end{subfigure}
 \hfill
 \begin{subfigure}[b]{0.65\columnwidth}
 \centering
 \captionsetup[subfigure]{oneside,margin={1.5cm,0cm}}
 \includegraphics[angle=-90, width=\columnwidth, trim={0 1.3cm 0 0}, clip]{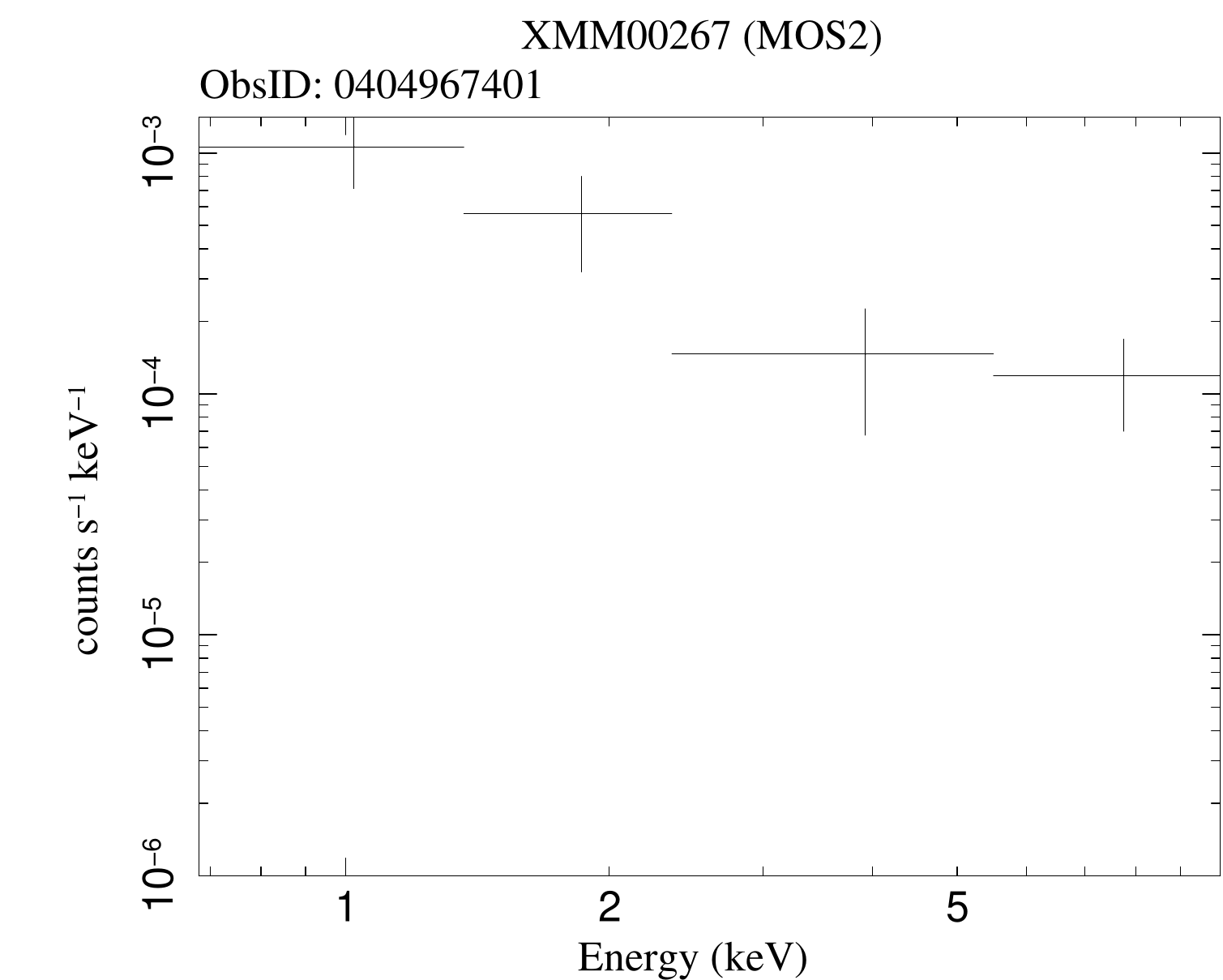}
 \end{subfigure}
 \hfill
  \begin{subfigure}[b]{0.65\columnwidth}
 \centering
 \captionsetup[subfigure]{oneside,margin={1.5cm,0cm}}
 \includegraphics[angle=-90, width=\columnwidth, trim={0 1.3cm 0 0}, clip]{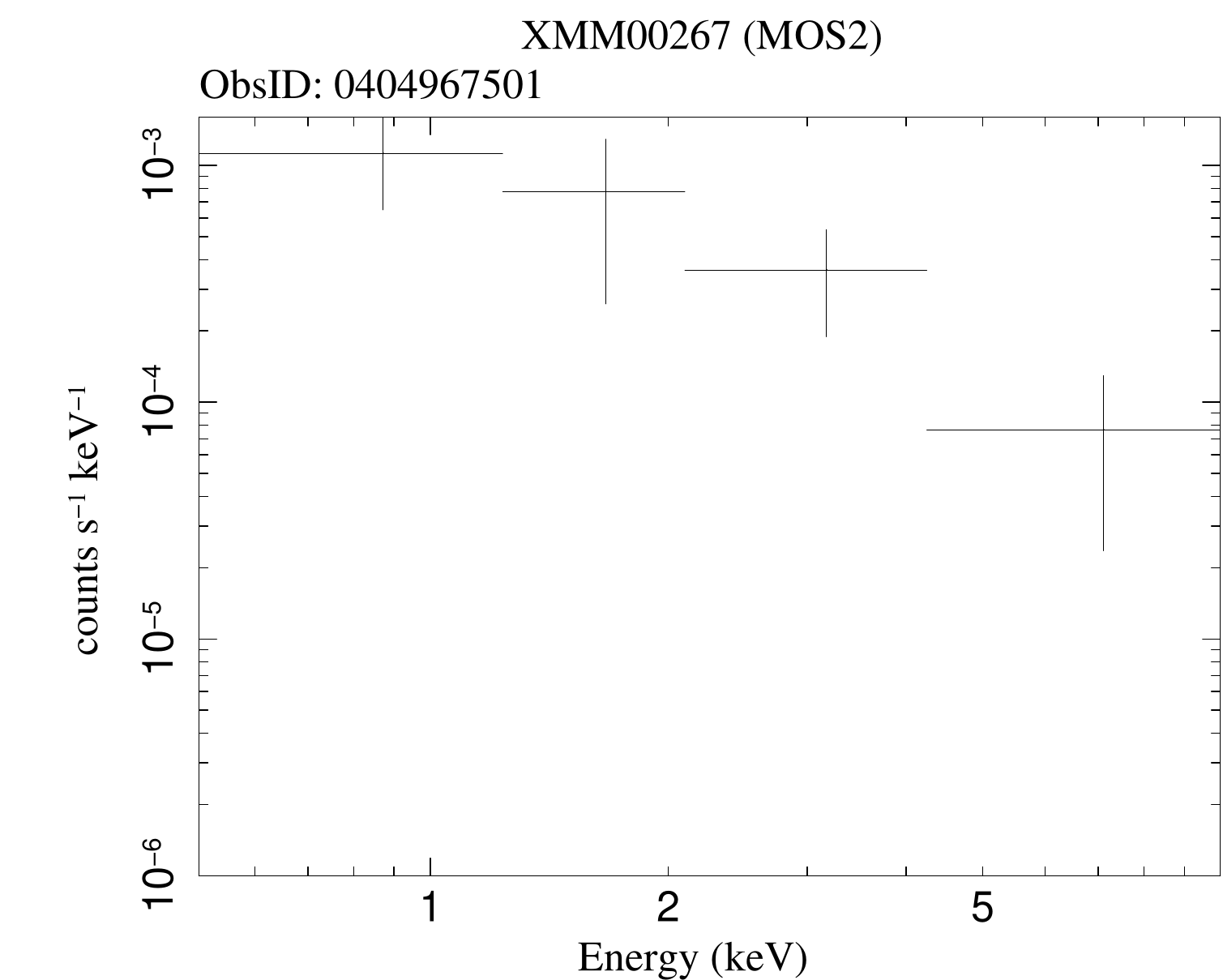}
 \end{subfigure}
  \hfill
 \begin{subfigure}[b]{0.65\columnwidth}
 \centering
 \captionsetup[subfigure]{oneside,margin={1.5cm,0cm}}
 \includegraphics[angle=-90, width=\columnwidth, trim={0 1.3cm 0 0}, clip]{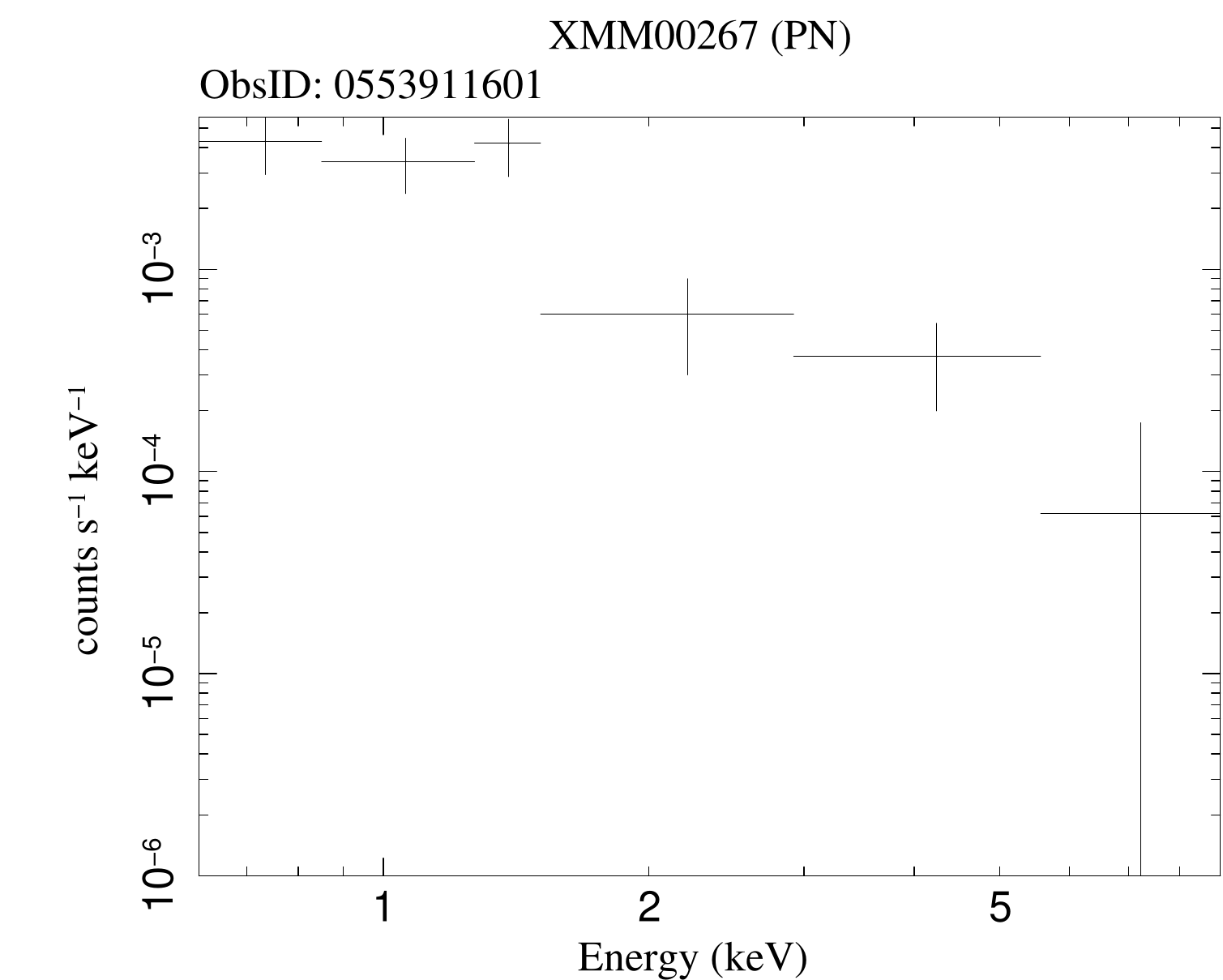}
 \end{subfigure}
 \hfill
   \begin{subfigure}[b]{0.65\columnwidth}
 \centering
 \captionsetup[subfigure]{oneside,margin={1.5cm,0cm}}
\includegraphics[angle=-90, width=\columnwidth, trim={0 1.3cm 0 0}, clip]{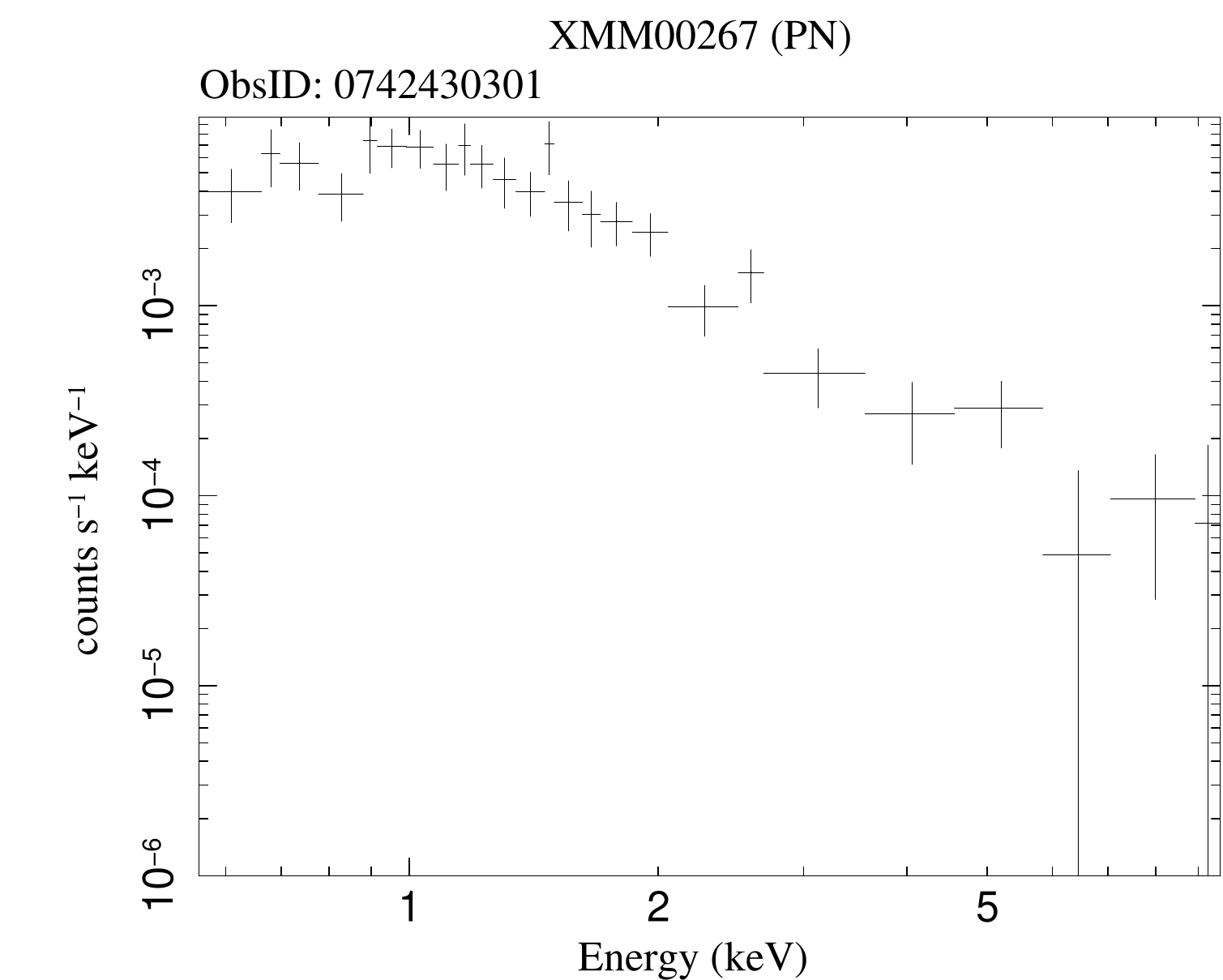}
 \end{subfigure}
\hfill
\begin{subfigure}[b]{0.65\columnwidth}
\centering
\captionsetup[subfigure]{oneside,margin={1.5cm,0cm}}
\includegraphics[angle=-90, width=\columnwidth, trim={0 1.3cm 0 0}, clip]{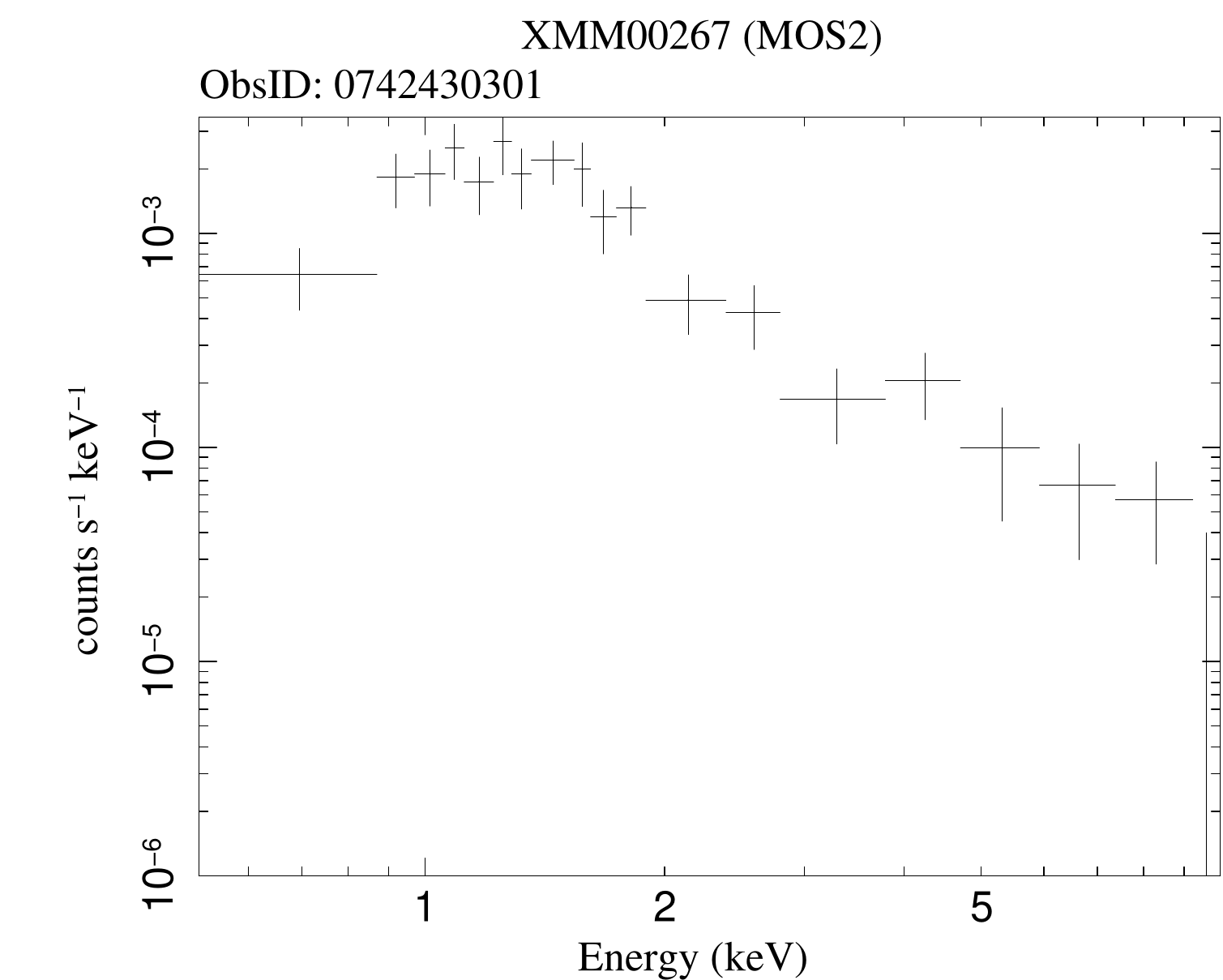}
\end{subfigure}
\caption{The 0.5$-$10 keV {\em XMM-Newton} pn and MOS background-subtracted source spectra of XMM00267 were extracted from the individual images
of different epochs (observation IDs). The corresponding images are shown in Figure \ref{fig:SourceBkgExtImgXMM00267}.
For better visualisation, all spectra are rebinned by 10 counts per bin. The XMM00267 spectra are shown here to present an example case.}
\label{fig:BestFitsXMM00267}
\end{figure*}
\begin{figure*}
\centering
\includegraphics[angle=-90, width=1\columnwidth, trim={0.0cm 0.0cm 0.0cm 0.0cm}, clip]{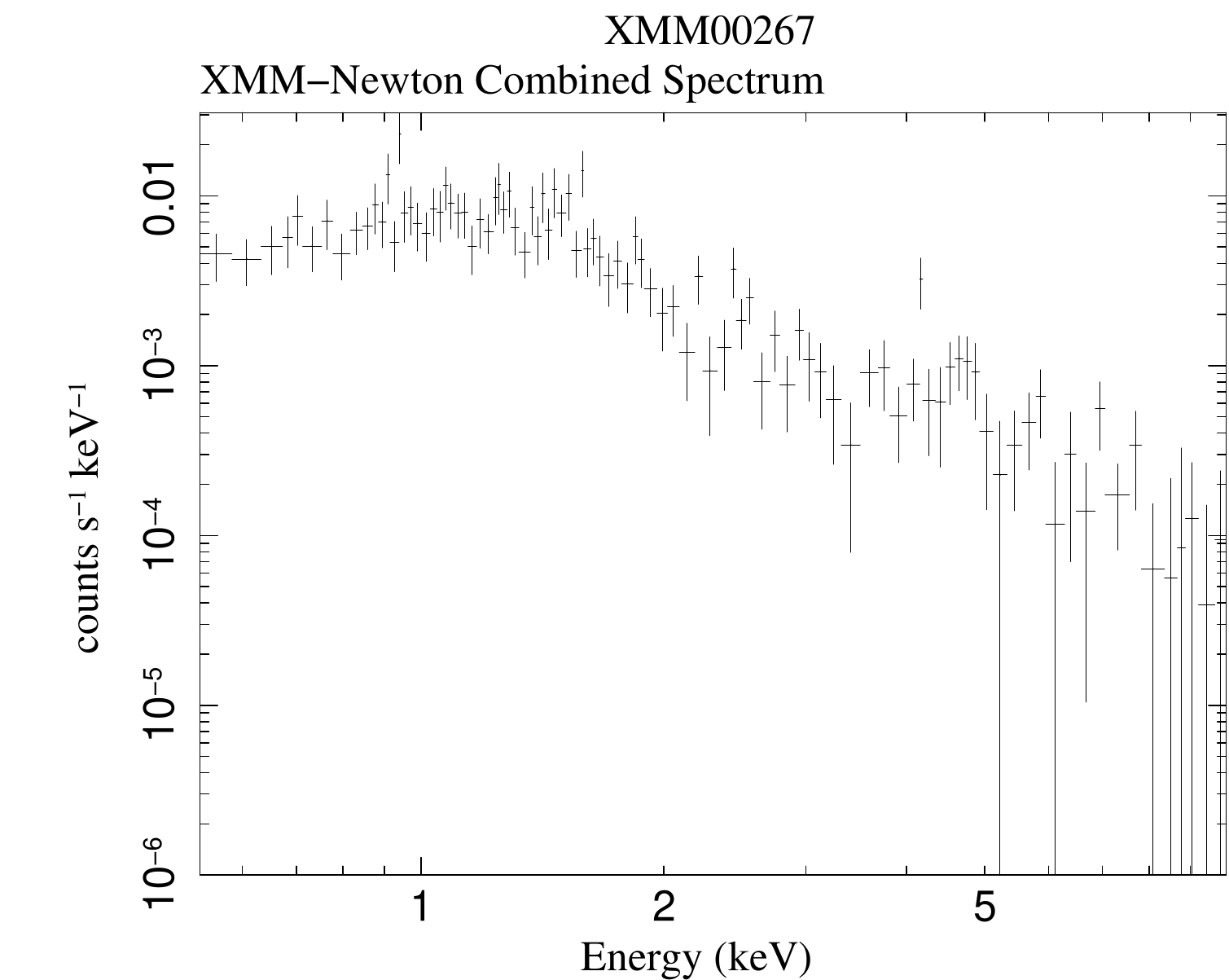}
\caption{The 0.5$-$10 keV {\em XMM-Newton} spectrum obtained by combining pn, MOS1 and MOS2 spectra of all epochs (observation IDs)
of XMM00267. The combined spectrum of XMM00267 is shown here as an example case and the same strategy is followed for other sample sources.}
\label{fig:CombinedSpecXMM00267}
\end{figure*}
\begin{figure*}
\centering
\includegraphics[angle=-90, width=5.5cm, trim={0.0cm 0.0cm 0.0cm 0.0cm}, clip]{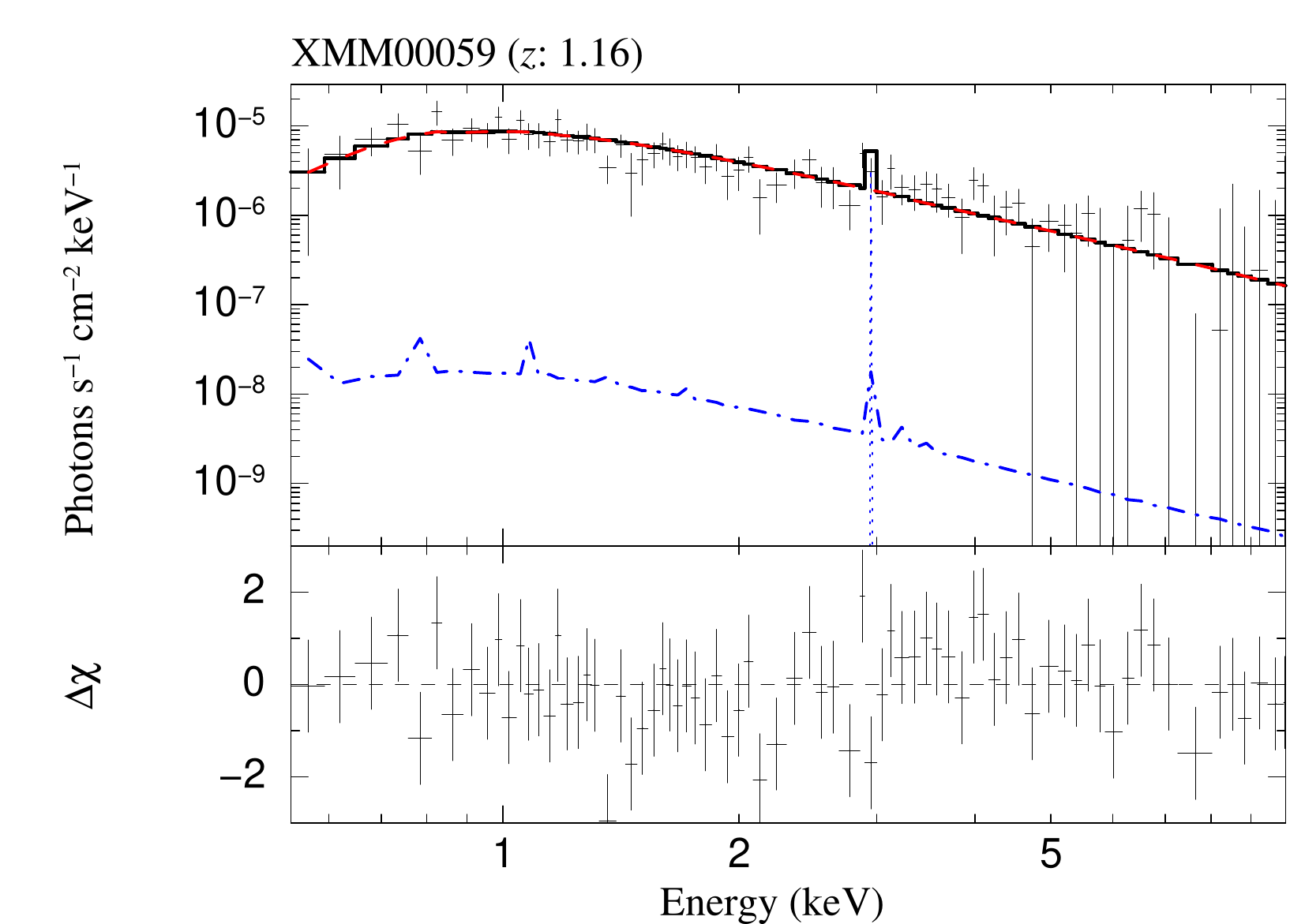}
\includegraphics[angle=-90, width=5.5cm, trim={0.0cm 0.0cm 0.0cm 0.0cm}, clip]{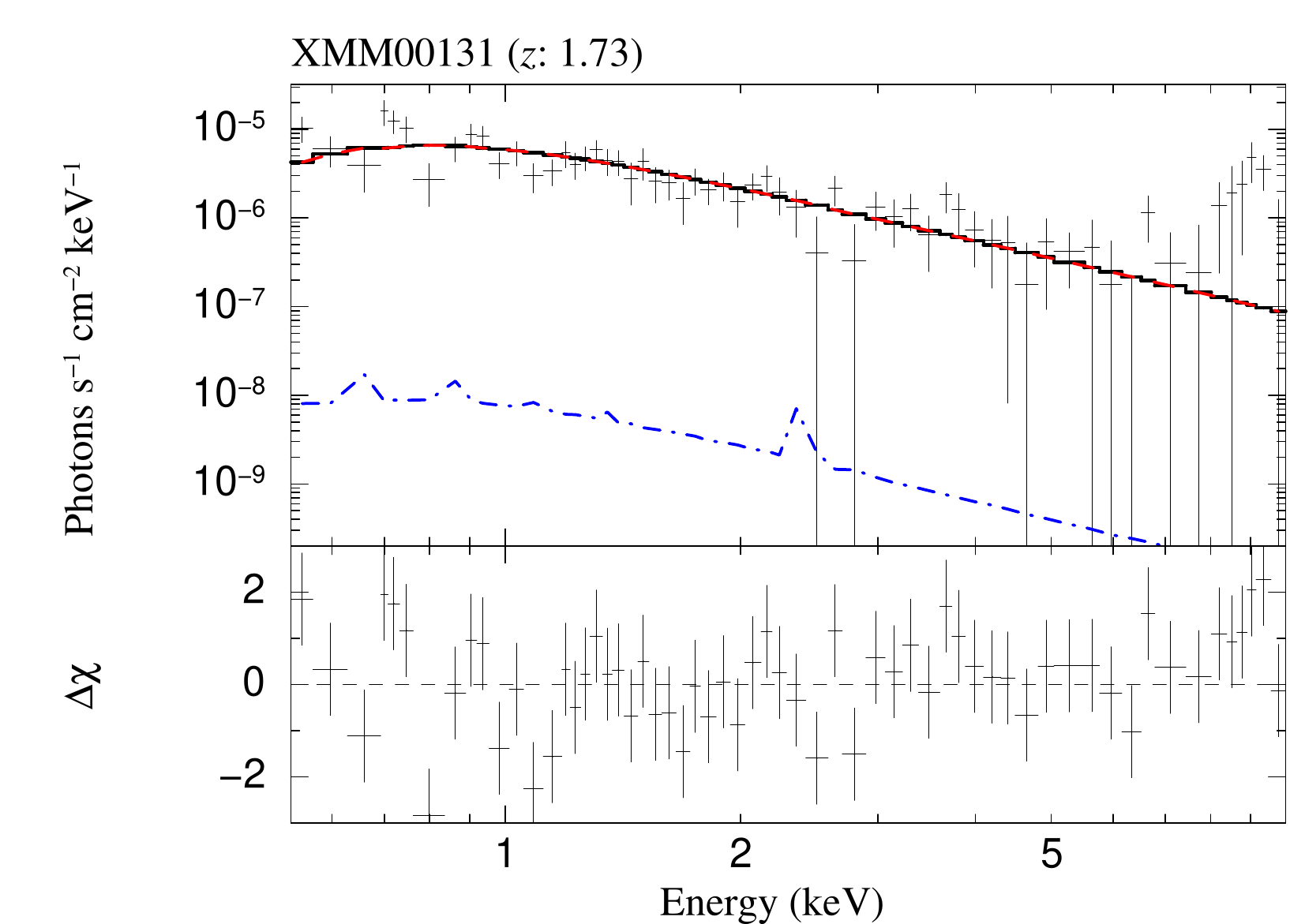}
\includegraphics[angle=-90, width=5.5cm, trim={0.0cm 0.0cm 0.0cm 0.0cm}, clip]{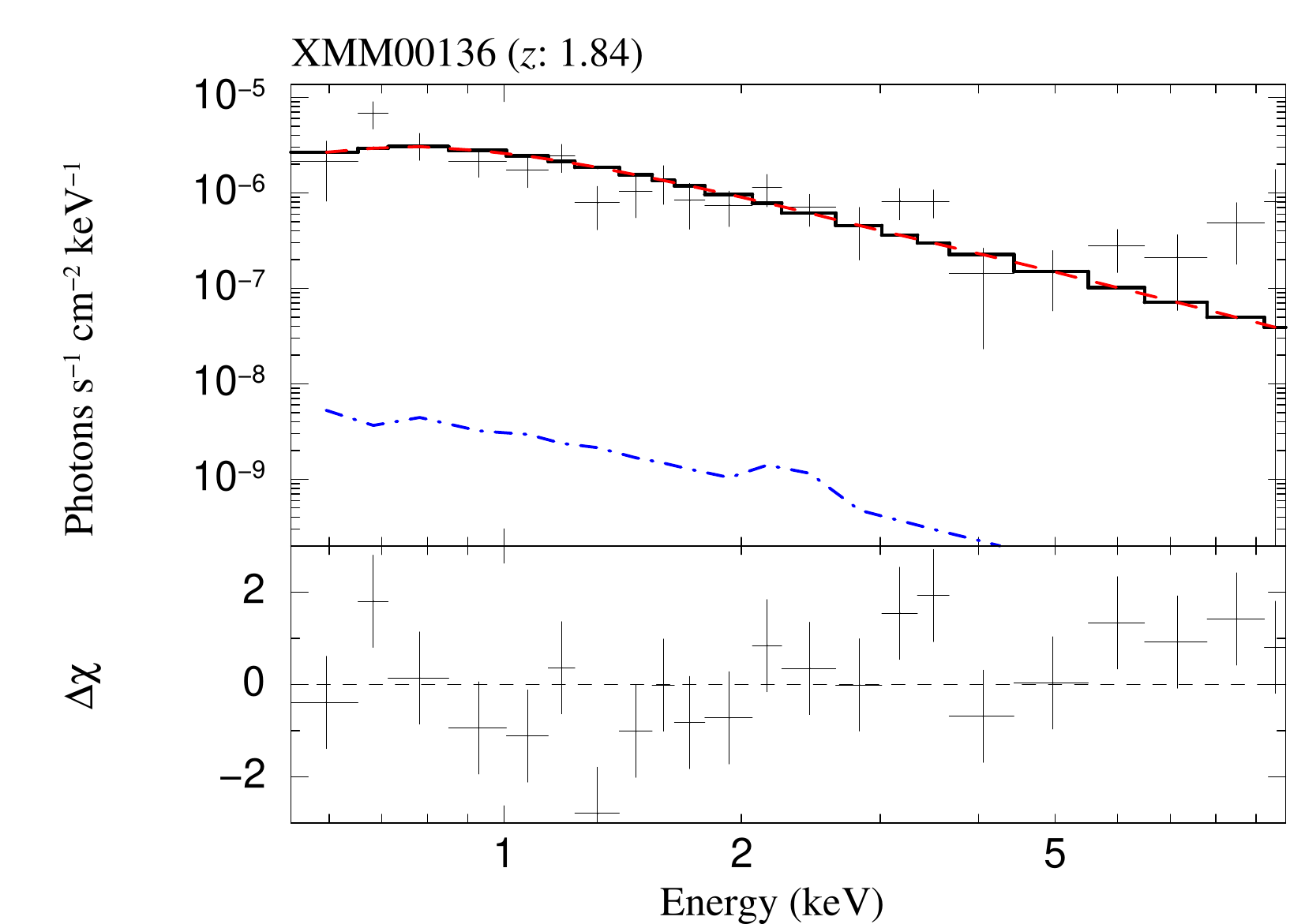}
\includegraphics[angle=-90, width=5.5cm, trim={0.0cm 0.0cm 0.0cm 0.0cm}, clip]{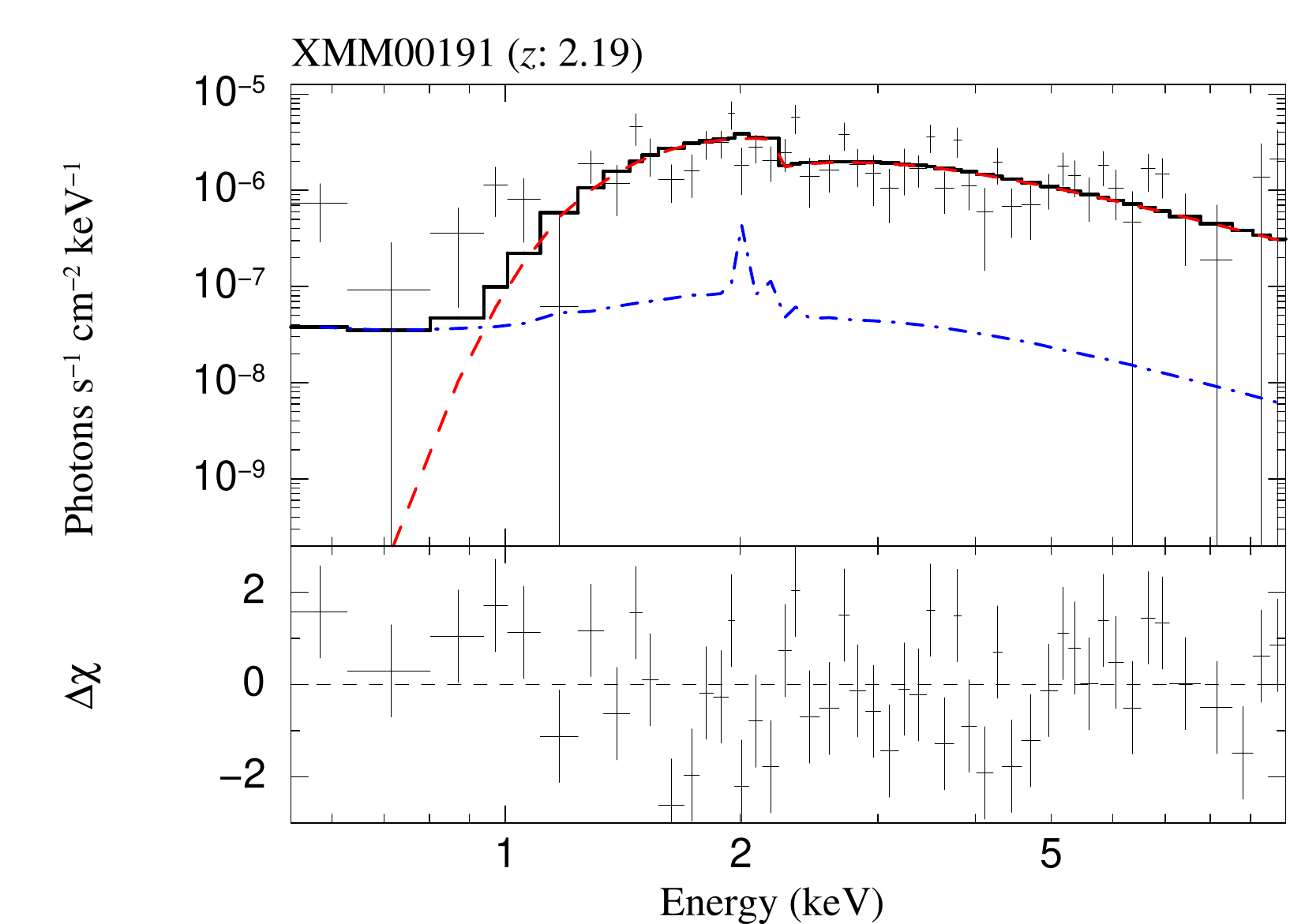}
\includegraphics[angle=-90, width=5.5cm, trim={0.0cm 0.0cm 0.0cm 0.0cm}, clip]{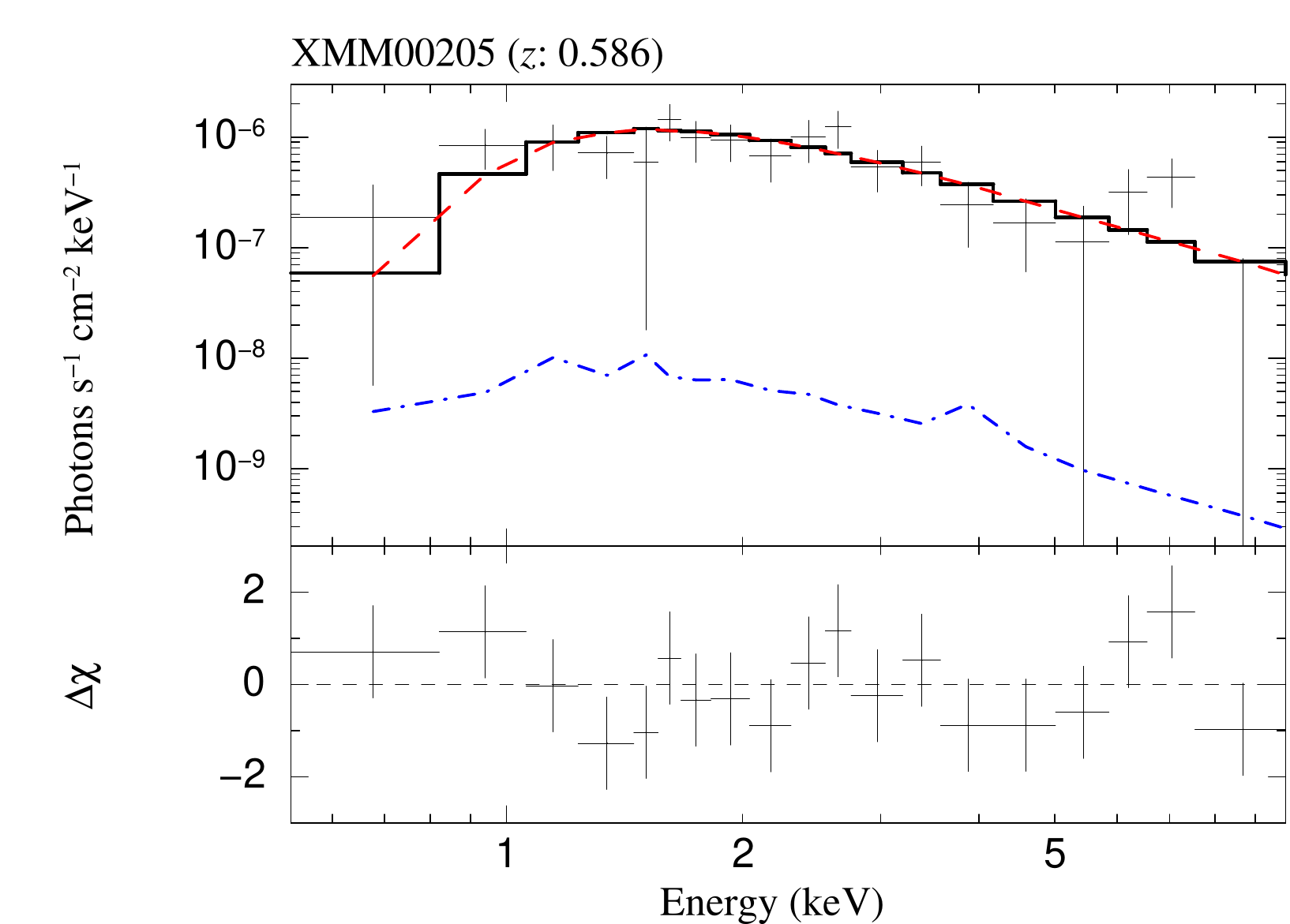}
\includegraphics[angle=-90, width=5.5cm, trim={0.0cm 0.0cm 0.0cm 0.0cm}, clip]{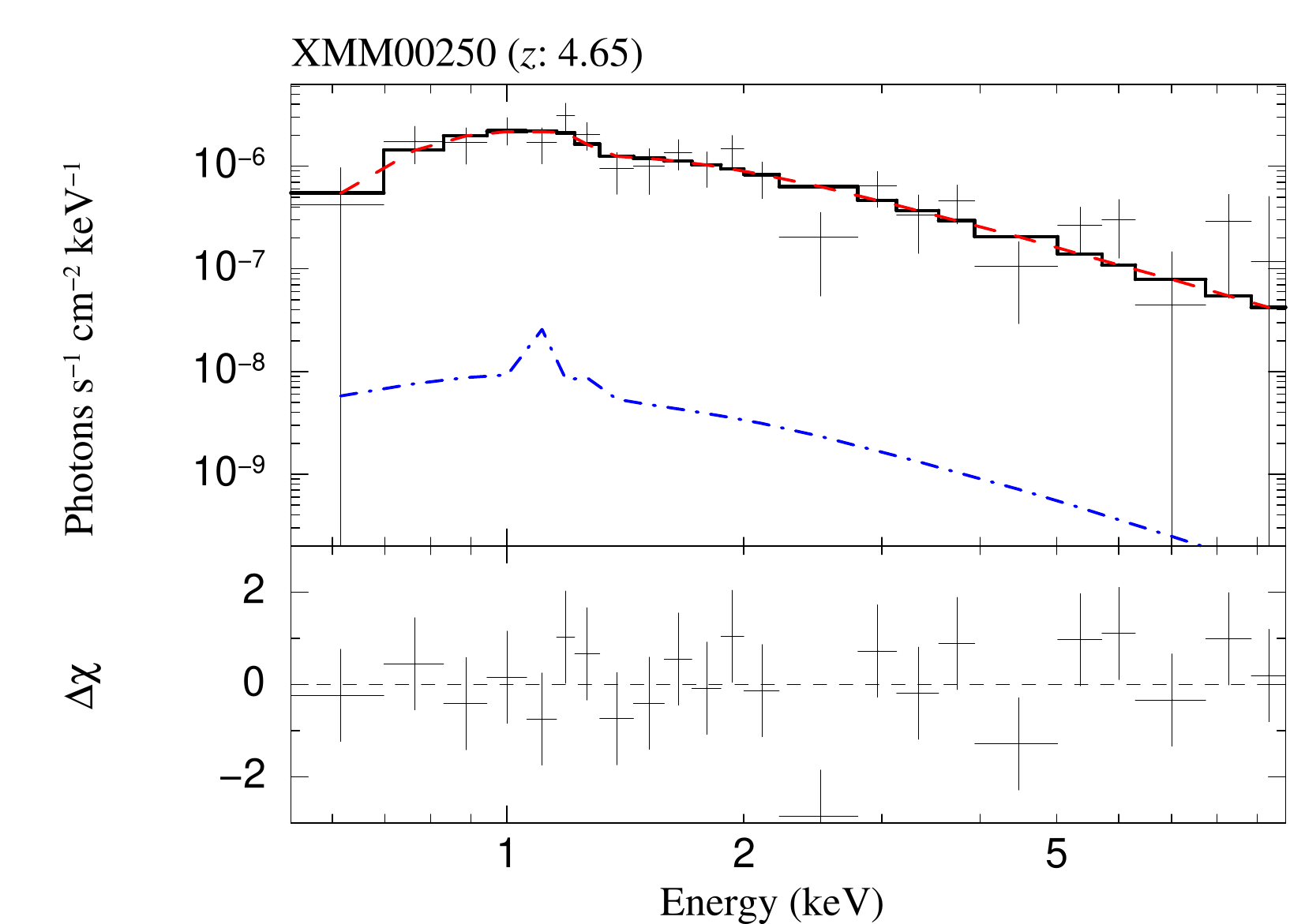}
\includegraphics[angle=-90, width=5.5cm, trim={0.0cm 0.0cm 0.0cm 0.0cm}, clip]{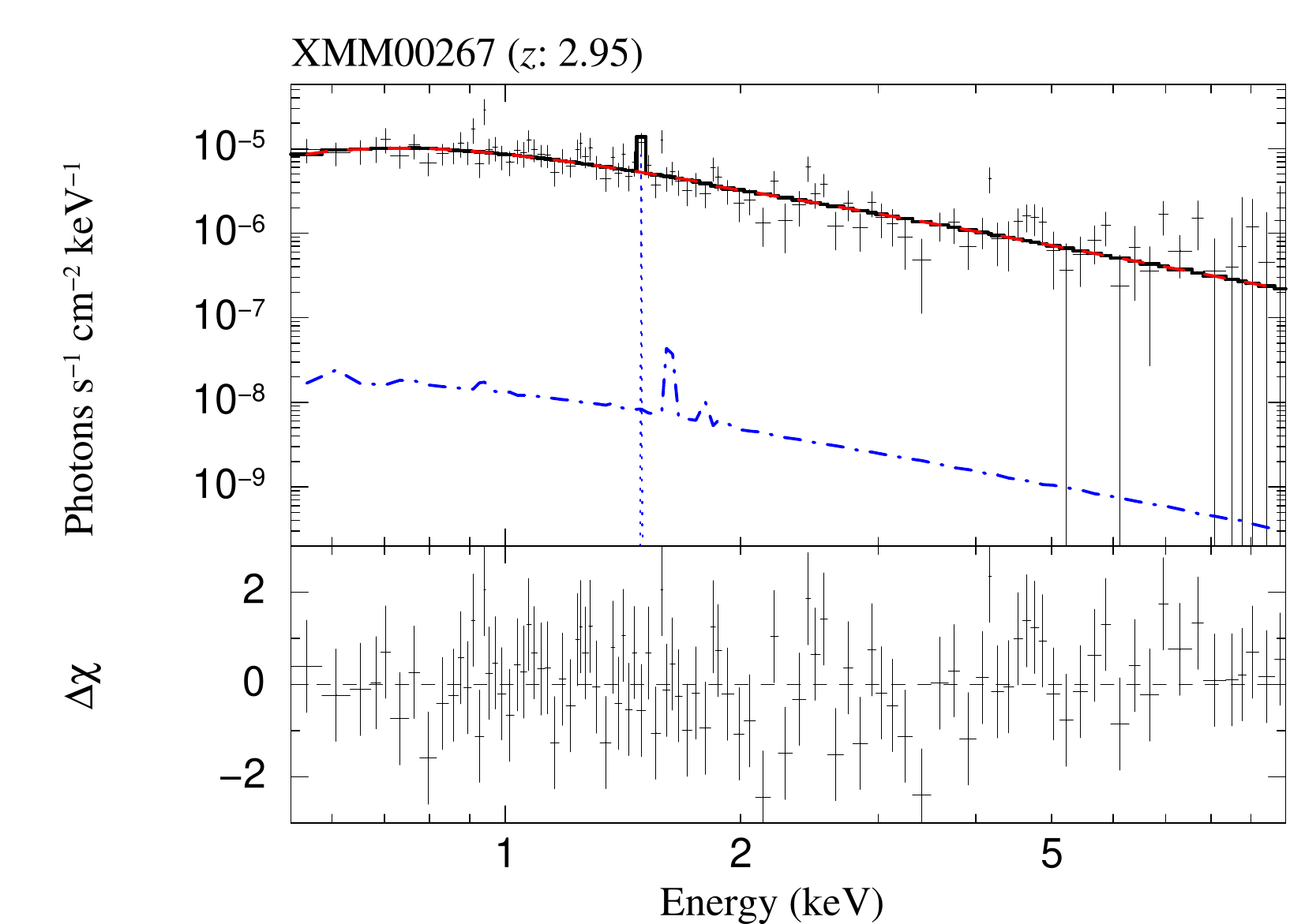}
\includegraphics[angle=-90, width=5.5cm, trim={0.0cm 0.0cm 0.0cm 0.0cm}, clip]{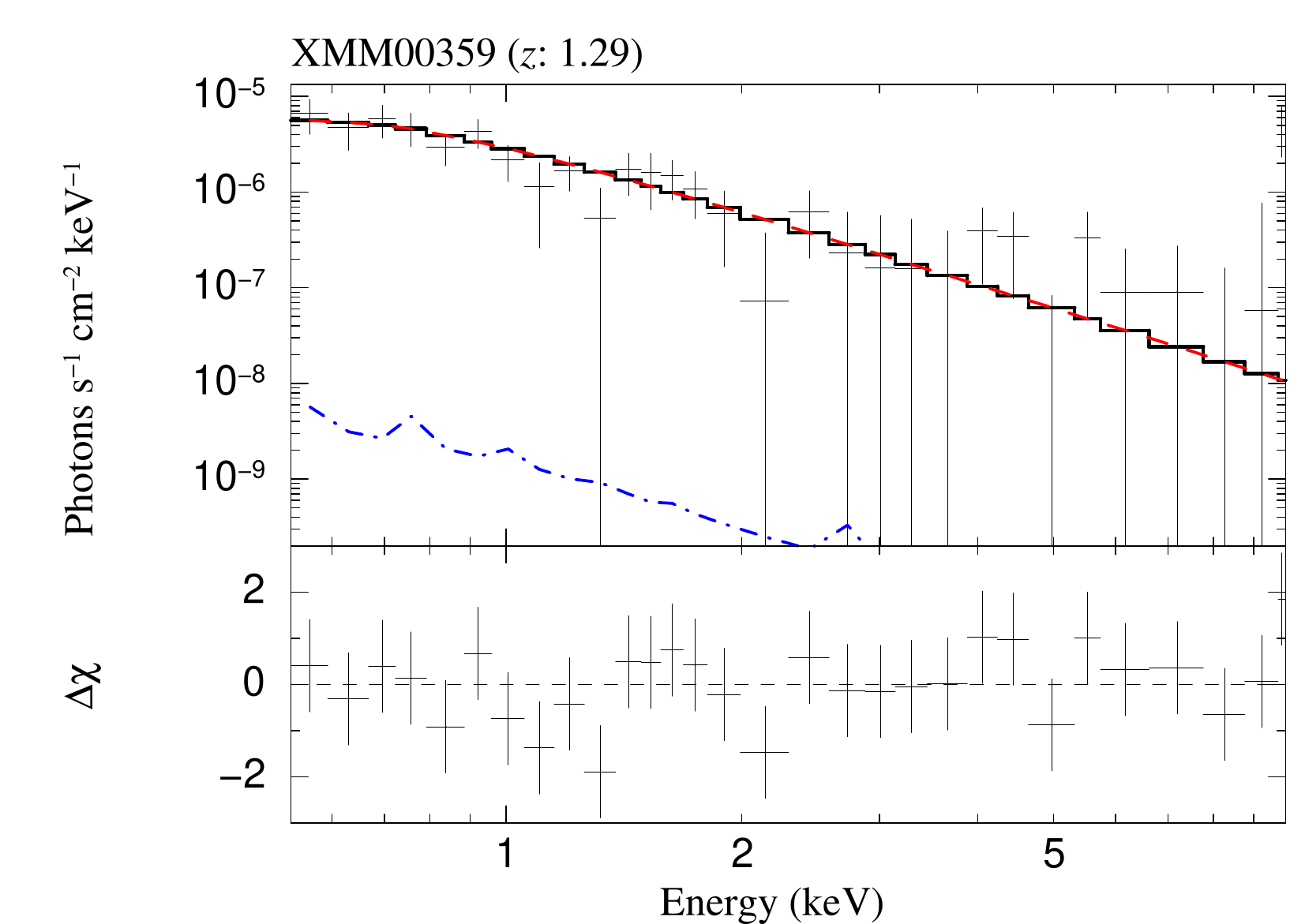}
\includegraphics[angle=-90, width=5.5cm, trim={0.0cm 0.0cm 0.0cm 0.0cm}, clip]{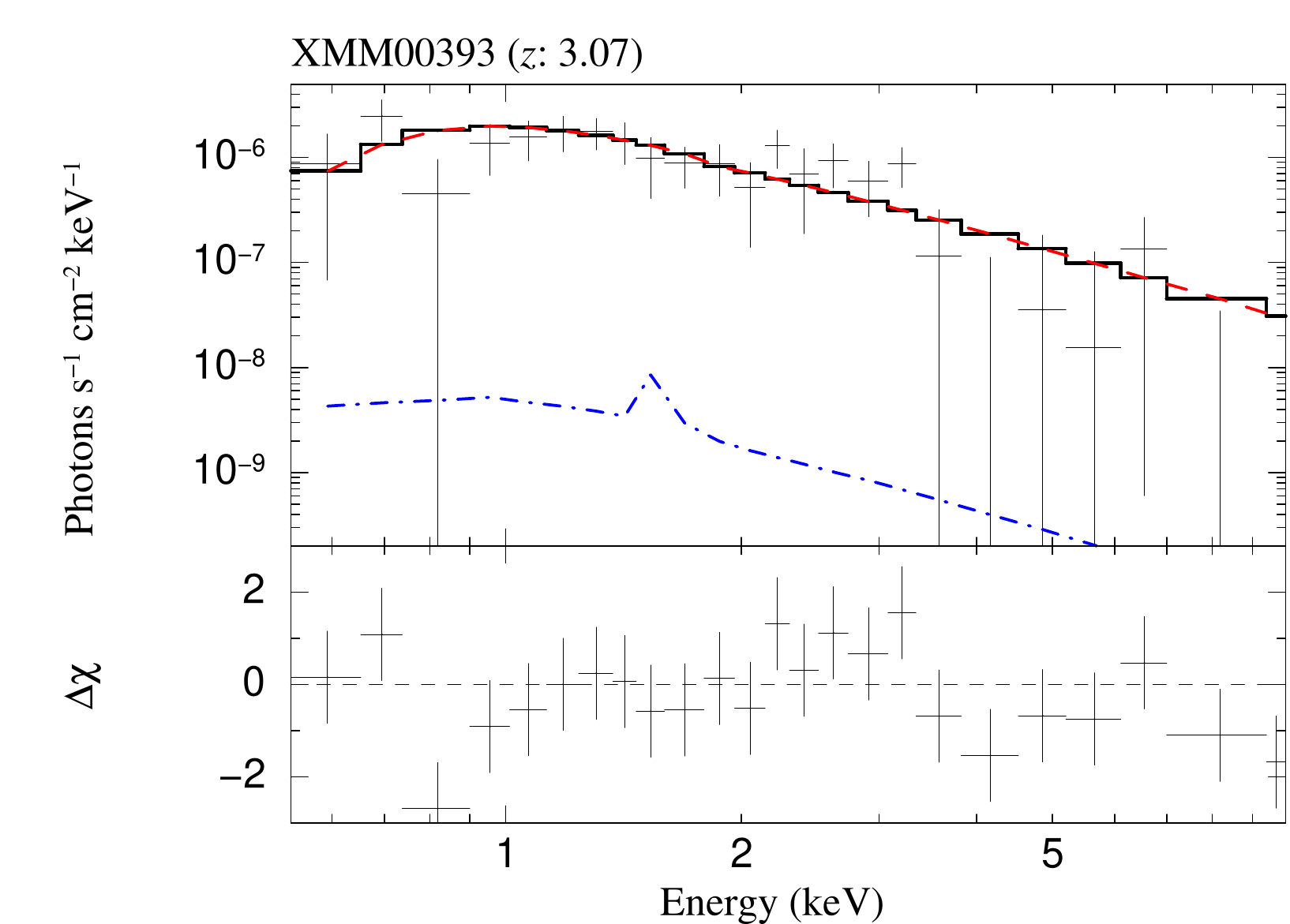}
\includegraphics[angle=-90, width=5.5cm, trim={0.0cm 0.0cm 0.0cm 0.0cm}, clip]{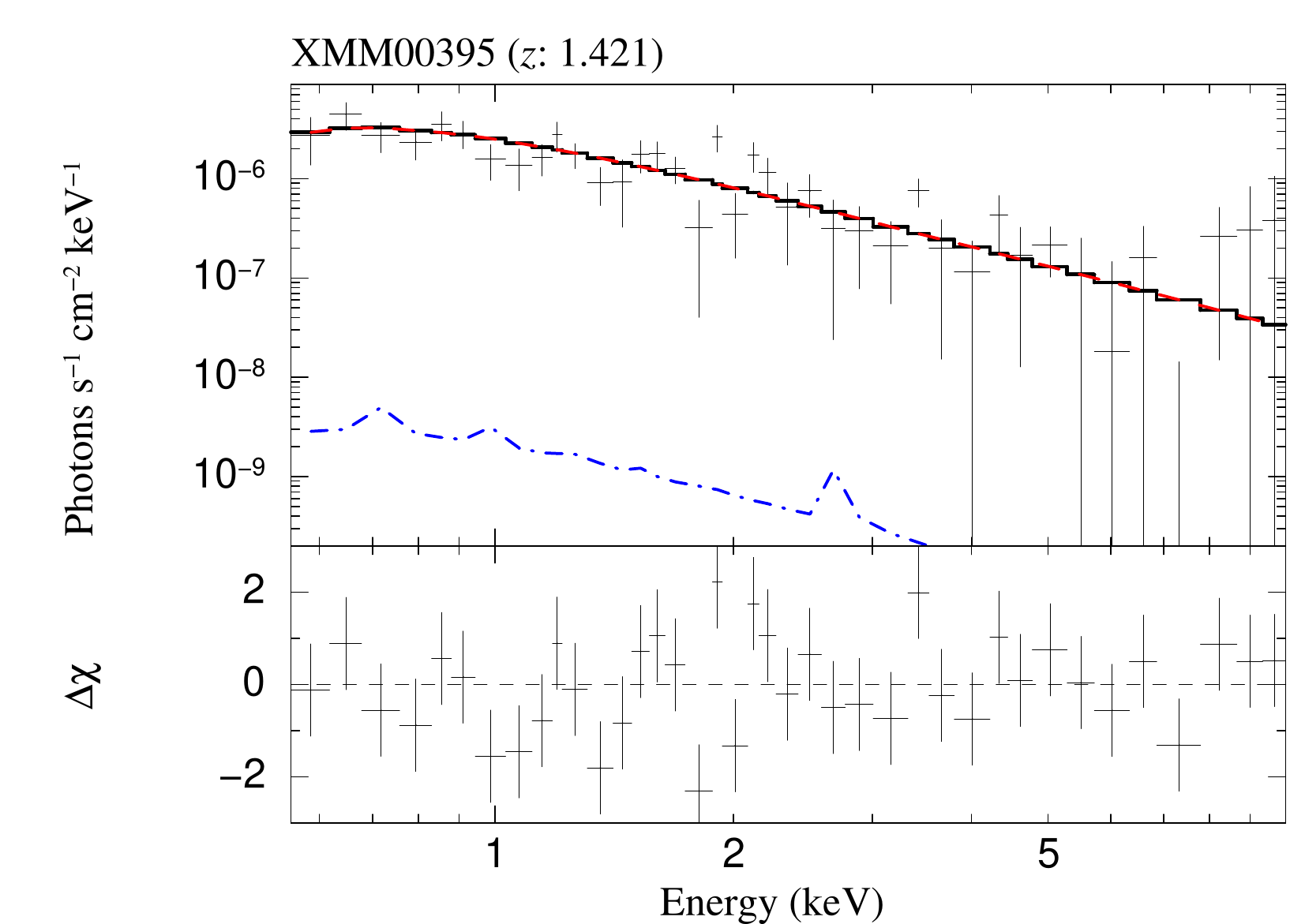}
\includegraphics[angle=-90, width=5.5cm, trim={0.0cm 0.0cm 0.0cm 0.0cm}, clip]{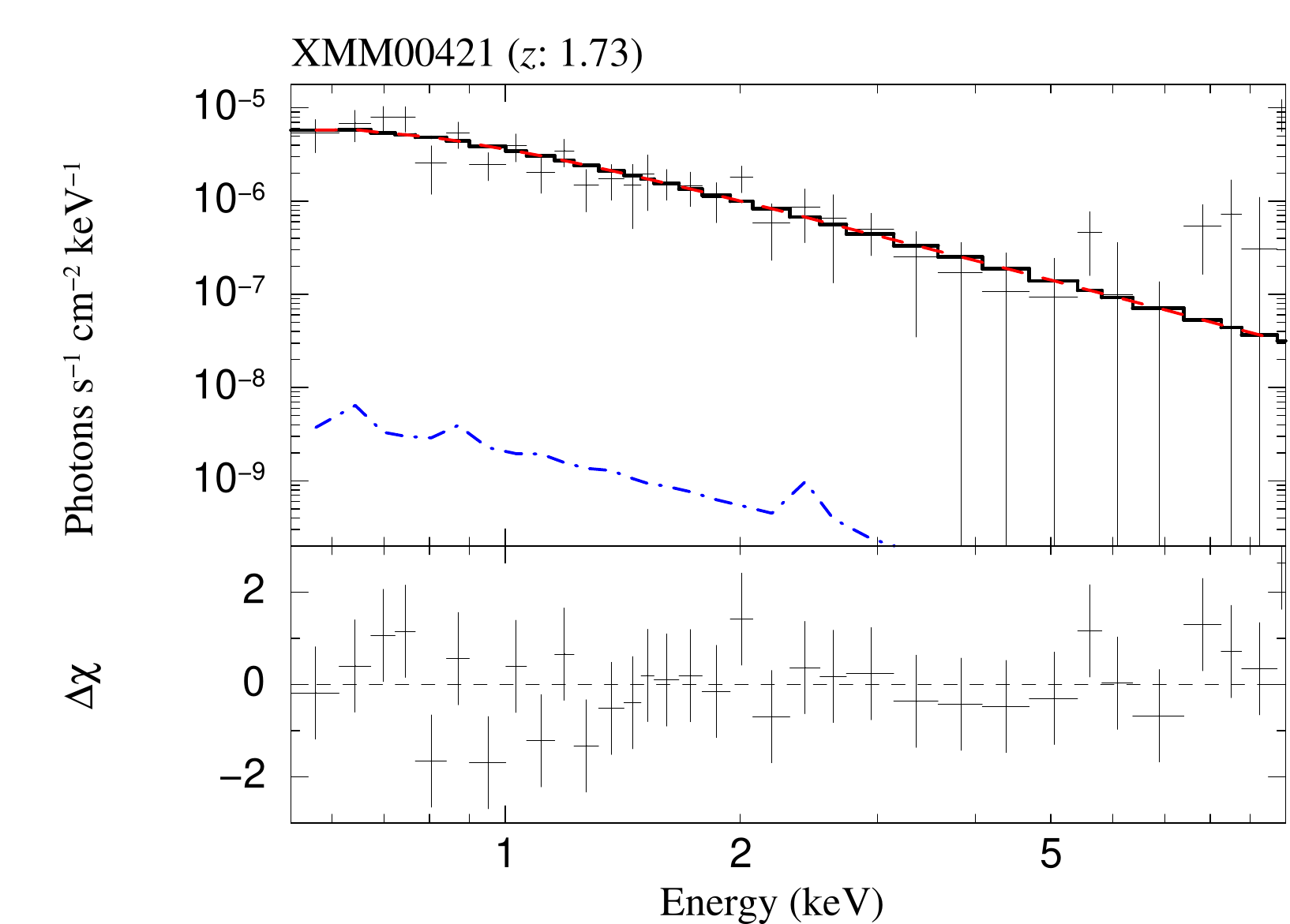}
\includegraphics[angle=-90, width=5.5cm, trim={0.0cm 0.0cm 0.0cm 0.0cm}, clip]{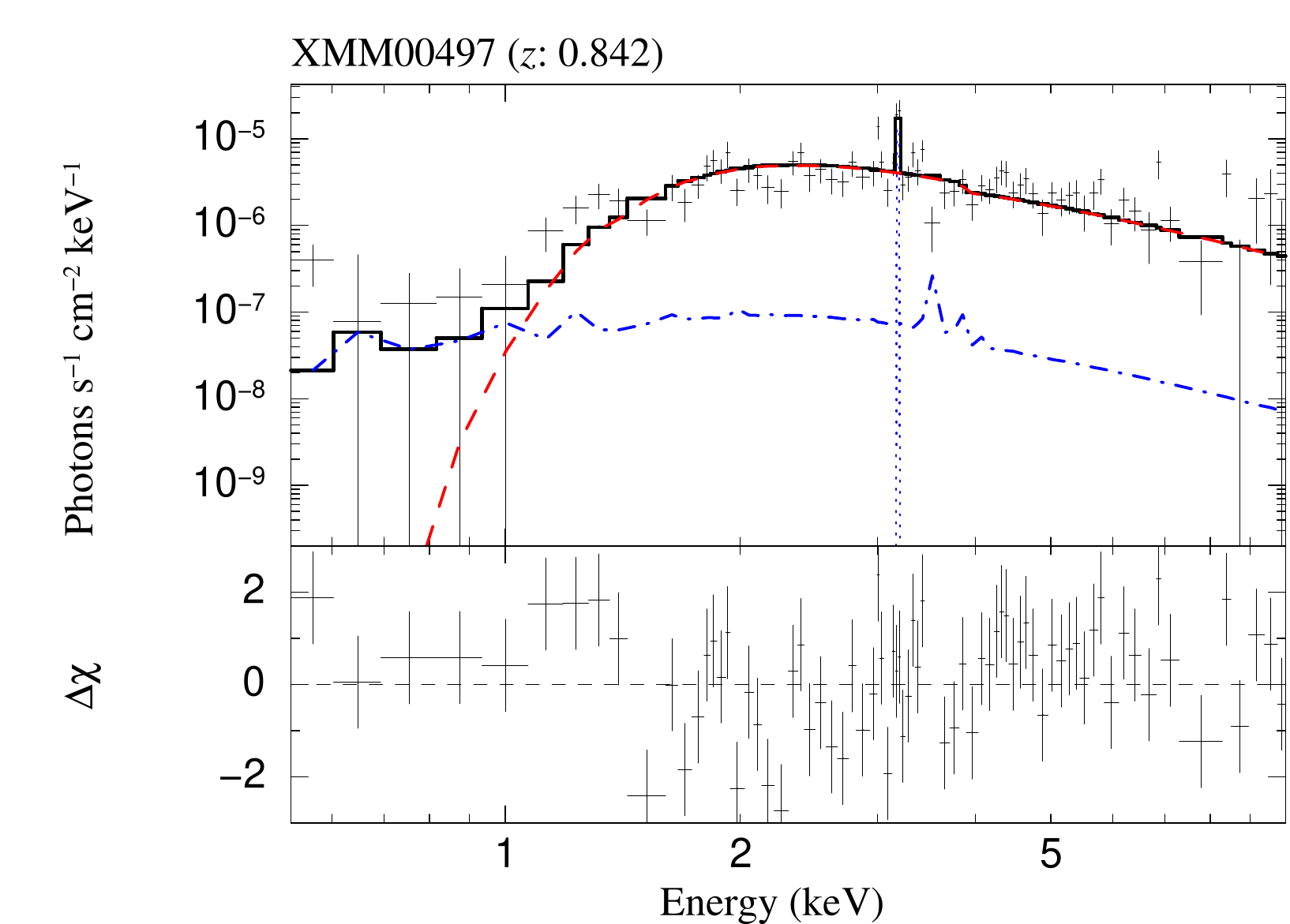}
\includegraphics[angle=-90, width=5.5cm, trim={0.0cm 0.0cm 0.0cm 0.0cm}, clip]{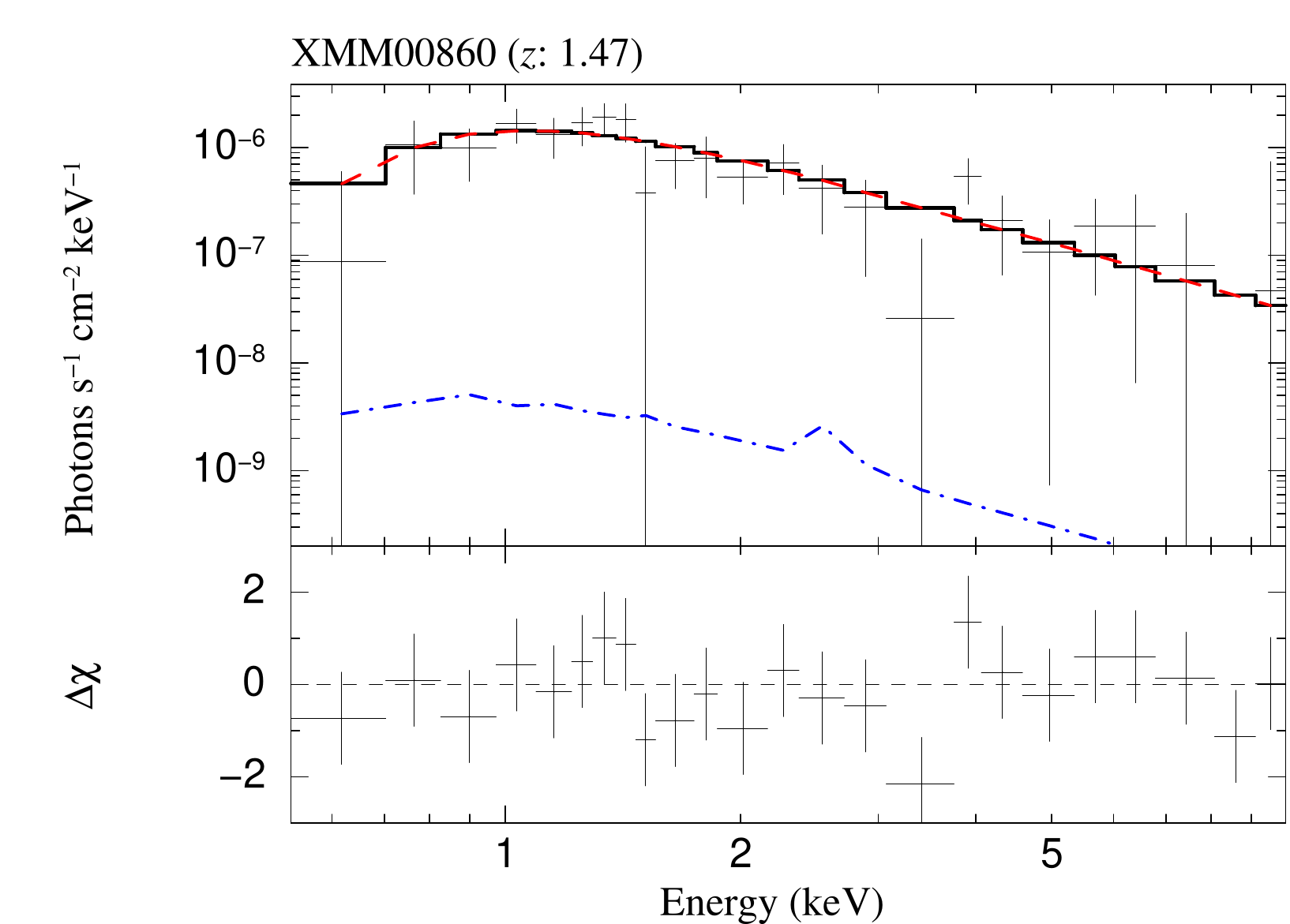}
\includegraphics[angle=-90, width=5.5cm, trim={0.0cm 0.0cm 0.0cm 0.0cm}, clip]{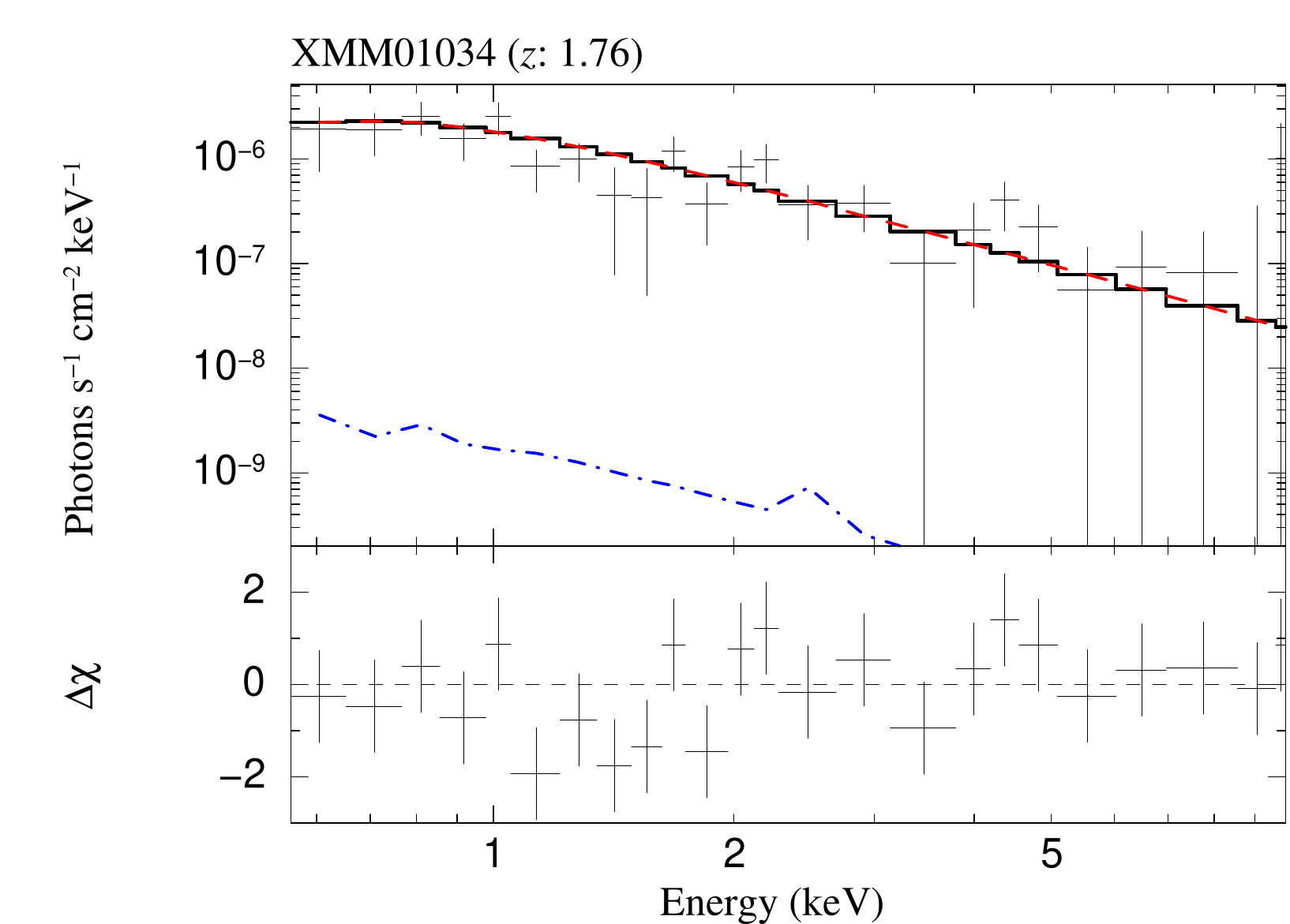}
\includegraphics[angle=-90, width=5.5cm, trim={0.0cm 0.0cm 0.0cm 0.0cm}, clip]{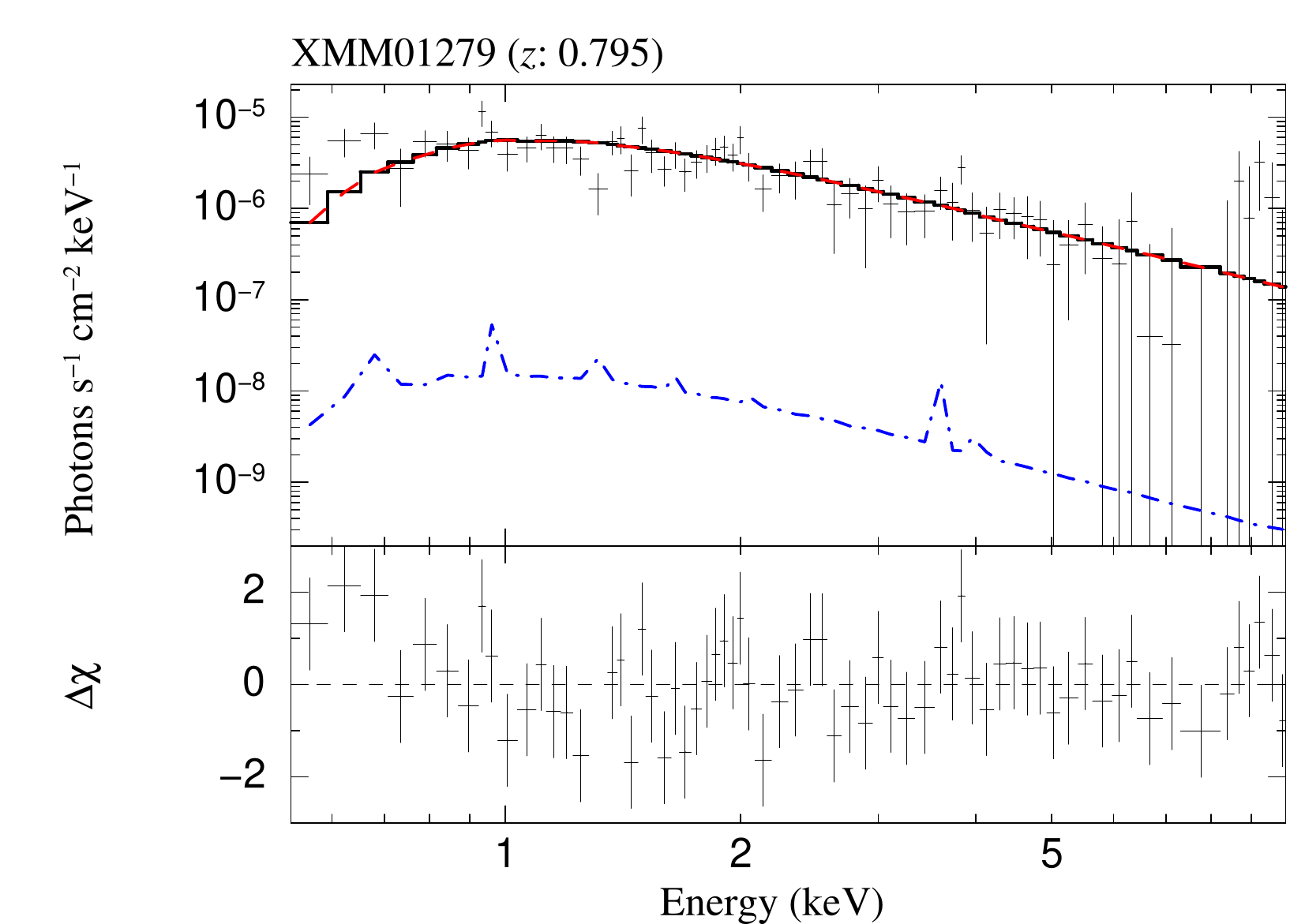}
\caption{{The $0.5-10$ keV unfolded X-ray spectra of our DOGs best-fitted with {\scshape borus02} model.
For better visualisation, spectra are rebinned with 10 counts per bin. The best-fitted model is shown with a solid black curve, and
model components, {\ie} cutoff power law, scattered cutoff power law and reflection component, are shown with a red dashed curve, cyan dash tipple-dotted curve and blue dash-dotted curve, respectively. The Fe K$\alpha$ line is shown by a blue narrow Gaussian.
The {XMM-Newton} and {\em Chandra} data points are shown with black crosses and purple empty circles, respectively. The bottom panels in each
sub-figure show residuals.}}
\label{fig:BestFits}
\end{figure*}
\addtocounter{figure}{-1}
\begin{figure*}
\centering
\includegraphics[angle=-90, width=5.5cm, trim={0.0cm 0.0cm 0.0cm 0.0cm}, clip]{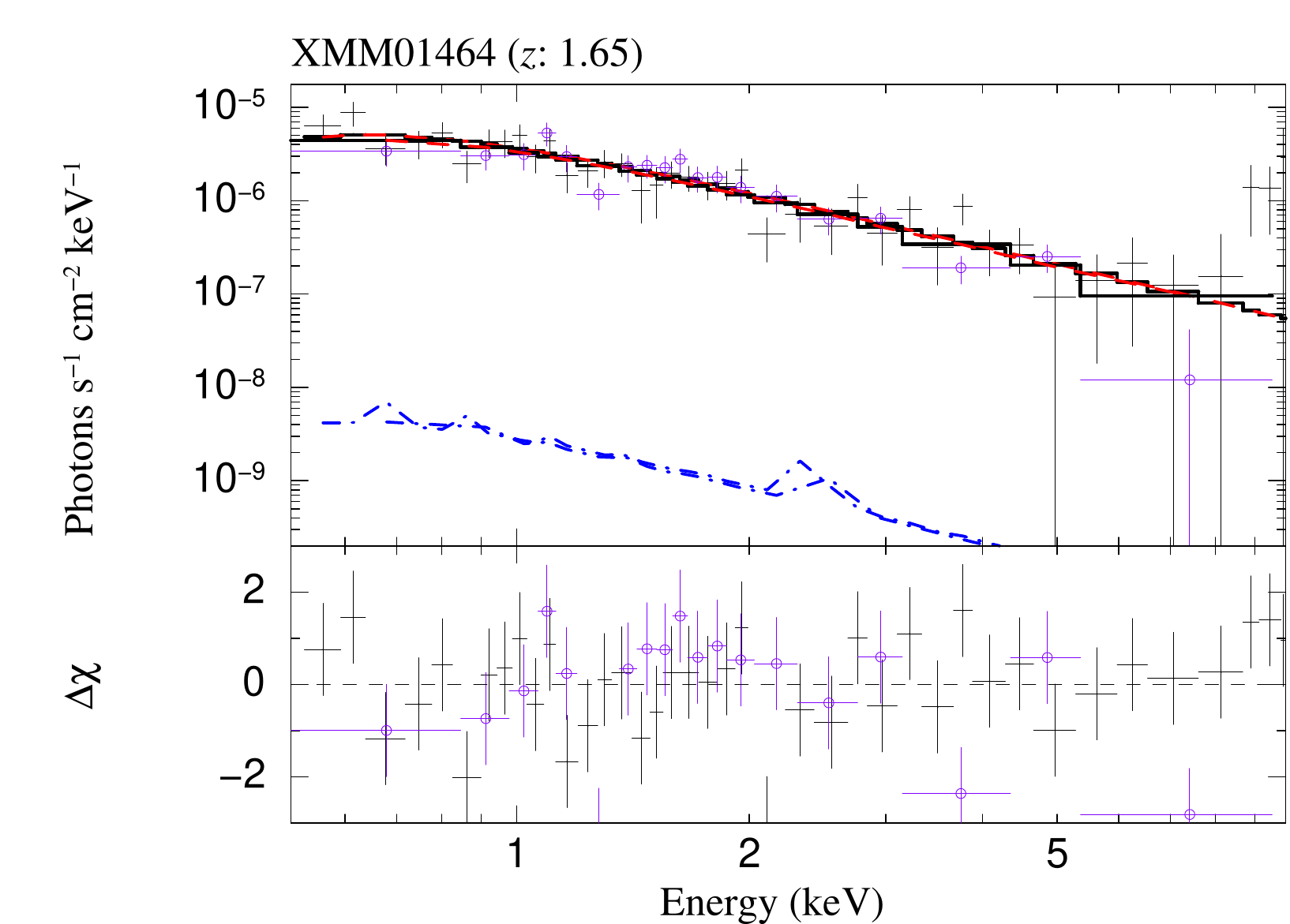}
\includegraphics[angle=-90, width=5.5cm, trim={0.0cm 0.0cm 0.0cm 0.0cm}, clip]{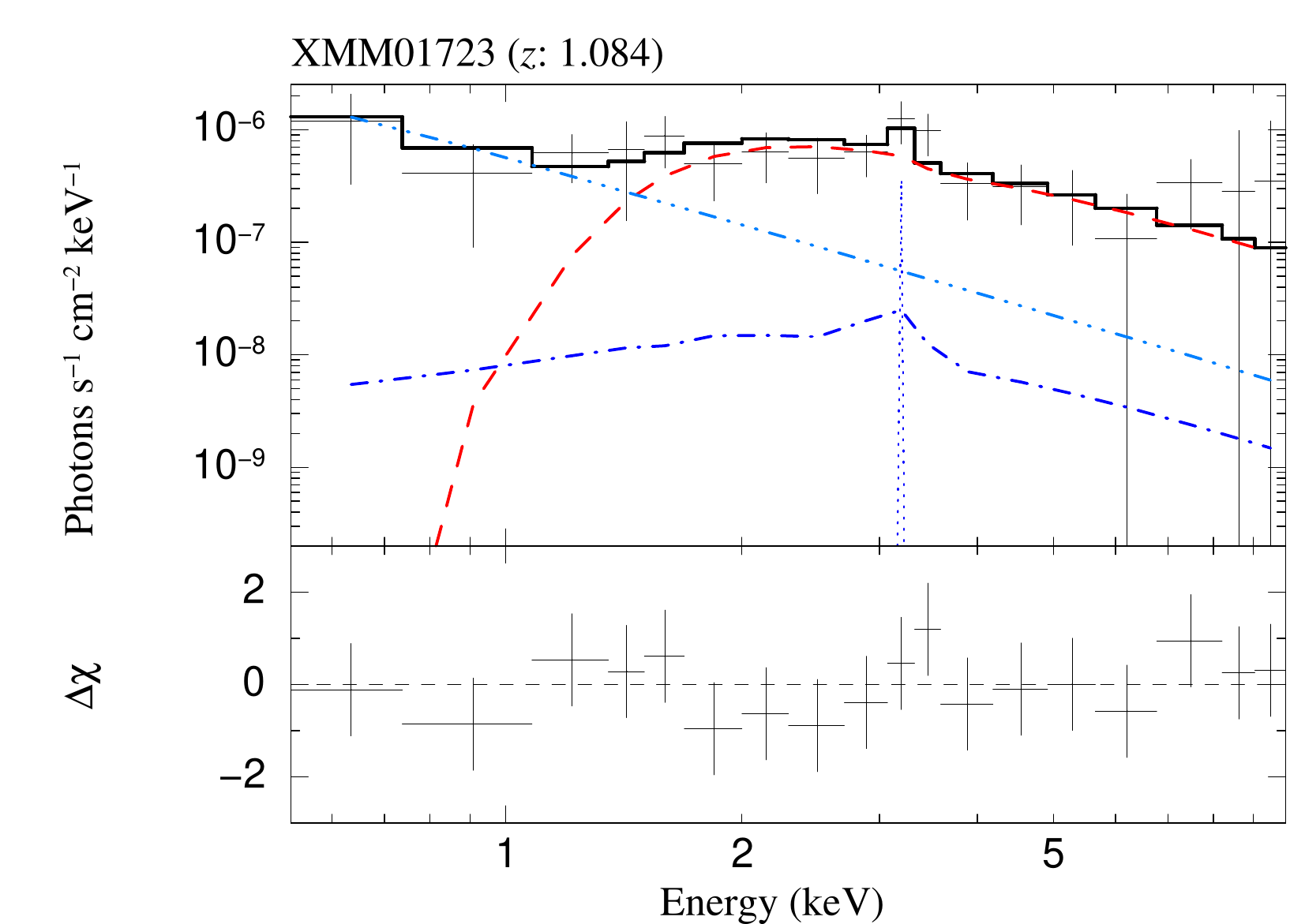}
\includegraphics[angle=-90, width=5.5cm, trim={0.0cm 0.0cm 0.0cm 0.0cm}, clip]{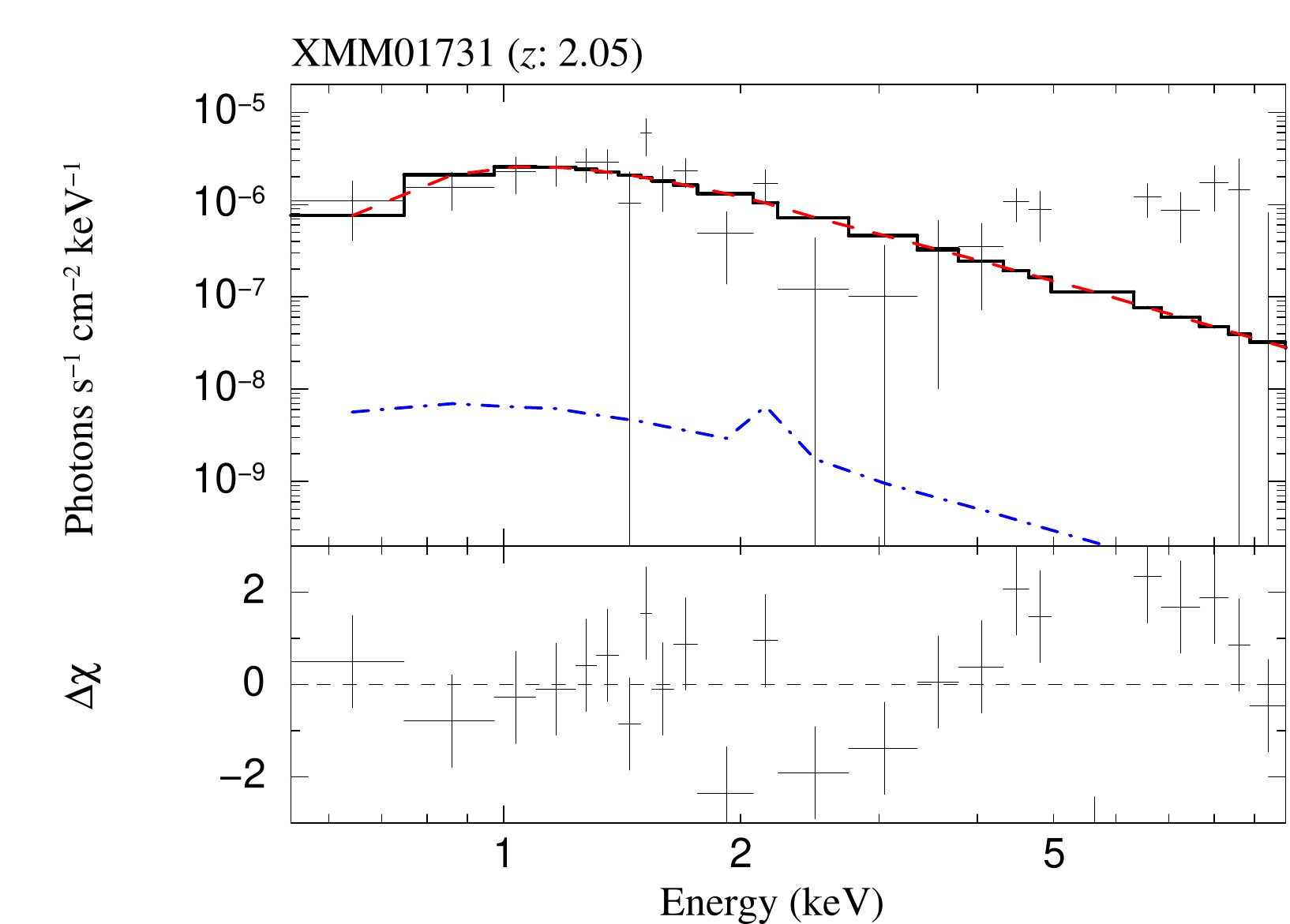}
\includegraphics[angle=-90, width=5.5cm, trim={0.0cm 0.0cm 0.0cm 0.0cm}, clip]{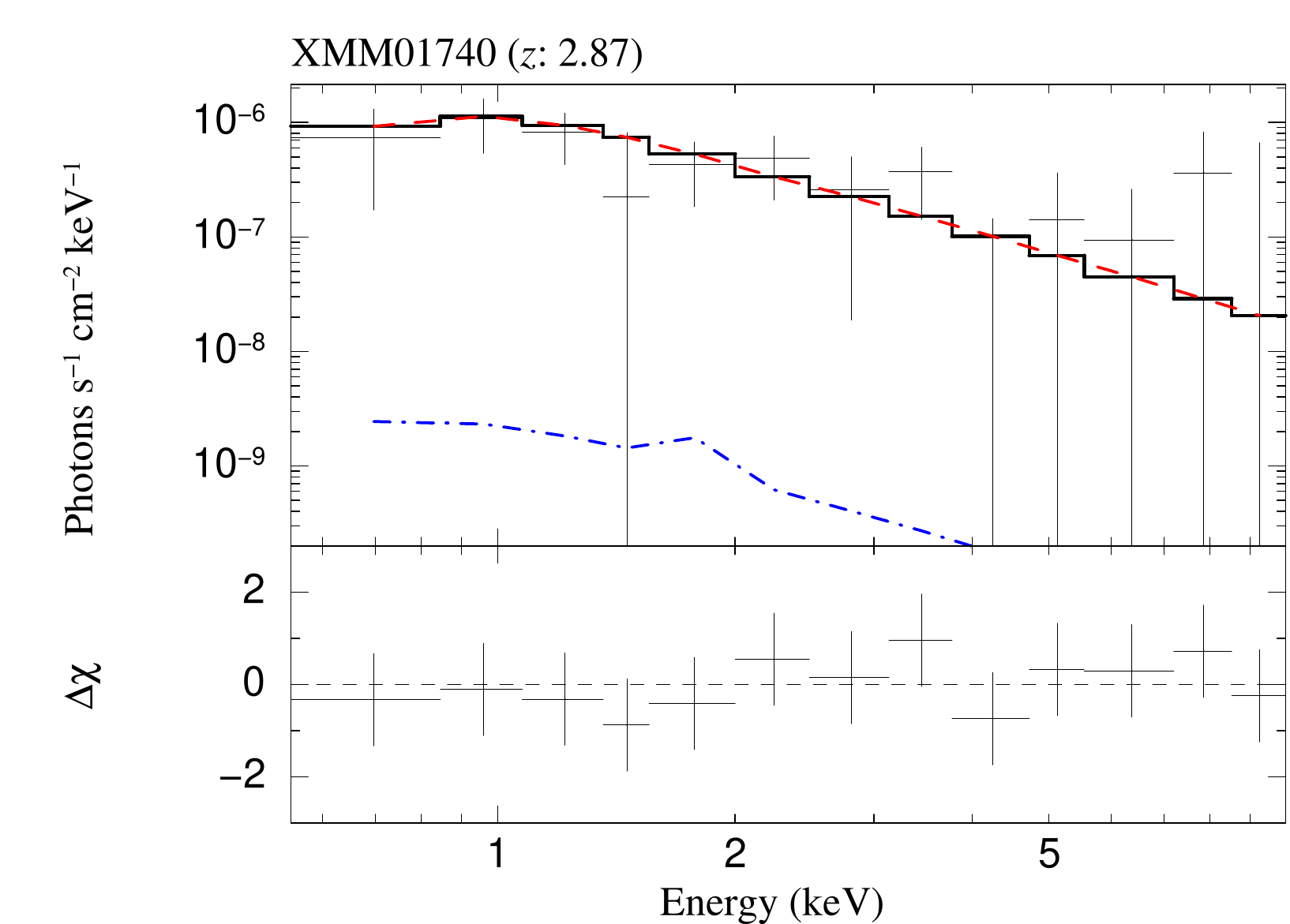}
\includegraphics[angle=-90, width=5.5cm, trim={0.0cm 0.0cm 0.0cm 0.0cm}, clip]{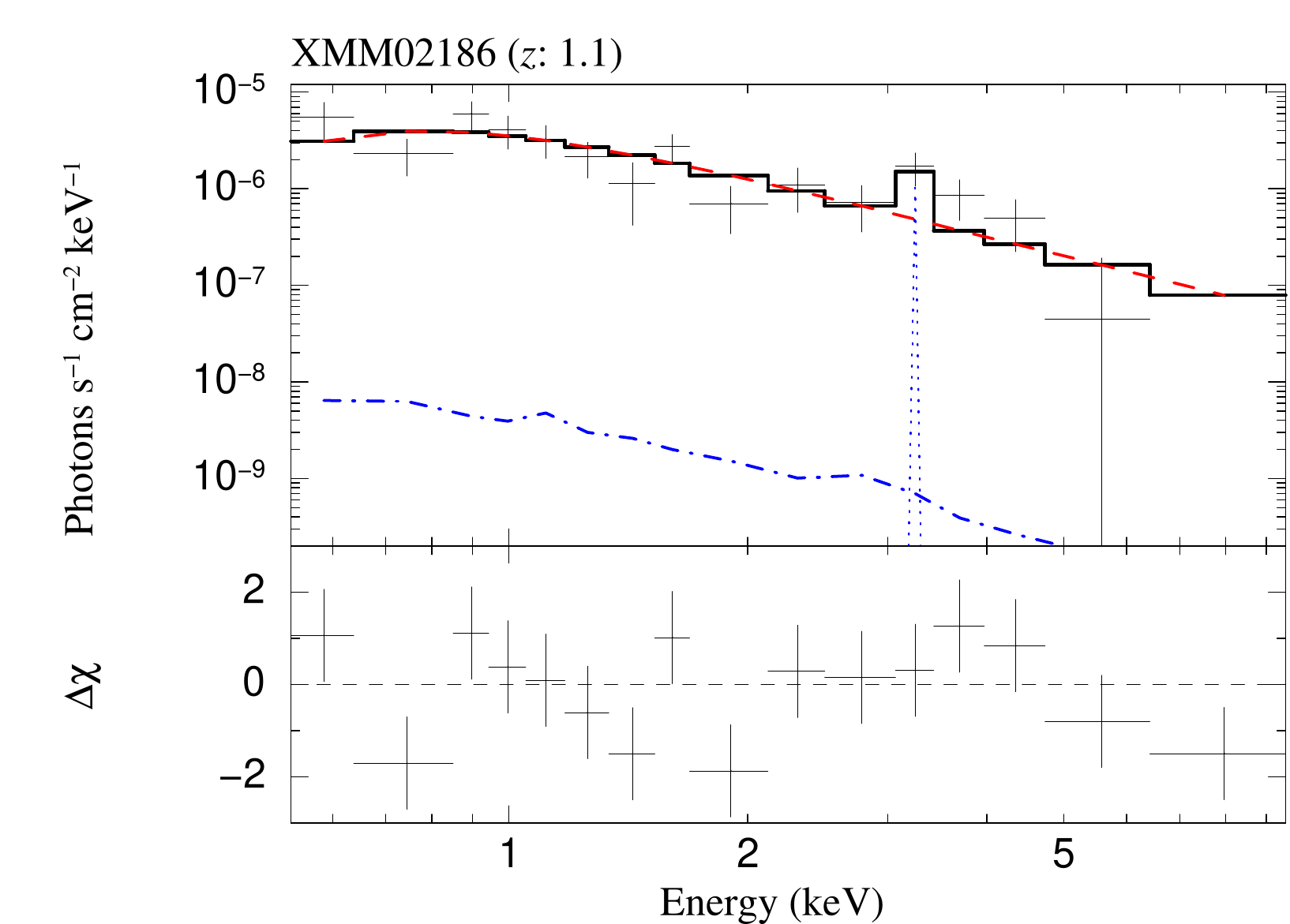}
\includegraphics[angle=-90, width=5.5cm, trim={0.0cm 0.0cm 0.0cm 0.0cm}, clip]{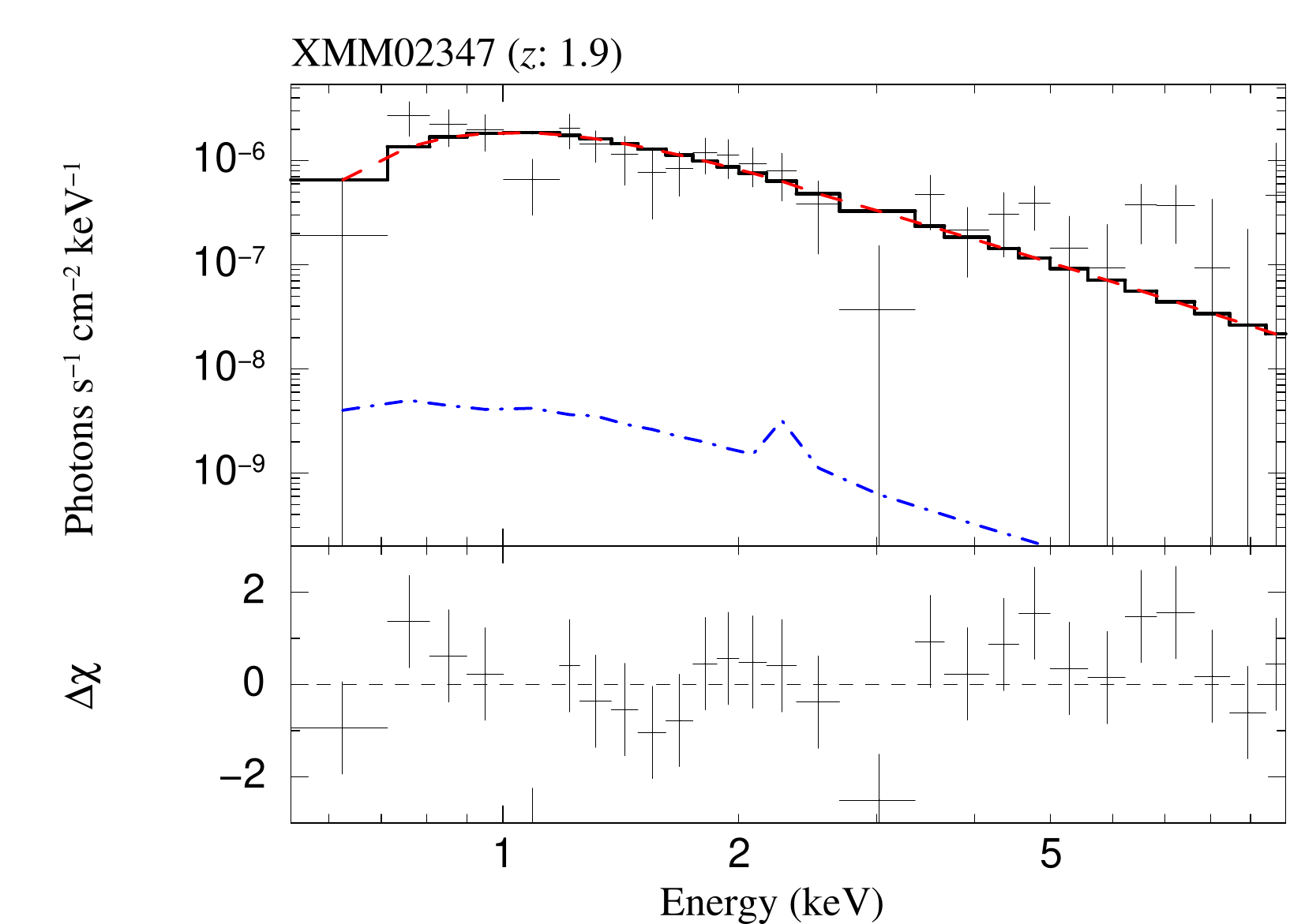}
\includegraphics[angle=-90, width=5.5cm, trim={0.0cm 0.0cm 0.0cm 0.0cm}, clip]{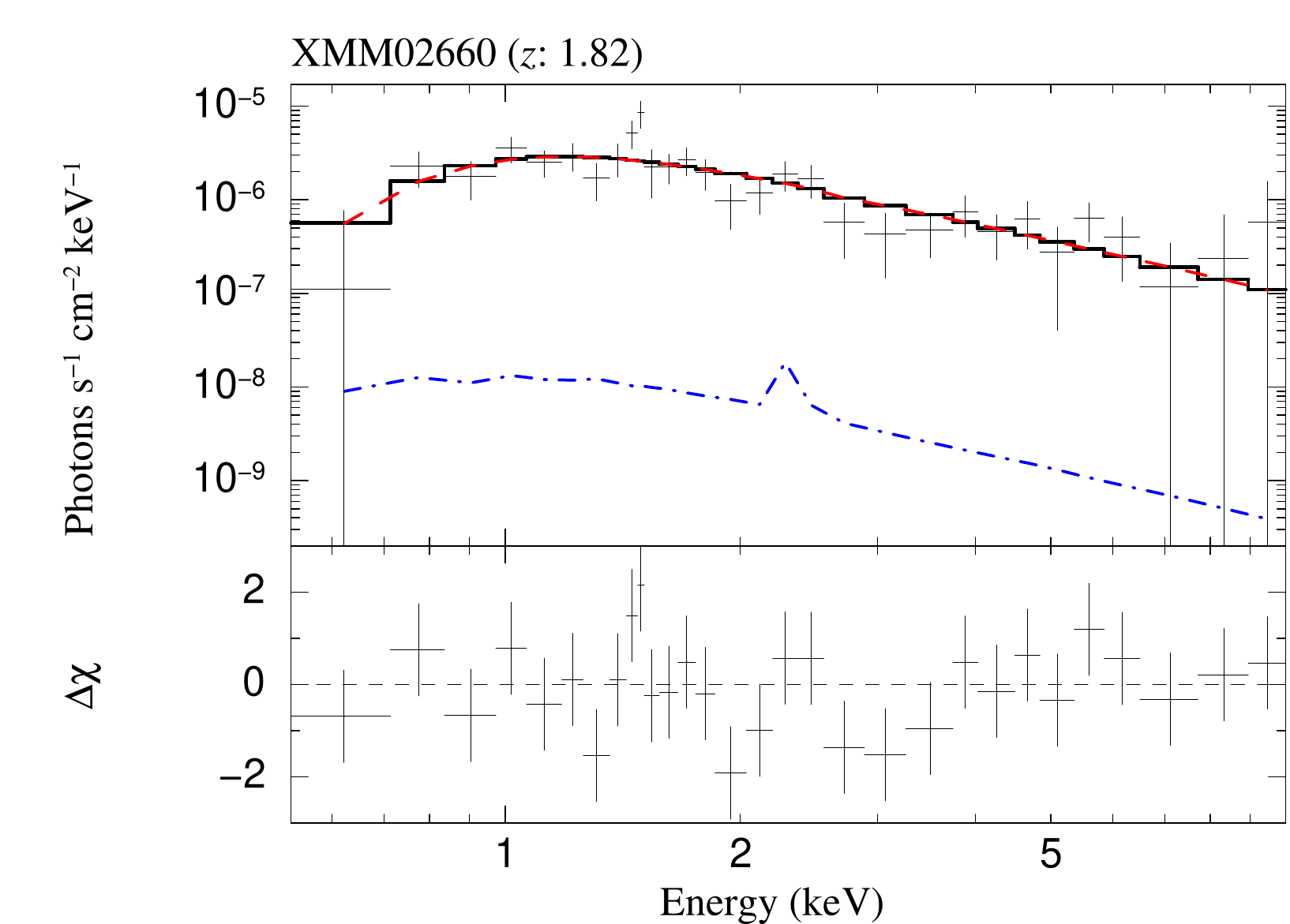}
\includegraphics[angle=-90, width=5.5cm, trim={0.0cm 0.0cm 0.0cm 0.0cm}, clip]{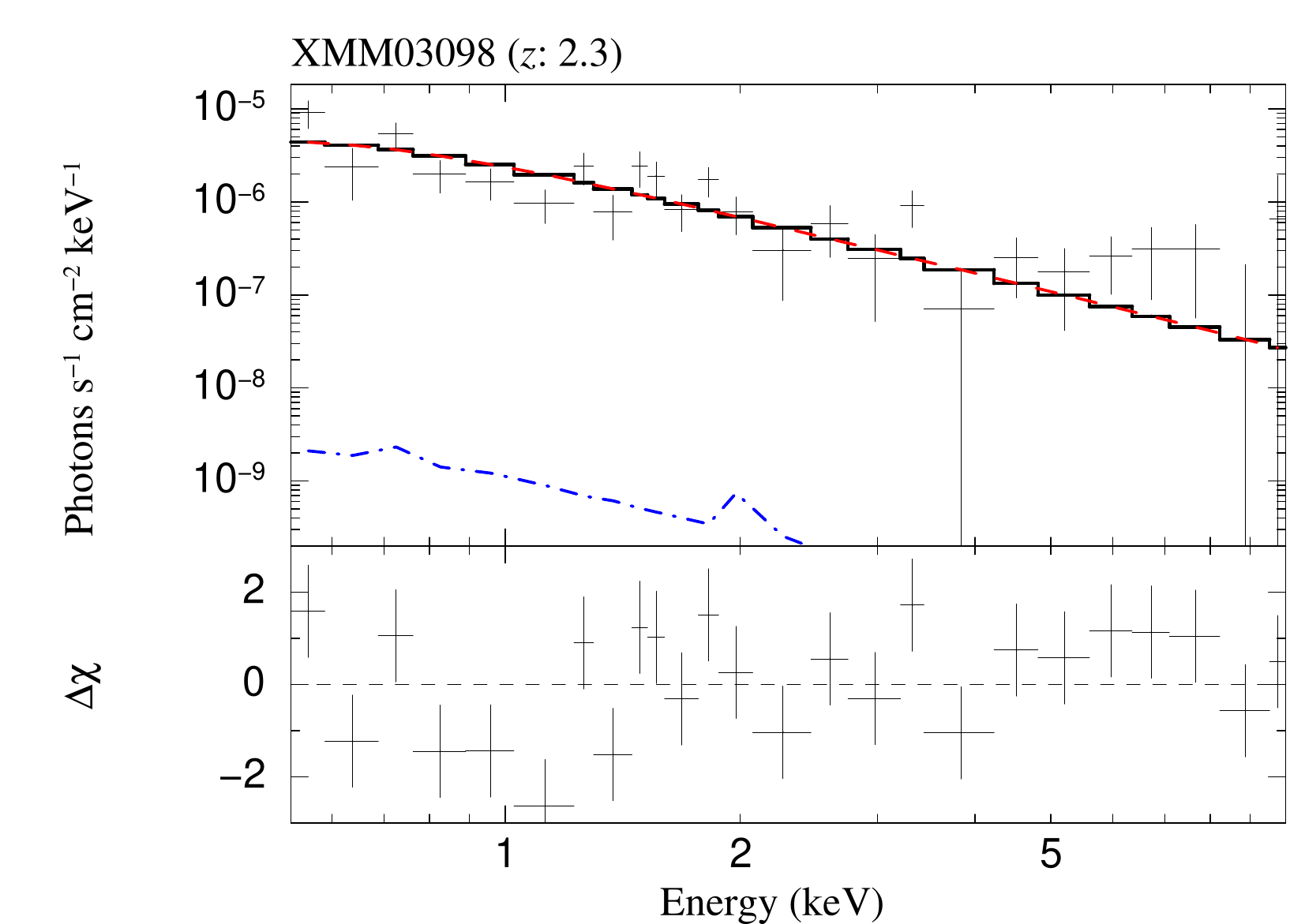}
\includegraphics[angle=-90, width=5.5cm, trim={0.0cm 0.0cm 0.0cm 0.0cm}, clip]{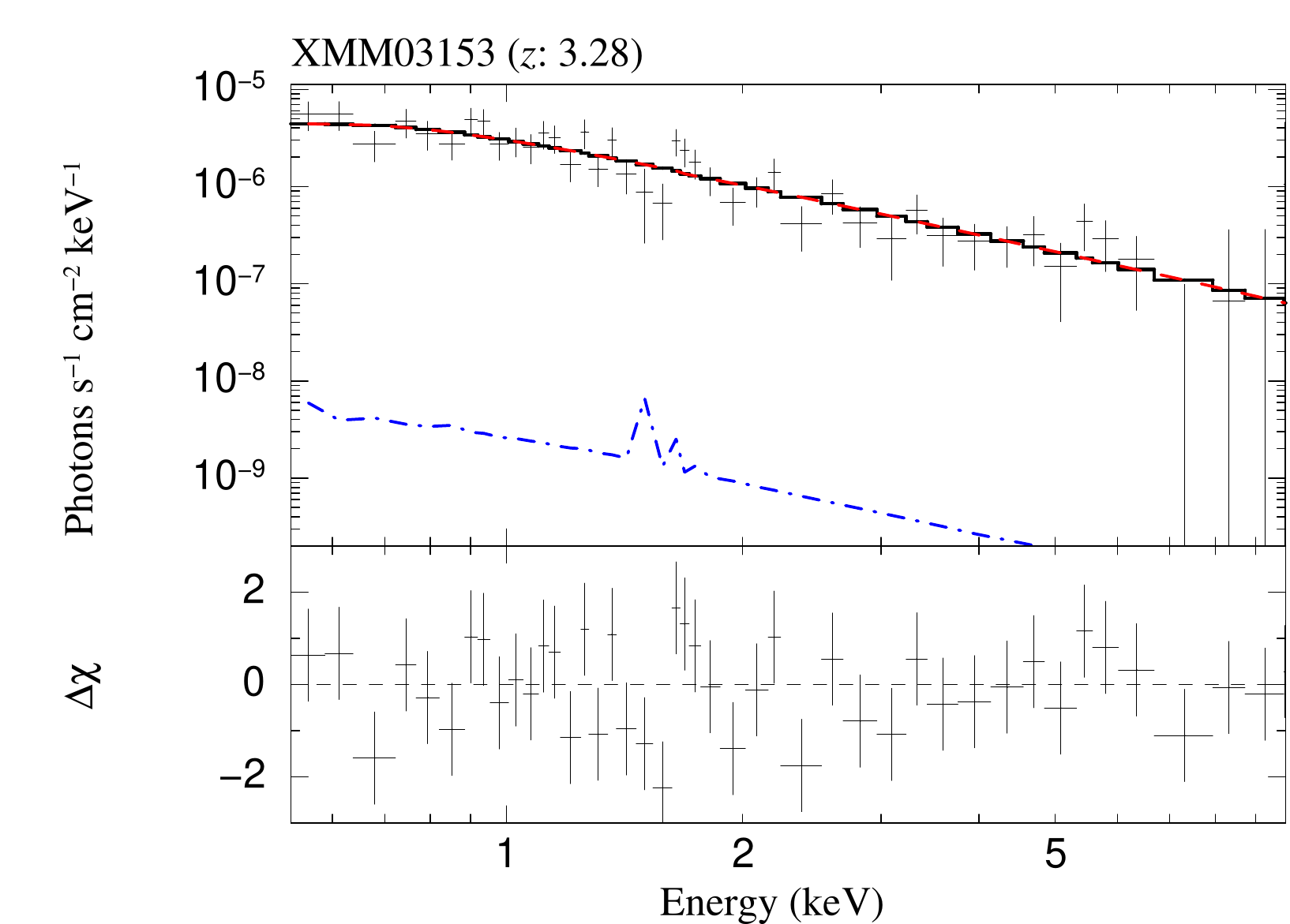}
\includegraphics[angle=-90, width=5.5cm, trim={0.0cm 0.0cm 0.0cm 0.0cm}, clip]{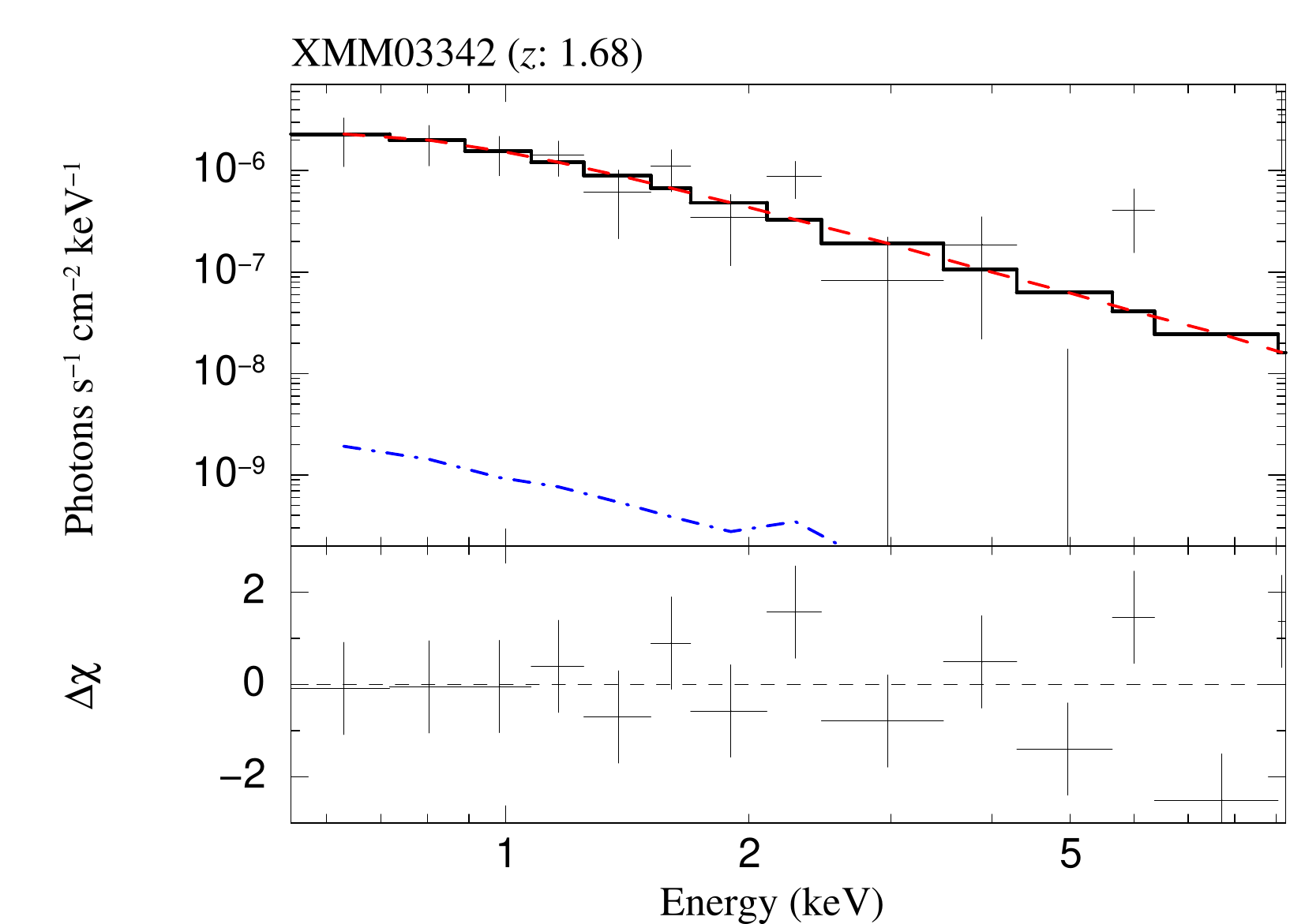}
\includegraphics[angle=-90, width=5.5cm, trim={0.0cm 0.0cm 0.0cm 0.0cm}, clip]{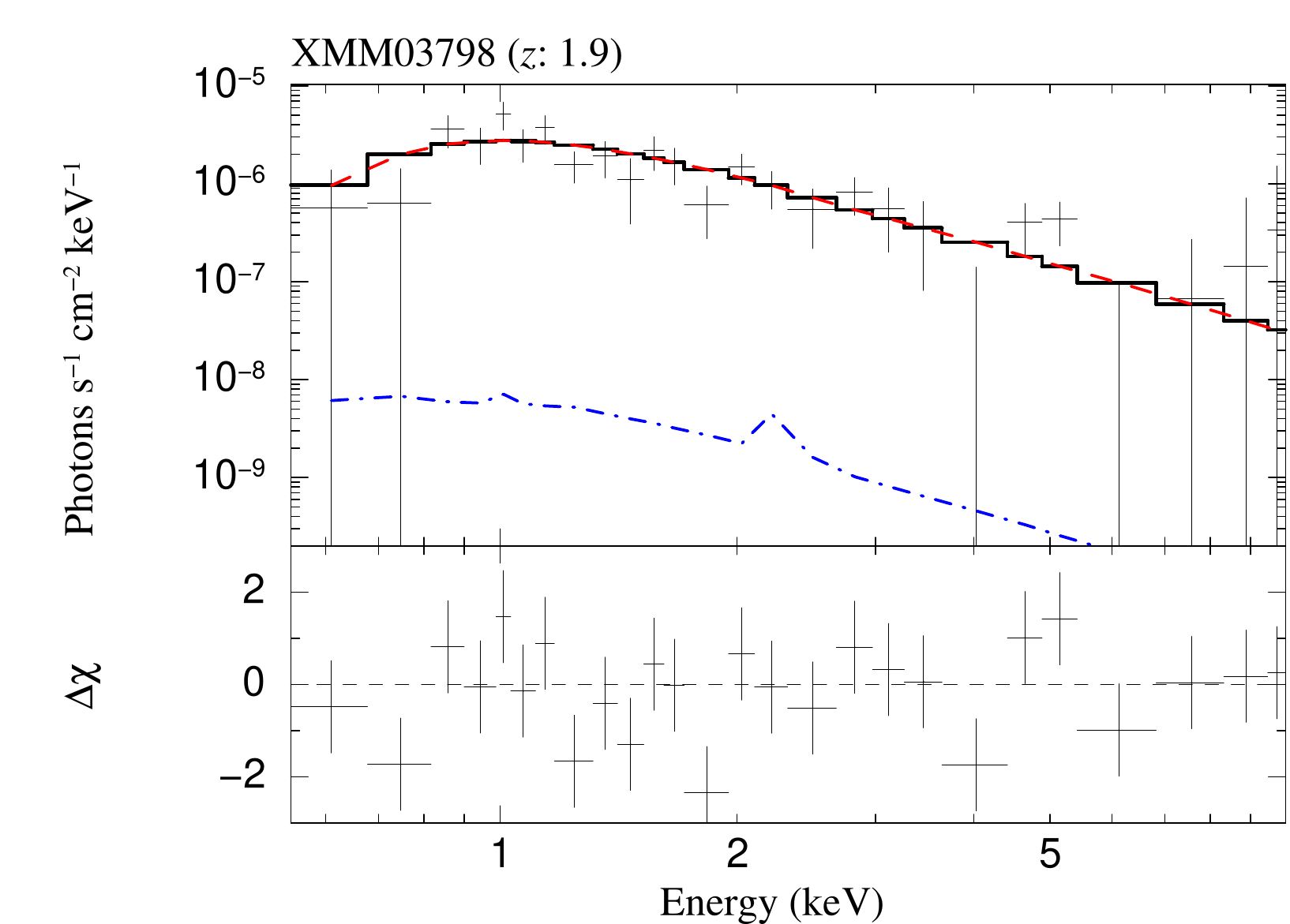}
\includegraphics[angle=-90, width=5.5cm, trim={0.0cm 0.0cm 0.0cm 0.0cm}, clip]{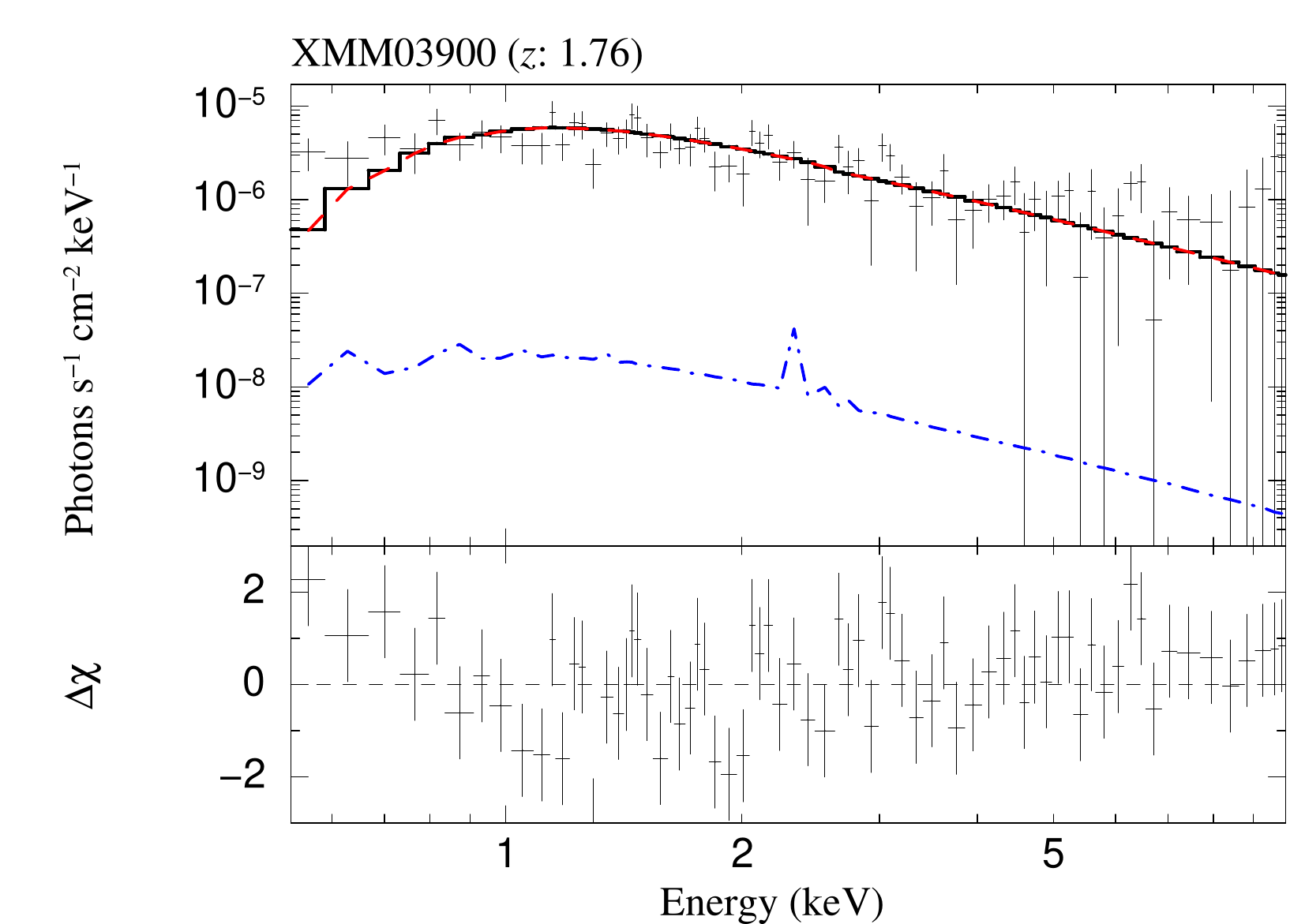}
\includegraphics[angle=-90, width=5.5cm, trim={0.0cm 0.0cm 0.0cm 0.0cm}, clip]{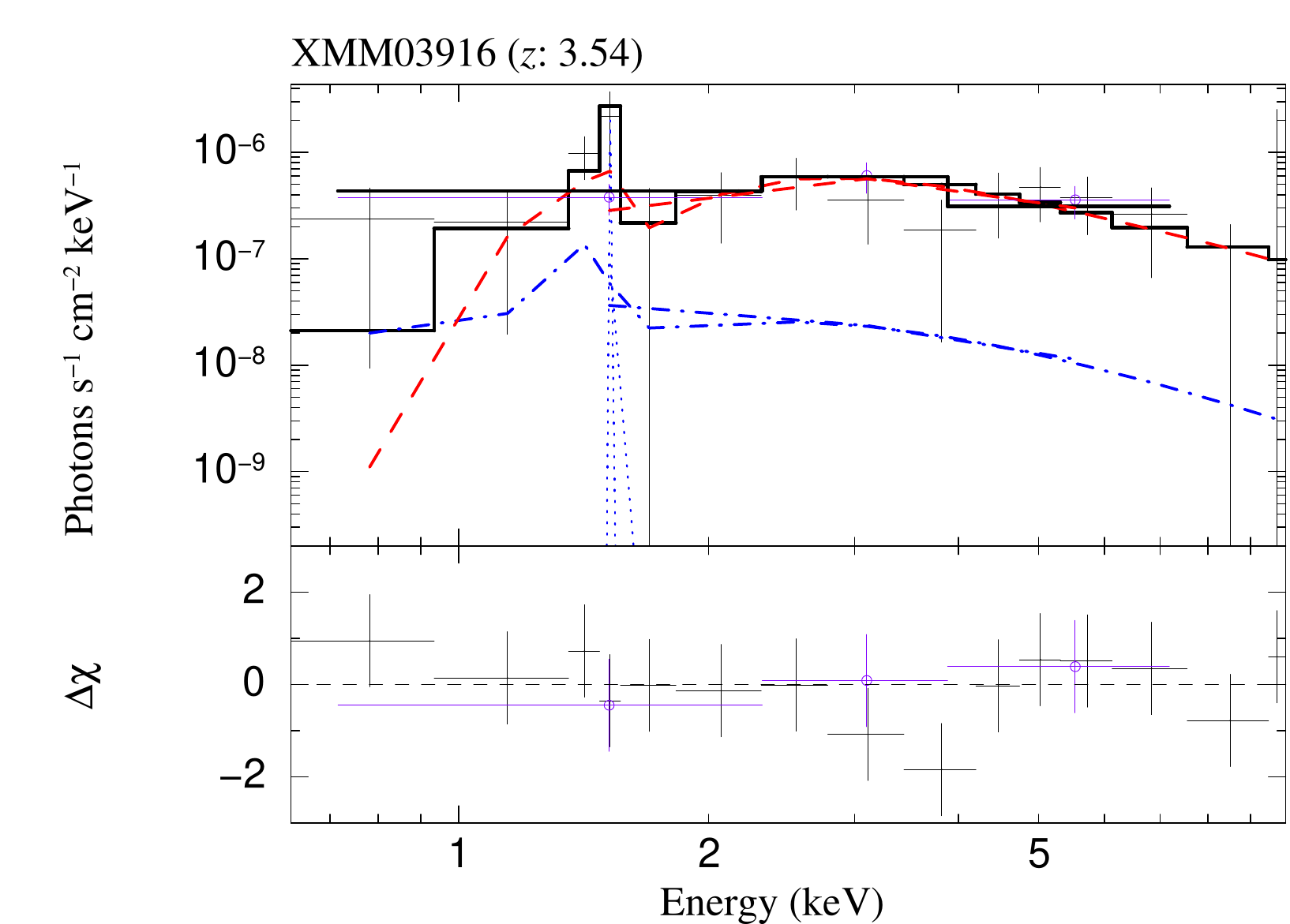}
\includegraphics[angle=-90, width=5.5cm, trim={0.0cm 0.0cm 0.0cm 0.0cm}, clip]{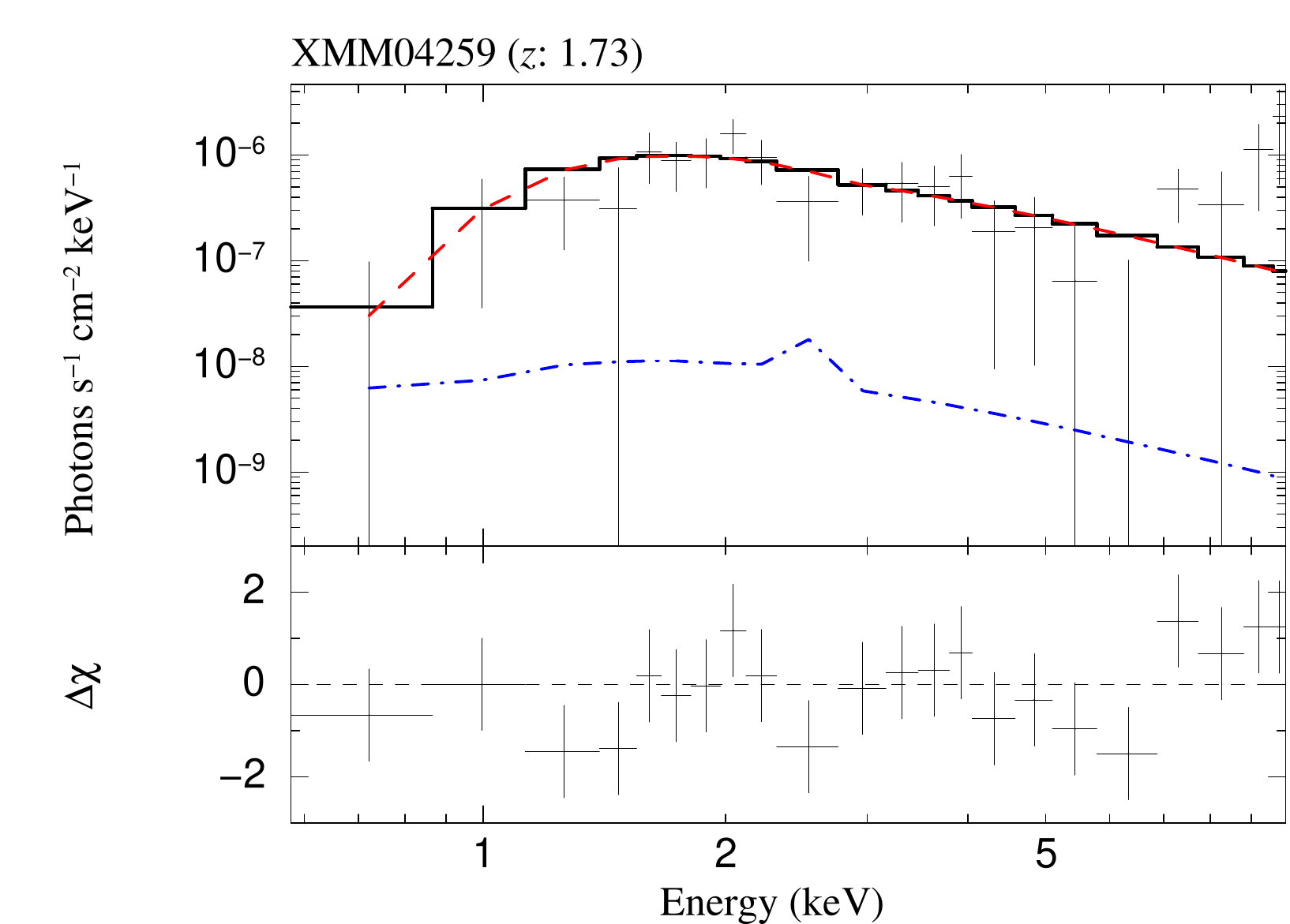}
\includegraphics[angle=-90, width=5.5cm, trim={0.0cm 0.0cm 0.0cm 0.0cm}, clip]{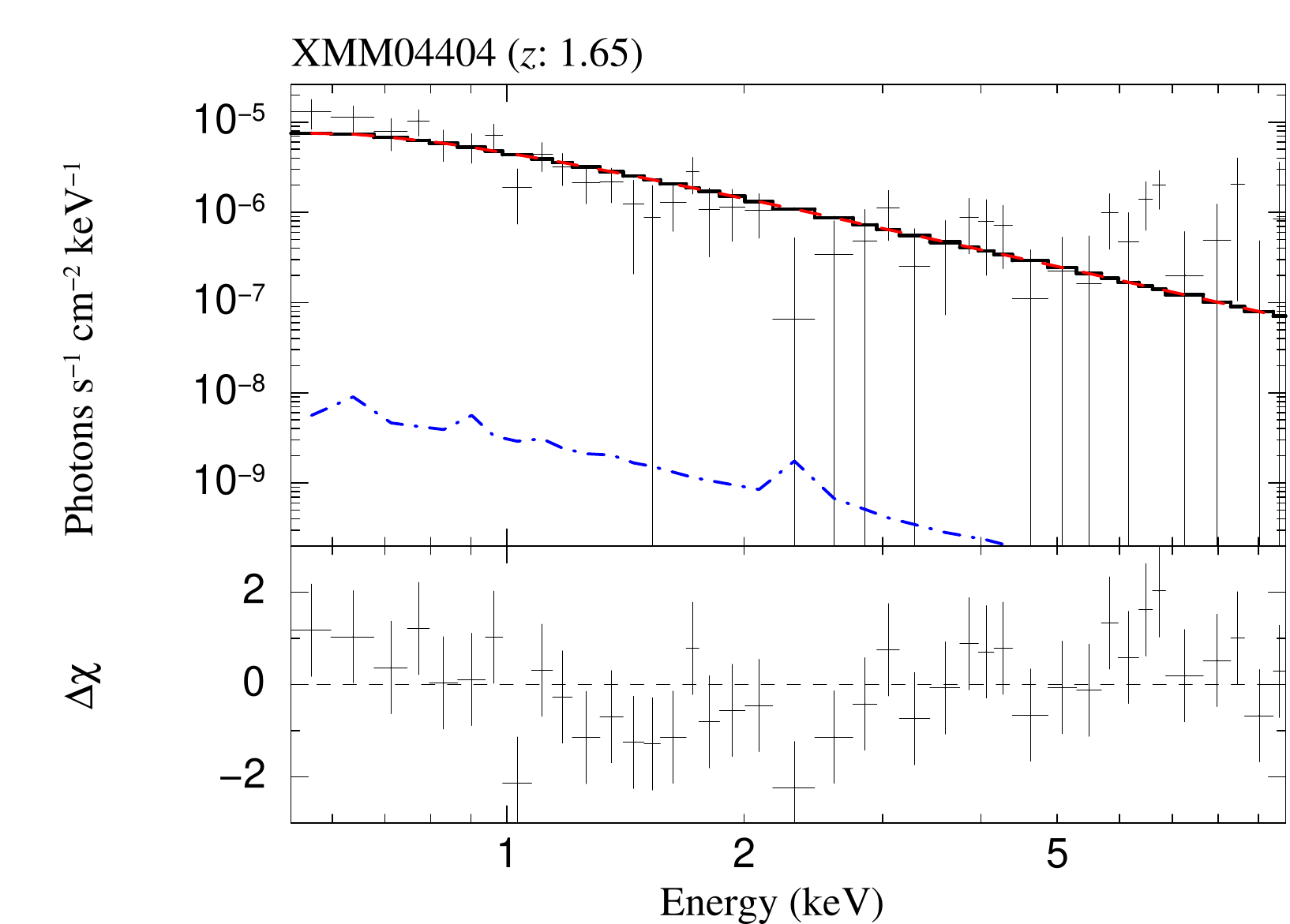}
\includegraphics[angle=-90, width=5.5cm, trim={0.0cm 0.0cm 0.0cm 0.0cm}, clip]{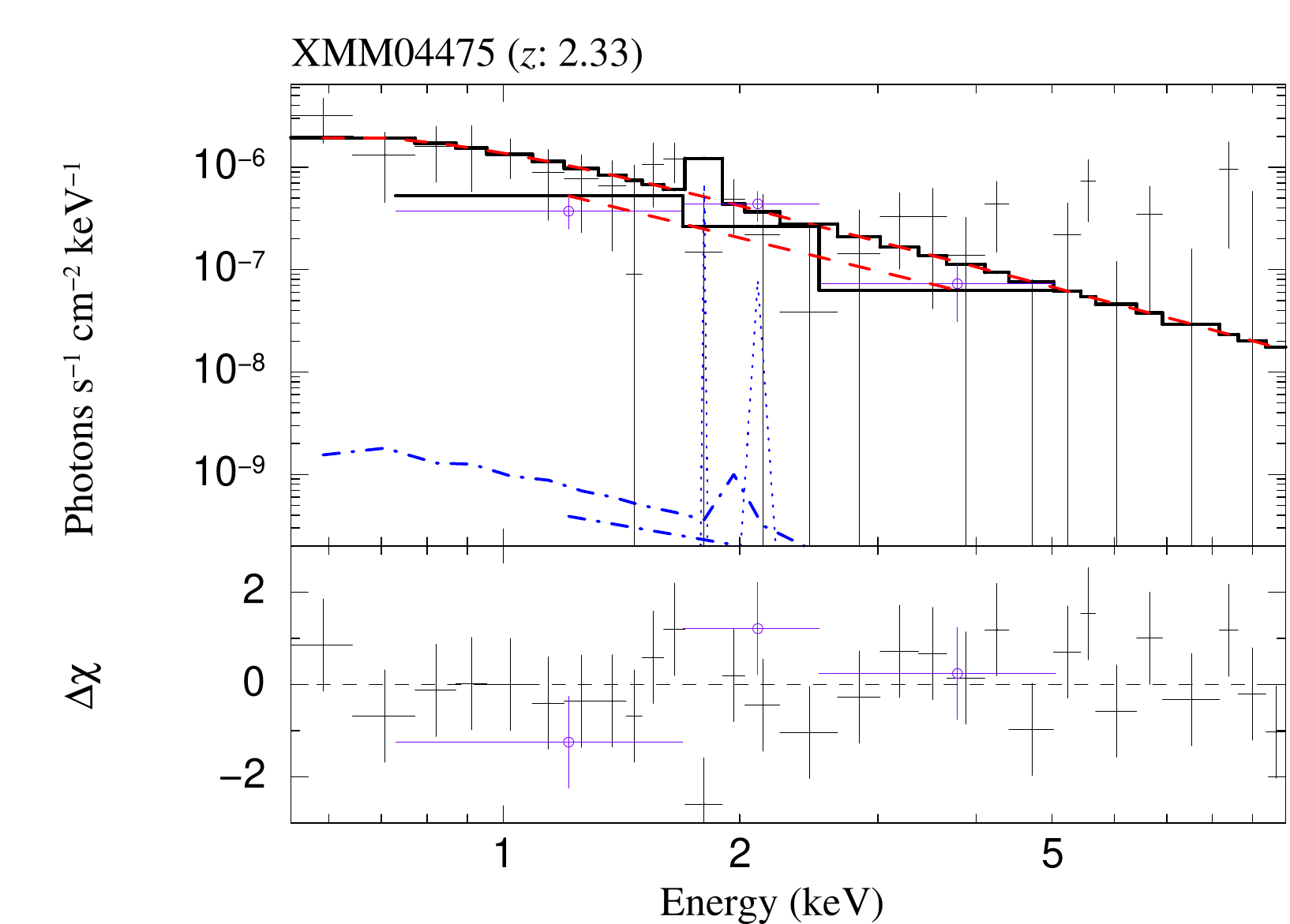}
\includegraphics[angle=-90, width=5.5cm, trim={0.0cm 0.0cm 0.0cm 0.0cm}, clip]{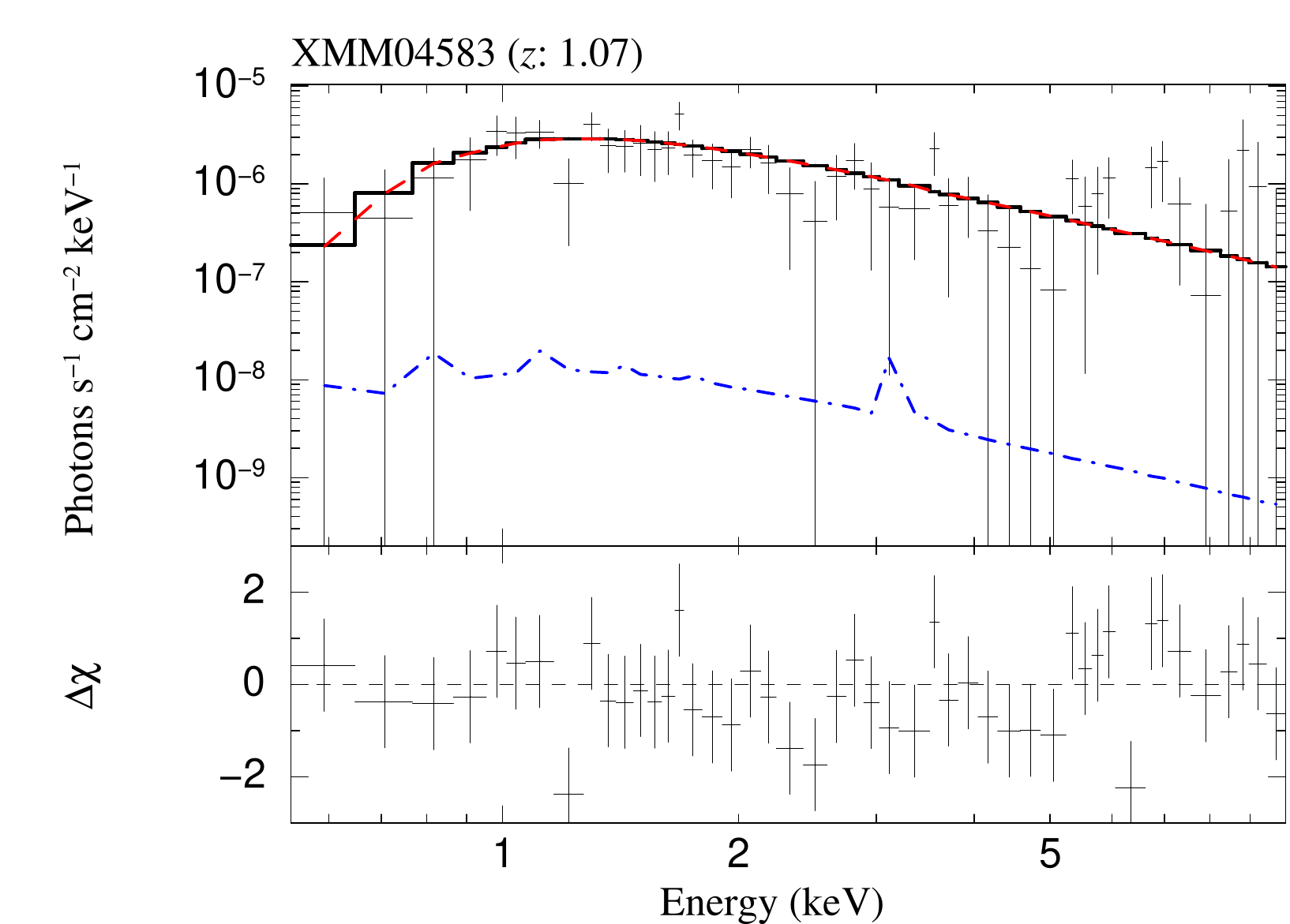}
\includegraphics[angle=-90, width=5.5cm, trim={0.0cm 0.0cm 0.0cm 0.0cm}, clip]{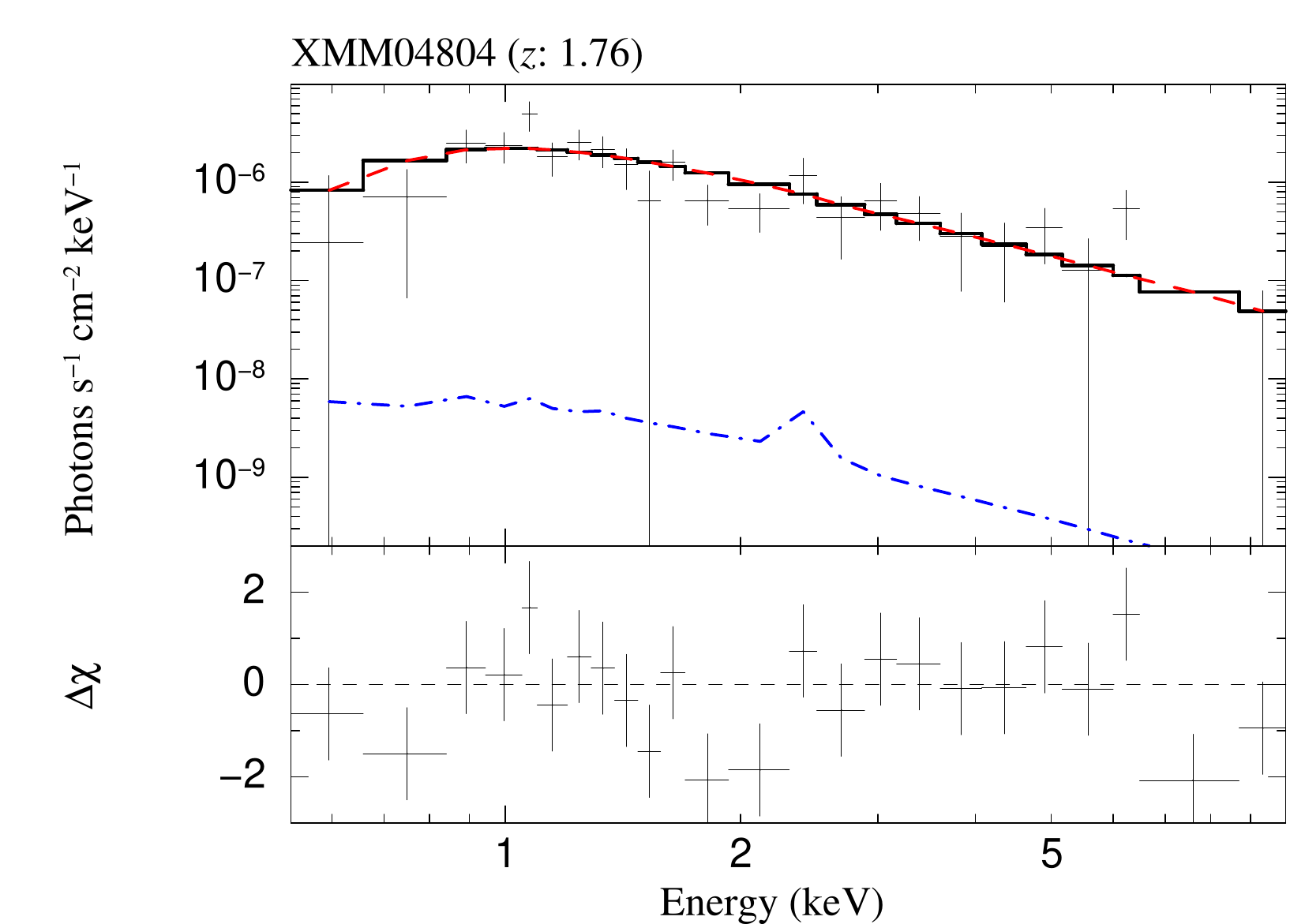}
\caption{{Continued.}}
\end{figure*}
\addtocounter{figure}{-1}
\begin{figure*}
\centering
\includegraphics[angle=-90, width=5.5cm, trim={0.0cm 0.0cm 0.0cm 0.0cm}, clip]{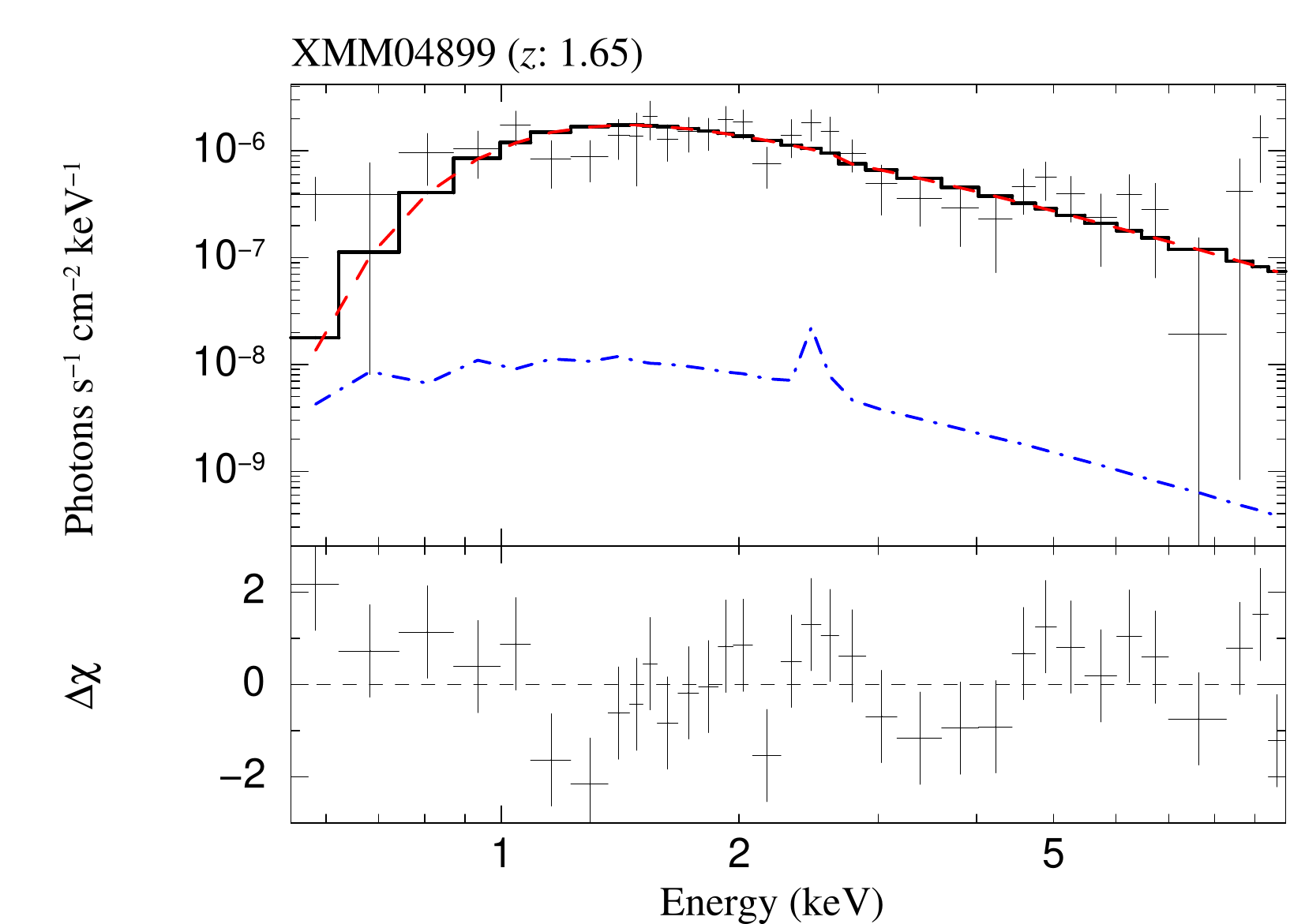}
\caption{{Continued.}}
\end{figure*}
\begin{figure*}
\centering
\includegraphics[angle=0, width=5.5cm, trim={0.0cm 0.0cm 0.0cm 0.0cm}, clip]{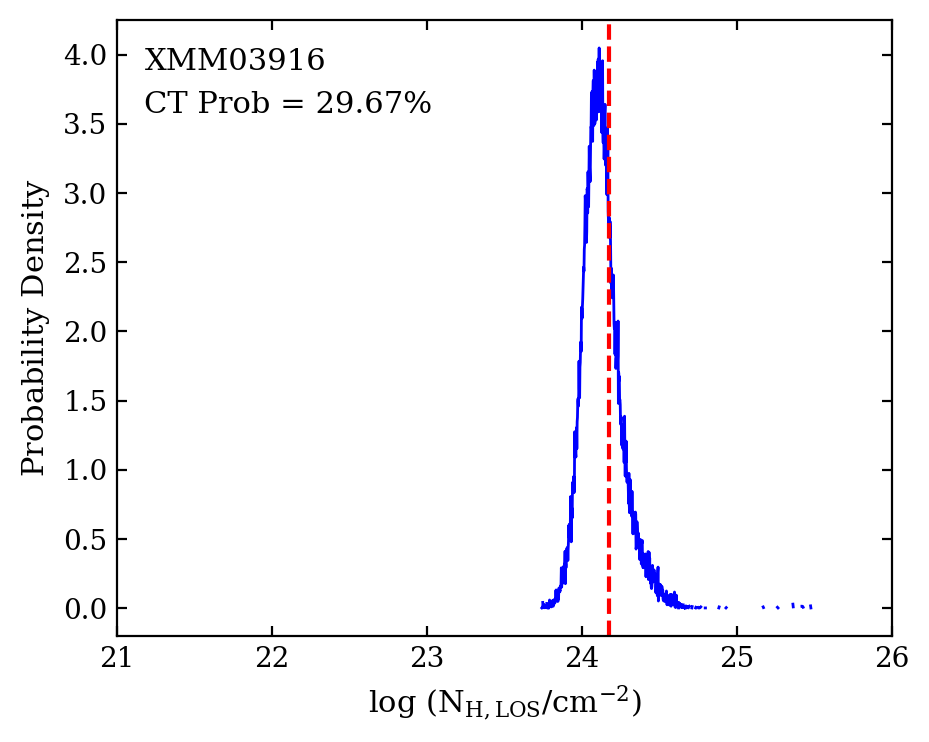}
\caption{The MCMC-based $N_{\rm H}$ probability distribution function (PDF) of the CT-AGN candidate XMM03916. The best fits using the {\scshape borus02} model are used to derive $N_{\rm H}$ PDFs.
The red dashed vertical line shows the Compton-thick limit of $N_{\rm H} = 1.5\times10^{24}$ cm$^{-2}$.}
\label{fig:PDFs}
\end{figure*}
\begin{table*}
 \centering
 \caption{The best-fitted X-ray fluxes and luminosities using absorbed {power law} and {\scshape borus02} models.}
 \renewcommand{\arraystretch}{1.2}
 \begin{adjustbox}{width=\textwidth}
 \begin{threeparttable}
\begin{tabular}{ccccccccc}
 \hline
XID  & \multicolumn{3}{c}{\scshape abspow} & \multicolumn{3}{c}{\scshape borus02} &  &  \\  \cline{2-4} \cline{5-7}
& log$F_{2-10~{\rm keV}}$ & log$L^{\rm obs}_{2-10~{\rm keV}}$ & log$L^{\rm int}_{2-10~{\rm keV}}$ & log$F_{2-10~{\rm keV}}$ & log$L^{\rm obs}_{2-10~{\rm keV}}$ & log$L^{\rm int}_{2-10~{\rm keV}}$ & log$L_{6~\mu{\rm m}}$ & $\lambda_{\rm Edd}$    \\
&   (erg cm$^{-2}$ s$^{-1}$)  &  (erg s$^{-1}$) & (erg s$^{-1}$) & (erg cm$^{-2}$ s$^{-1}$) &  (erg s$^{-1}$) & (erg s$^{-1}$) & (erg s$^{-1}$) &  \\
(1) & (2) & (3) & (4) & (5) & (6) & (7) & (8) & (9)   \\
\hline
XMM00059 & ${-13.34^{+0.04}_{-0.04}}$ & ${44.48^{+0.04}_{-0.04}}$ & ${44.54^{+0.05}_{-0.05}}$ & ${-13.36^{+0.04}_{-0.04}}$ & $44.48^{+0.04}_{-0.04}$ & $44.51^{+0.03}_{-0.03}$ & & ${0.18^{+0.39}_{-0.12}}$  \\
XMM00131 & ${-13.64^{+0.06}_{-0.06}}$ & ${44.61^{+0.04}_{-0.05}}$ & ${44.70^{+0.06}_{-0.06}}$ & ${-13.65^{+0.06}_{-0.06}}$ & $44.61^{+0.04}_{-0.06}$ & $44.65^{+0.04}_{-0.04}$ & ${45.11}$ & ${0.27^{+0.58}_{-0.18}}$ \\
XMM00136 & ${-14.02^{+0.08}_{-0.09}}$ & ${44.30^{+0.07}_{-0.07}}$ & ${44.39^{+0.09}_{-0.09}}$ & ${-14.03^{+0.08}_{-0.08}}$ & $44.30^{{+0.07}}_{{-0.07}}$ & $44.34^{+0.06}_{-0.07}$ & ${45.70}$ & ${0.11^{+0.25}_{-0.08}}$ \\
XMM00191 & ${-13.21^{+0.06}_{-0.07}}$ & ${44.77^{+0.05}_{-0.06}}$ & ${45.49^{+0.08}_{-0.09}}$ & ${-13.22^{+0.08}_{-0.06}}$ & $44.78^{+0.04}_{-0.06}$ & ${45.59^{+0.05}_{-0.05}}$ & & ${4.59^{+9.92}_{-3.14}}$ \\
XMM00205 & ${-13.86^{+0.11}_{-0.10}}$ & $43.24^{{+0.09}}_{{-0.09}}$ & ${43.30^{+0.13}_{-0.11}}$ & ${-13.86^{+0.11}_{-0.10}}$ & ${43.24^{+0.09}_{-0.09}}$ & ${43.30^{+0.13}_{-0.11}}$ & ${44.84}$ & ${0.01^{+0.02}_{-0.01}}$ \\
XMM00250 & ${-13.96^{+0.09}_{-0.10}}$ & $44.89^{{+0.08}}_{{-0.09}}$ & ${45.43^{+0.10}_{-0.10}}$ & ${-13.99^{+0.15}_{-0.17}}$ & $44.89^{{+0.08}}_{{-0.09}}$ & ${45.52^{+0.06}_{-0.07}}$ & ${46.22}$ & ${3.67^{+7.93}_{-2.51}}$ \\
XMM00267 & ${-13.35^{+0.07}_{-0.08}}$ & $45.27^{+0.11}_{-0.03}$ & ${45.38^{+0.06}_{-0.06}}$ & ${-13.35^{+0.07}_{-0.08}}$ & $45.27^{+0.03}_{-0.03}$ & $45.39^{{+0.07}}_{{-0.06}}$ & ${45.57}$ & ${2.43^{+5.25}_{-1.66}}$ \\
XMM00359 & $-14.36^{+0.28}_{-0.12}$ & $43.80^{+0.14}_{-0.10}$ & $43.84^{+0.13}_{-0.10}$ & $-14.36^{+0.19}_{-0.11}$ & $43.85^{+0.08}_{-0.15}$ & $43.86^{+0.07}_{-0.10}$ & $45.13$ & $0.03^{+0.07}_{-0.02}$\\
XMM00393 & ${-14.09^{+0.20}_{-0.24}}$ & $44.63^{+0.07}_{-0.16}$ & ${44.94^{+0.27}_{-0.20}}$ & ${-14.10^{+0.20}_{-0.24}}$ & $44.63^{+0.06}_{-0.23}$ & $44.90^{{+0.31}}_{{-0.22}}$ & ${45.62}$ & ${0.55^{+1.18}_{-0.37}}$ \\
XMM00395 & ${-14.05^{+0.07}_{-0.07}}$ & ${44.01^{+0.06}_{-0.06}}$ & ${44.07^{+0.07}_{-0.08}}$ & ${-14.04^{+0.06}_{-0.04}}$ & $44.04^{+0.06}_{-0.07}$ & $44.06^{+0.05}_{-0.05}$ & ${44.87}$ & ${0.05^{+0.12}_{-0.04}}$ \\
XMM00421 & ${-13.99^{+0.08}_{-0.08}}$ & $44.32^{+0.07}_{{-0.07}}$ & ${44.34^{+0.08}_{-0.09}}$ & ${-14.03^{+0.16}_{-0.20}}$ & ${44.32^{+0.07}_{-0.13}}$ & ${44.34^{+0.07}_{-0.07}}$ & ${44.53}$ & ${0.11^{+0.25}_{-0.08}}$ \\
XMM00497 & ${-12.94^{+0.04}_{-0.04}}$ & ${44.38^{+0.03}_{-0.08}}$ & ${44.70^{+0.05}_{-0.06}}$ & ${-12.96^{+0.05}_{-0.03}}$ & $44.38^{+0.03}_{-0.03}$ & ${44.70^{+0.03}_{-0.03}}$ & ${45.24}$ & ${0.31^{+0.66}_{-0.21}}$ \\
XMM00860 & ${-14.08^{+0.23}_{-0.14}}$ & $43.97^{{+0.09}}_{{-0.10}}$ & ${44.11^{+0.15}_{-0.13}}$ & ${-14.09^{+0.23}_{-0.25}}$ & $43.97^{{+0.09}}_{-0.27}$ & $44.06^{+0.08}_{-0.09}$ & ${45.03}$ & ${0.05^{+0.12}_{-0.04}}$ \\
XMM01034 & ${-14.20^{+0.10}_{-0.11}}$ & $44.09^{{+0.08}}_{{-0.09}}$ & ${44.15^{+0.10}_{-0.11}}$ & ${-14.21^{+0.07}_{-0.07}}$ & ${44.07^{+0.07}_{-0.08}}$ & $44.10^{+0.07}_{-0.08}$ & ${44.87}$ & ${0.06^{+0.13}_{-0.04}}$ \\
XMM01279 & ${-13.49^{+0.05}_{-0.06}}$ & ${43.96^{+0.05}_{-0.05}}$ & ${44.01^{+0.06}_{-0.06}}$ & ${-13.45^{+0.04}_{-0.04}}$ & ${43.99^{+0.04}_{-0.04}}$ & ${44.02^{+0.04}_{-0.04}}$ & ${44.46}$ & ${0.05^{+0.11}_{-0.03}}$ \\
XMM01464 & ${-13.87^{+0.10}_{-0.10}}$ & $44.33^{{+0.05}}_{{-0.05}}$ & $44.37^{{+0.06}}_{{-0.06}}$ & ${-13.92^{+0.09}_{-0.09}}$ & $44.33^{{+0.05}}_{{-0.05}}$ & $44.43^{+0.06}_{-0.07}$ & ${44.78}$ & ${0.14^{+0.31}_{-0.10}}$ \\
XMM01723 & ${-13.73^{+0.12}_{-0.14}}$ & ${43.82^{+0.09}_{-0.10}}$ & ${44.19^{+0.26}_{-0.24}}$ & ${-13.75^{+0.11}_{-0.14}}$ & ${43.83^{+0.09}_{-0.10}}$ & $44.19^{+0.12}_{-0.15}$ & ${43.17}$ & ${0.08^{+0.16}_{-0.05}}$ \\
XMM01731 & ${-14.00^{+0.35}_{-0.48}}$ & ${44.46^{+0.09}_{-0.11}}$ & ${44.73^{+0.37}_{-0.20}}$ & ${-14.01^{+0.35}_{-0.19}}$ & $44.46^{+0.10}_{-0.30}$ & $44.69^{+0.08}_{-0.09}$ & ${45.76}$ & ${0.30^{+0.65}_{-0.20}}$ \\
XMM01740 & ${-14.33^{+0.20}_{-0.23}}$ & $44.33^{{+0.14}}_{{-0.17}}$ & ${44.55^{+0.22}_{-0.24}}$ & ${-14.34^{+0.14}_{-0.17}}$ & $44.33^{{+0.14}}_{{-0.17}}$ & $44.51^{+0.14}_{-0.17}$ & ${45.67}$ & ${0.18^{+0.39}_{-0.12}}$ \\
XMM02186 & ${-13.86^{+0.11}_{-0.11}}$ & ${43.93^{+0.09}_{-0.10}}$ & ${43.98^{+0.11}_{-0.12}}$ & ${-13.86^{+0.07}_{-0.08}}$ & ${43.94^{+0.08}_{-0.08}}$ & ${43.98^{+0.08}_{-0.08}}$ & & ${0.04^{+0.10}_{-0.03}}$ \\
XMM02347 & ${-14.03^{+0.09}_{-0.10}}$ & $44.25^{{+0.07}}_{{-0.08}}$ & ${44.41^{+0.10}_{-0.10}}$ & ${-14.04^{+0.09}_{-0.10}}$ & ${44.25^{+0.07}_{-0.08}}$ & ${44.37^{+0.10}_{-0.11}}$ & ${44.77}$ & ${0.12^{+0.27}_{-0.08}}$ \\
XMM02660 & ${-13.70^{+0.07}_{-0.08}}$ & ${44.52^{+0.06}_{-0.06}}$ & ${44.70^{+0.08}_{-0.09}}$ & ${-13.64^{+0.05}_{-0.06}}$ & ${44.52^{+0.06}_{-0.06}}$ & ${44.68^{+0.06}_{-0.06}}$ & ${45.44}$ & ${0.29^{+0.63}_{-0.20}}$ \\
XMM03098 & ${-14.15^{+0.10}_{-0.11}}$ & $44.43^{+0.07}_{{-0.08}}$ & $44.49^{{+0.10}}_{{-0.11}}$ & ${-14.15^{+0.10}_{-0.06}}$ & ${44.43^{+0.07}_{-0.07}}$ & $44.46^{+0.07}_{-0.07}$ & ${45.46}$ & ${0.16^{+0.34}_{-0.11}}$ \\
XMM03153 & ${-13.87^{+0.11}_{-0.12}}$ & $44.88^{+0.04}_{-0.05}$ & ${44.99^{+0.10}_{-0.09}}$ & ${-13.87^{+0.10}_{-0.12}}$ & $44.88^{+0.03}_{-0.13}$ & $44.92^{+0.04}_{-0.04}$ & ${46.17}$ & ${0.58^{+1.26}_{-0.40}}$  \\
XMM03342 & ${-14.41^{+0.38}_{-0.42}}$ & $43.91^{+0.13}_{-0.36}$ & ${43.98^{+0.25}_{-0.20}}$ & ${-14.49^{+0.25}_{-0.18}}$ & $43.91^{{+0.14}}_{{-0.17}}$ & ${43.94^{+0.11}_{-0.13}}$ & ${45.40}$ & ${0.04^{+0.09}_{-0.03}}$ \\
XMM03798 & ${-13.99^{+0.19}_{-0.23}}$ & $44.40^{+0.06}_{-0.13}$ & ${44.61^{+0.15}_{-0.13}}$ & ${-13.99^{+0.19}_{-0.15}}$ & $44.40^{+0.05}_{-0.17}$ & $44.56^{+0.06}_{-0.07}$ & ${45.72}$ & ${0.21^{+0.45}_{-0.14}}$ \\
XMM03900 & ${-13.40^{+0.05}_{-0.05}}$ & ${44.78^{+0.04}_{-0.04}}$ & ${44.97^{+0.06}_{-0.06}}$ & ${-13.41^{+0.05}_{-0.05}}$ & ${44.78^{+0.04}_{-0.04}}$ & ${44.93^{+0.04}_{-0.04}}$ & ${45.26}$ & ${0.60^{+1.29}_{-0.41}}$ \\
XMM03916 & ${-13.74^{+0.13}_{-0.14}}$ & ${44.29^{+0.20}_{-0.26}}$ & ${45.48^{+0.14}_{-0.16}}$ & ${-13.75^{+0.12}_{-0.14}}$ & $44.30^{+0.13}_{-0.20}$ & $45.79^{+0.12}_{-0.14}$ & ${46.32}$ & ${8.77^{+18.96}_{-6.00}}$ \\
XMM04259 & ${-13.81^{+0.18}_{-0.22}}$ & $44.15^{+0.18}_{-0.09}$ & ${44.48^{+0.33}_{-0.25}}$ & ${-13.81^{+0.22}_{-0.15}}$ & $44.15^{{+0.09}}_{-0.10}$ & $44.46^{+0.09}_{-0.10}$ & ${45.31}$ & ${0.16^{+0.34}_{-0.11}}$ \\
XMM04404 & $-13.77^{+0.06}_{-0.07}$ & $44.42^{+0.08}_{-0.09}$ & $44.45^{+0.08}_{-0.09}$ & $-13.80^{+0.06}_{-0.07}$ & $44.42^{+0.06}_{-0.07}$ & $44.45^{+0.06}_{-0.07}$ & $45.33$ & $0.15^{+0.33}_{-0.10}$ \\
XMM04475 & ${-14.38^{+0.13}_{-0.15}}$ & $44.26^{+0.09}_{-0.14}$ & ${44.28^{+0.17}_{-0.18}}$ & ${-14.37^{+0.12}_{-0.11}}$ & $44.26^{+0.07}_{-0.18}$ & $44.29^{+0.10}_{-0.12}$ & ${44.96}$ & ${0.10^{+0.21}_{-0.07}}$ \\
XMM04583 & ${-13.58^{+0.08}_{-0.09}}$ & ${44.12^{+0.07}_{-0.07}}$ & ${44.25^{+0.08}_{-0.09}}$ & ${-13.59^{+0.08}_{-0.09}}$ & ${44.11^{+0.07}_{-0.07}}$ & ${44.21^{+0.06}_{-0.06}}$ & & ${0.08^{+0.17}_{-0.05}}$ \\
XMM04804 & ${-13.95^{+0.18}_{-0.22}}$ & $44.28^{+0.07}_{-0.07}$ & ${44.44^{+0.14}_{-0.12}}$ & ${-13.95^{+0.06}_{-0.07}}$ & ${44.28^{+0.07}_{-0.07}}$ & ${44.39^{+0.06}_{-0.07}}$ & ${43.18}$ & ${0.13^{+0.28}_{-0.09}}$ \\
XMM04899 & ${-13.76^{+0.08}_{-0.08}}$ & ${44.29^{+0.06}_{-0.05}}$ & ${44.56^{+0.08}_{-0.09}}$ & ${-13.77^{+0.08}_{-0.07}}$ & ${44.29^{+0.05}_{-0.06}}$ & ${44.53^{+0.05}_{-0.06}}$ & ${44.88}$ & ${0.19^{+0.41}_{-0.13}}$ \\
 \hline
\end{tabular}
\begin{tablenotes}
\item Notes -
{The X-ray fluxes and errors are obtained using the `{\scshape cflux}' command in the {\scshape xspec}. The observed and absorption-corrected X-ray luminosities and errors are obtained using the `{\scshape clumin}' command in the {\scshape xspec}.
The $6.0~{\mu}m$ luminosities are gleaned from \cite{Zou22}, who derived them from the broad-band SED modelling.
The large error bars introduced in ${\lambda}_{\rm Edd}$ correspond to the assumed one dex spread (10$^{8}$ $-$ 10$^{9}$ $M_{\odot}$)
in $M_{\rm BH}$.}
\end{tablenotes}
\end{threeparttable}
\end{adjustbox}
\label{tab:XrayFluxLumin}
\end{table*}
%
%
%
\label{lastpage}
\end{document}